\documentclass[twocolumn,showpacs,amsmath,amssymb,prd,floatfix]{revtex4}
\usepackage{graphicx}
\usepackage{dcolumn}
\usepackage{bm}
\usepackage{epsf}
\usepackage{float}

\usepackage{multirow}
\usepackage{mathrsfs}

\def\agt{\mathrel{\raise.3ex\hbox{$>$}\mkern-14mu\lower0.6ex\hbox{$\sim$}}}
\def\alt{\mathrel{\raise.3ex\hbox{$<$}\mkern-14mu\lower0.6ex\hbox{$\sim$}}}

\newcommand{\beq}{\begin{equation}}
\newcommand{\eeq}{\end{equation}}
\newcommand{\beqn}{\begin{eqnarray}}
\newcommand{\eeqn}{\end{eqnarray}}

\begin{document}

\title{Remnant massive neutron stars of binary neutron star mergers: \\
Evolution process and gravitational waveform}

\author{Kenta Hotokezaka$^1$}
\author{Kenta Kiuchi$^2$}
\author{Koutarou Kyutoku$^3$}
\author{Takayuki Muranushi$^{2,4}$}
\author{Yu-ichiro Sekiguchi$^2$}
\author{Masaru Shibata$^2$}
\author{Keisuke Taniguchi$^5$}

\affiliation{
$^1$Department of Physics, Kyoto University, Kyoto 606-8502, Japan \\
$^2$Yukawa Institute for Theoretical Physics, Kyoto University,
Kyoto 606-8502, Japan \\
$^3$Department of Physics, Universitiy of Wisconsin-Milwaukee, P.O. Box 413,
Milwaukee, Wisconsin 53201, USA\\
$^4$Hakubi Center for Advanced Research,  Kyoto University,
Kyoto 606-8501 Japan \\
$^5$Graduate School of Arts and Sciences, The University of Tokyo,
Tokyo 153-8902, Japan
}

\begin{abstract}
Massive (hypermassive and supramassive) neutron stars are likely to be
often formed after the merger of binary neutron stars. We explore the
evolution process of the remnant massive neutron stars and
gravitational waves emitted by them, based on numerical-relativity
simulations for binary neutron star mergers employing a variety of
equations of state and choosing a plausible range of the neutron-star
mass of binaries. We show that the lifetime of remnant hypermassive
neutron stars depends strongly on the total binary mass and also on
the equations of state.  Gravitational waves emitted by the remnant
massive neutron stars universally have a quasiperiodic nature of an
approximately constant frequency although the frequency varies with
time. We also show that the frequency and time-variation feature of
gravitational waves depend strongly on the equations of state.  We
derive a fitting formula for the quasiperiodic gravitational
waveforms, which may be used for the data analysis of a
gravitational-wave signal.
\end{abstract}

\maketitle

\section{Introduction} \label{secI}

Coalescence of binary neutron stars is one of the most promising
sources for next-generation kilo-meter-size gravitational-wave
detectors such as advanced LIGO, advanced VIRGO, and KAGRA
(LCGT)~\cite{LIGOVIRGO}. The first detection of gravitational waves
will be achieved in the next $\sim 5$\,years by observing a chirp
signal of gravitational waves emitted in the so-called inspiral stage
in which binary neutron stars are in quasicircular orbits with orbital
radius 30\,--\,700\,km (the gravitational-wave frequency in the range
$\approx 10$\,--\,$10^3$\,Hz).  Statistical studies have predicted
that the detection rate of gravitational waves for this signal will be
$\sim 1$\,--\,100 per year (e.g., ~\cite{Rate,RateLIGO}).
After the first detection of gravitational waves from coalescence of
a binary neutron star, it will be a great challenge to extract the
information of matter effects from a gravitational-wave signal.
For instance, a tidal deformability parameter will be measurable
with the gravitational-wave signal during the inspiral stage~\cite{tidal}. 

After the merger of a binary neutron star sets in, there are two
possible fates: If the total mass is large enough, a black hole is
promptly formed, while if not, a massive neutron star (MNS) is formed.
(Here, a massive neutron star means hypermassive or supramassive
neutron star: see~\cite{BSS00} and \cite{CST92,CST94} for their
definitions, respectively; see also Sec.~\ref{sec:MNS}).
Numerical-relativity simulations have shown that the threshold mass
depends strongly on the equation of state (EOS) of the neutron-star
matter~\cite{STU,hotoke,skks2011a,skks2011b,hotoke2012}, which is sill
poorly known to date~\cite{lattimerprakash}. However, the latest
discoveries of high-mass neutron stars with mass $1.97 \pm
0.04M_{\odot}$~\cite{twosolar} and $2.01\pm0.04M_{\odot}$~\cite{twosolar2}
constrain that the maximum mass of
(cold) spherical neutron stars for a given hypothetical EOS has to be
larger than $\sim 2M_{\odot}$. This suggests that the EOS of neutron
stars has to be stiff; the pressure above the nuclear-matter density
$\sim 2.8 \times 10^{14}\,{\rm g/cm^3}$ has to be sufficiently high.
Motivated by this fact, we performed numerical-relativity simulations
for a variety of stiff EOSs in previous
papers~\cite{hotoke,hotoke2012}, and found that a MNS is the universal
outcome for the binary of total mass smaller than the typical mass
$\sim 2.6$\,--\,$2.8M_{\odot}$.  The purpose of this paper is to
summarize our latest more systematic studies for the evolution process
of the MNSs and quasiperiodic gravitational waves emitted by the MNSs
formed after the merger of binary neutron stars. 


In the past decade, the numerical simulation for the merger of binary
neutron stars in full general relativity, which is the unique approach
of the rigorous theoretical study for this subject, has been
extensively performed since the first success in 2000~\cite{SU00}
(see, e.g.,~\cite{Duez,FR2012} for a review of this field).  However,
most of the simulations have been performed with simple polytropic
EOSs (but see, e.g.,~\cite{PEOS1,PEOS2,PEOS3} for the latest
progress).  For the detailed and physical study of the merger remnants
and gravitational waves emitted by remnant MNSs, we have to employ
physical EOSs. In the past two years, we have performed a number of
simulations using piecewise polytropic EOSs~\cite{hotoke,hotoke2012}
and tabulated finite-temperature EOSs taking into account a neutrino
cooling process~\cite{skks2011a,skks2011b,kkss2013} for a variety of
masses of binary systems (see also~\cite{BJ2012}).  We now have a
number of numerical results; a variety of the sample for remnant MNSs
and possible gravitational waves emitted by them. By analyzing these
samples, we can now summarize possible evolution processes of the
remnant MNSs and the resulting gravitational waveforms.  Furthermore,
a variety of numerical gravitational waveforms enable us to construct
an analytic model for such gravitational waveforms. In this paper, we
report the results of our exploration for these issues. 

The paper is organized as follows: In Sec.~II, we summarize the EOSs
employed in our latest numerical-relativity simulations, and models of
binary neutron stars for which MNSs are formed.  In Sec.~III, we
describe the properties and possible evolution processes of the MNSs.
Section IV summarizes the properties of gravitational waves emitted by
MNSs.  Section~V is devoted to deriving analytic formulas for modeling
gravitational waves emitted by the MNSs. Section~VI is devoted to a
summary and discussion.  Throughout this paper, we employ the
geometrical units $c=1=G$ where $c$ and $G$ are the speed of light and
gravitational constant, respectively, although we recover $c$ when we
need to clarify the units \footnote{We assume general relativity is
the correct theory of gravity on the scale of neutron stars.}.

\section{Equations of state and chosen models} \label{secII}

\subsection{Equations of state} \label{sec:EOS}

In this section, we summarize the EOSs employed in our latest
studies~\cite{skks2011a,skks2011b,hotoke2012,kkss2013} and in the
present work. 

\begin{table*}[t]
\caption{Parameters and key quantities for five piecewise polytropic
EOSs and finite-temperature (Shen) EOSs employed so far.  $P_2$ is
shown in units of ${\rm dyn/cm}^2$.  $M_{\rm max}$ is the maximum-mass
along the sequences of cold spherical neutron stars.  ($R_{1.35},
\rho_{1.35}$), ($R_{1.5}, \rho_{1.5}$), ($R_{1.6}, \rho_{1.6}$), 
and ($R_{1.8}, \rho_{1.8}$) are the circumferential radius in units of
km and the central density in units of ${\rm g/cm^3}$ for
$1.35M_{\odot}$, $1.5M_{\odot}$, $1.6M_{\odot}$, and $1.8M_{\odot}$
neutron stars, respectively.  We note that the values of the mass,
radius, and density listed for the piecewise polytropic EOSs are
slightly different from those obtained in the original tabulated EOSs
(see the text for the reason).  MS1 is referred to as this name
in~\cite{rlof2009}, but in other references
(e.g.,~\cite{lattimerprakash}), it is referred to as MS0. We
follow~\cite{rlof2009} in this paper.  The fitted parameters
$(\log(P_2), \Gamma_1, \Gamma_2, \Gamma_3)$ are taken
from~\cite{rlof2009}.  }
{\begin{tabular}{ccccccccccc} \hline
EOS &  $(\log(P_2), \Gamma_1, \Gamma_2, \Gamma_3)$ & $M_{\rm max}(M_{\odot})$ &
$~R_{1.35}~$ & $~\rho_{1.35}~$ &
$~R_{1.5} ~$ & $~\rho_{1.5} ~$ &
$~R_{1.6} ~$ & $~\rho_{1.6} ~$ &
$~R_{1.8} ~$ & $~\rho_{1.8} ~$
\\ \hline
APR4 & $(34.269, 2.830, 3.445, 3.348)$ & 2.20
& 11.1 & $8.9 \times 10^{14}$ & 11.1 & $9.6 \times 10^{14}$ 
& 11.1 & $10.1 \times 10^{14}$ 
& 11.0 & $11.4 \times 10^{14}$ \\
SLy  & $(34.384, 3.005, 2.988, 2.851)$ & 2.06 
& 11.5 & $8.6 \times 10^{14}$ & 11.4 & $9.5 \times 10^{14}$ 
& 11.4 & $10.2 \times 10^{14}$ 
& 11.2 & $12.0 \times 10^{14}$  \\
ALF2 & $(34.616, 4.070, 2.411, 1.890)$ & 1.99
& 12.4 & $6.4 \times 10^{14}$ & 12.4 & $7.2 \times 10^{14}$ 
& 12.4 & $7.8 \times 10^{14}$ 
& 12.2 & $9.5 \times 10^{14}$ \\
H4   & $(34.669, 2.909, 2.246, 2.144)$ & 2.03
& 13.6 & $5.5 \times 10^{14}$ & 13.5 & $6.3 \times 10^{14}$ 
& 13.5 & $6.9 \times 10^{14}$ 
& 13.1 & $8.7 \times 10^{14}$ \\
MS1  & $(34.858, 3.224, 3.033, 1.325)$ & 2.77
& 14.4 & $4.2 \times 10^{14}$ & 14.5 & $4.5 \times 10^{14}$
& 14.6 & $4.7 \times 10^{14}$ 
& 14.6 & $5.1 \times 10^{14}$ \\
Shen      & (34.717, ---, ---, ---) & 2.20 
& 14.5 & $4.4 \times 10^{14}$ & 14.4 & $4.9\times 10^{14}$ 
& 14.4 & $5.8 \times 10^{14}$  
& 14.2 & $6.7 \times 10^{14}$  \\
\hline
\end{tabular}
}
\label{table:EOS}
\end{table*}

The exact EOS for the high-density nuclear matter is still
unknown~\cite{lattimerprakash}, and hence, a numerical simulation
employing a single particular EOS, which might not be correct, would
not yield a scientific result in this field. Simulations
systematically employing a wide variety of possible hypothetical EOSs
are required for exploring the merger of binary neutron
stars. Nevertheless, the latest discoveries of a high-mass neutron star
PSR J1614-2230 with mass $1.97 \pm 0.04 M_{\odot}$~\cite{twosolar}
and PSR J0348+0432 with $2.01 \pm 0.04M_{\odot}$~\cite{twosolar2}
significantly constrain the hypothetical EOSs to be chosen, because
these suggest that the maximum allowed mass of (cold) spherical neutron
stars for a given EOS (hereafter denoted by $M_{\rm max}$) has to be
larger than $\sim 2M_{\odot}$.  This indicates that the EOS should be
rather stiff, although there are still many candidate
EOSs~\cite{lattimerprakash}.

To this time, we have employed two types of EOSs. One is a piecewise
polytropic EOS proposed by Read and her collaborators~\cite{rlof2009}
(described below) and the other is a tabulated EOS in which
finite-temperature effects together with the effects associated with
the electron fraction per baryon is taken into account.  In this
study, we analyze numerical results obtained in the so-called Shen
EOS~\cite{skks2011a} which was derived from a relativistic mean-field
theory~\cite{Shen} assuming that neutron stars are composed of normal
nuclear matter such as protons and neutrons. This EOS is stiff and
yields $M_{\rm max}=2.2M_{\odot}$ (see Table~\ref{table:EOS}).  We
also employed the hyperonic version of Shen EOS~\cite{ShenHyp}
in~\cite{skks2011b}. However, this EOS is rather soft with $M_{\rm
max}=1.75M_{\odot}$, and hence, we do not adopt the numerical results
for this EOS in this paper. 

The piecewise polytropic EOS is described assuming that neutron stars
are cold (in a zero-temperature state); the rest-mass density, $\rho$,
determines all other thermodynamical quantities.  In the prescription
of~\cite{rlof2009}, there are the following four parameters; $P_2$:
the pressure at $\rho:=\rho_2=10^{14.7}\,{\rm g/cm^3}$ and $(\Gamma_1,
\Gamma_2, \Gamma_3)$: the adiabatic indices that characterize the EOS
for the nuclear matter (see, e.g.,~\cite{hotoke2012} for the detail).
Table~\ref{table:EOS} lists parameters of the five piecewise
polytropic EOSs which we employed and which are representative EOSs
derived in nuclear theories.  The values of ($P_2,\Gamma_1, \Gamma_2,
\Gamma_3$) are taken from~\cite{rlof2009}. The properties of these
EOSs are described in~\cite{hotoke2012}
(see~\cite{APR4,SLy,ALF2,H4,MS1} for APR4, SLy, ALF2, H4, and MS1,
respectively.)

\begin{figure}[t]
\includegraphics[width=90mm,clip]{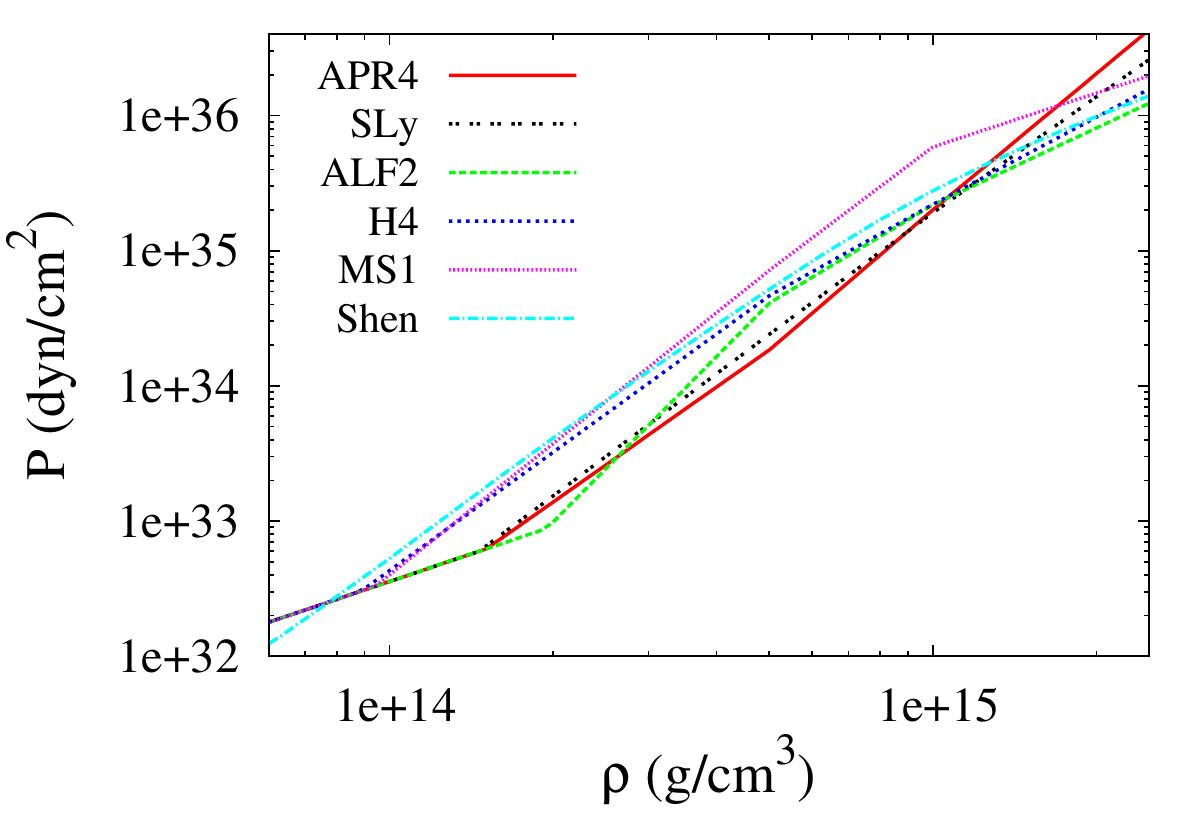}
\vspace{-7mm}
\caption{Pressure as a function of the rest-mass density for seven
EOSs listed in Table~\ref{table:EOS}.  }
\label{fig1}
\end{figure}

Figure~\ref{fig1} plots the pressure as a function of the rest-mass
density for five piecewise polytropic EOSs as well as for Shen
EOS. APR4 and SLy have relatively small pressure for $\rho \alt
\rho_3:=10^{15}\,{\rm g/cm^3}$, while they have high pressure for
$\rho \agt \rho_3$.  By contrast, H4, MS1, and Shen have pressure
higher than APR4 and SLy for $\rho \alt \rho_3$, while they become
softer for a high-density region $\rho \agt \rho_3$.  MS1 has the
highest pressure for $\rho \alt \rho_3$ (i.e., the highest value of
$P_2$) among many other EOSs, and thus, it is the stiffest EOS as far
as the canonical neutron stars are concerned.  ALF2 has small pressure
for $\rho \leq \rho_2$ as in the case of APR4, but for $\rho_2 \alt
\rho \leq \rho_3$, the pressure is higher than that for APR4. For
$\rho \geq \rho_2$ the pressure of ALF2 is as high as that for H4. The
profile of Shen is similar to H4 although Shen has slightly higher
pressure than H4 for a given value of the density.  Also, APR4 and SLy
are similar EOSs, but the slight difference between two EOSs results
in a significant difference in the merger remnants for canonical-mass
binary neutron stars (see Sec.~III). 

All the properties mentioned above are reflected in the radius,
$R_{1.35}$, and central density, $\rho_{1.35}$, of (cold) spherical
neutron stars with the canonical mass $M=1.35M_{\odot}$ \cite{Stairs}
where $M$ is
the gravitational (Arnowitt-Deser-Misner; ADM) mass of the cold
spherical neutron stars in isolation: see Table~\ref{table:EOS}. The
pressure at $\rho=\rho_2$ (i.e., $P_2$) is correlated well with this
radius and central density~\cite{lattimerprakash2001} (see also
below).

In the numerical simulation, we used a modified version of the
piecewise polytropic EOS to approximately take into account thermal
effects. In this EOS, the pressure and specific internal energy are
decomposed into cold and thermal parts as
\begin{equation}
 P = P_{\rm cold} + P_{\rm th} \; , \; \varepsilon = \varepsilon_{\rm
  cold} + \varepsilon_{\rm th} .
\end{equation}
The cold parts of both variables are calculated using the original
piecewise polytropic EOS from the primitive variable $\rho$, and then
the thermal part of the specific internal energy is defined from
$\varepsilon$ as $\varepsilon_{\rm th} = \varepsilon -
\varepsilon_{\rm cold}(\rho)$. Because $\varepsilon_{\rm th}$ vanishes in
the absence of shock heating, it is regarded as the finite-temperature
part determined by the shock heating in the present context.  For the
thermal part of the pressure and specific internal energy, a
$\Gamma$-law ideal-gas EOS was adopted as
\begin{equation}
 P_{\rm th} = ( \Gamma_{\rm th} - 1 ) \rho \varepsilon_{\rm th}.
\end{equation}
Following the conclusion of a detailed study in~\cite{BJO2010},
$\Gamma_{\rm th}$ is chosen in the range 1.6\,--\,2.0 with the canonical
value 1.8. For several models, simulations were performed
varying the value of $\Gamma_{\rm th}$ (see Table~\ref{table:ID}). 

\begin{table*}
\caption{List of the simulation models in which a MNS is formed.  The
model is referred to as the name ``EOS''-``$m_1$''``$m_2$''; e.g., the
model employing APR4, $m_1=1.3M_{\odot}$, and $m_2=1.4M_{\odot}$ is
referred to as model APR4-130140.  Second\,--\,fourth columns show the
adiabatic index for the thermal pressure for the piecewise polytropic
EOS and masses of two components.  The last three columns show the
numerical results; approximate lifetime of the MNS that was found in
our simulation time, the rest mass of disks surrounding the remnant
black hole, and final gravitational mass of the system. --- denotes
that the lifetime of the MNS is much longer than 30~ms and we did not
find black-hole formation in our simulation time. The disk mass is
measured at 10\,ms after the formation of the black hole. We note that
a black hole is formed soon after the onset of the merger with $m \geq
2.9M_{\odot}$ for APR4 and with $m \geq 2.8M_{\odot}$ for SLy. For
ALF2 with $m=2.9M_{\odot}$, a MNS is formed after the merger but its
lifetime is quite short, $< 5$~ms.}
{\begin{tabular}{ccccccc} \hline
Model & $\Gamma_{\rm th}$ & $m_1 (M_{\odot})$ & $m_2 (M_{\odot})$ 
& Lifetime (ms) & disk mass ($M_{\odot}$)
& final mass ($M_{\odot}$)\\ \hline \hline
APR4-130150 & 1.8 & 1.30 & 1.50  & 30 & 0.12 &
2.69 \\
APR4-140140 & 1.8 & 1.30 & 1.50  & 35 & 0.12 &
2.69 \\
APR4-120150 & 1.6, 1.8, 2.0 & 1.20 & 1.50 & ---
& --- & 2.60, 2.59, 2.59 \\
APR4-125145 & 1.8 & 1.25 & 1.45 & --- & --- &
2.60 \\
APR4-130140 & 1.8 & 1.30 & 1.40 & --- & --- &
2.60 \\
APR4-135135 & 1.6, 1.8, 2.0 & 1.35 & 1.35 & --- 
& --- & 2.59, 2.61, 2.60 \\
APR4-120140 & 1.8 & 1.20 & 1.40 & --- & --- &
2.52 \\
APR4-125135 & 1.8 & 1.25 & 1.35 & --- & --- &
2.53 \\
APR4-130130 & 1.8 & 1.30 & 1.30 & --- & --- &
2.53 \\ 
\hline
SLy-120150& 1.8  & 1.20 & 1.50 & 10 & 0.12 & 2.60\\
SLy-125145& 1.8  & 1.25 & 1.45 & 15 & 0.14 & 2.60 \\
SLy-130140& 1.8  & 1.30 & 1.40 & 15 & 0.11 & 2.60 \\
SLy-135135& 1.8  & 1.35 & 1.35 & 10 & 0.08 & 2.58 \\
SLy-130130& 1.8  & 1.30 & 1.30 & --- & --- & 2.51 \\
\hline
ALF2-145145&1.8 & 1.45 & 1.45 & 2  & 0.04 &2.84 \\
ALF2-140140&1.8 & 1.40 & 1.40 & 5  & 0.07 &2.72 \\
ALF2-120150&1.8 & 1.20 & 1.50 & 45 & 0.31 &2.63 \\
ALF2-125145&1.8 & 1.25 & 1.25 & 40 & 0.23 &2.63 \\
ALF2-130140&1.8 & 1.30 & 1.40 & 10 & 0.12 &2.63 \\
ALF2-135135&1.8 & 1.35 & 1.35 & 15 & 0.17 &2.62 \\
ALF2-130130&1.8 & 1.30 & 1.30 & --- & --- &2.54 \\
\hline
H4-130160  & 1.8  & 1.30 & 1.60 & 5  & 0.12 & 2.83 \\
H4-145145  & 1.8  & 1.45 & 1.45 & 5  & 0.03 & 2.81  \\
H4-130150  & 1.8  & 1.30 & 1.50 & 20 & 0.25 & 2.72 \\
H4-140140  & 1.8  & 1.40 & 1.40 & 10 & 0.03 & 2.72 \\
H4-120150  & 1.6, 1.8, 2.0  & 1.20 & 1.50 & --- &
--- & 2.65, 2.64, 2.64\\
H4-125145  & 1.8  & 1.25 & 1.25 & --- & ---& 2.63\\
H4-130140  & 1.8  & 1.30 & 1.40 & --- & ---& 2.62 \\
H4-135135  & 1.6, 1.8, 2.0  & 1.35 & 1.35 & 15, 25,
35 & 0.08, 0.08, 0.08 & 2.62, 2.62, 2.62\\
H4-120140  & 1.8  & 1.30 & 1.30 & --- & --- & 2.54 \\
H4-125135  & 1.8  & 1.30 & 1.30 & --- & --- & 2.55 \\
H4-130130  & 1.8  & 1.30 & 1.30 & --- & --- & 2.53 \\ 
\hline
MS1-130160  & 1.8  & 1.30 & 1.60 & --- & --- &2.85\\
MS1-145145  & 1.8  & 1.45 & 1.45 & --- & --- &2.85\\
MS1-140140  & 1.8  & 1.40 & 1.40 & --- & --- &2.75\\
MS1-120150  & 1.8  & 1.20 & 1.50 & --- & --- &2.65\\
MS1-125145  & 1.8  & 1.25 & 1.25 & --- & --- &2.66\\
MS1-130140  & 1.8  & 1.30 & 1.40 & --- & --- &2.66\\
MS1-135135  & 1.8  & 1.35 & 1.35 & --- & --- &2.65\\
MS1-130130  & 1.8  & 1.30 & 1.30 & --- & --- &2.56\\
\hline
Shen-120150 & ---  & 1.20 & 1.50 & --- & ---&
2.64 \\
Shen-125145 & ---  & 1.25 & 1.45 & --- & ---&
2.61 \\
Shen-130140 & ---  & 1.30 & 1.40 & --- & ---&
2.63 \\
Shen-135135 & ---  & 1.35 & 1.35 & --- & ---&
2.62 \\
Shen-140140 & ---  & 1.40 & 1.40 & --- & ---&
2.74 \\
Shen-150150 & ---  & 1.50 & 1.50 & --- & ---&
2.95 \\
Shen-160160 & ---  & 1.60 & 1.60 & 10 & 0.10&
3.12 \\
\hline
\end{tabular}
}
\label{table:ID}
\end{table*}

\subsection{Models}\label{sec:ID}

Numerical simulations were performed for a variety of EOSs and for
many sets of masses of binary neutron stars.  Because the mass of each
neutron star in the observed binary systems is in a narrow range
between $\sim 1.23$\,--\,$1.45M_{\odot}$~\cite{Stairs}, we basically
choose the neutron-star mass $1.20$, $1.25$, $1.30$, $1.35$, $1.40$,
$1.45$, and $1.5M_{\odot}$ with the canonical mass $1.35M_{\odot}$
(the canonical total mass $m=m_1+m_2=2.7M_{\odot}$).  Also, the mass
ratio of the observed system $q:=m_1/m_2 (\leq 1)$ where $m_1$ and
$m_2$ denote the mass of lighter and heavier neutron stars,
respectively, is in a narrow range $\sim 0.85$\,--\,1. Thus, we choose
$q$ as $0.8 \leq q \leq 1$. The models employed in the present
analysis are listed in Table~\ref{table:ID}. In the following, we
specify the model by the names listed in this table. We note that for
APR4 and SLy, a black hole is formed promptly after the onset of the
merger for $m \geq 2.9M_{\odot}$ and $m \geq 2.8M_{\odot}$,
respectively. (Although they are not listed in Table~\ref{table:ID},
we performed a simulation for $m_1=m_2=1.4M_\odot$ with SLy and three
simulations for $m=2.9M_\odot$ with APR4.) 

We found that for models with $m \leq 2.7M_{\odot}$, MNSs were always
formed irrespective of the EOSs employed. Even for $m = 2.8M_{\odot}$,
MNSs were formed for all the EOSs except for SLy.  For stiffer EOSs,
MNSs can be formed even for $m=3M_{\odot}$ (e.g., for Shen and MS1
EOSs). Thus, for H4, MS1, and Shen, we performed simulations for
higher-mass models with $m_2=1.6M_{\odot}$.  The models in which MNSs
are formed are summarized in Table~\ref{table:ID}. In Secs.~\ref{sec3}
and~\ref{sec:GW}, we will analyze the evolution process of the MNSs
and the waveform of emitted gravitational waves, derived for these
models.

Numerical simulations with the piecewise polytropic EOSs were
performed using an adaptive-mesh refinement code {\tt
SACRA}~\cite{yst2008}. For these simulations, the semi-major diameter
of neutron stars is initially covered by $\approx 100$ grid points (we
refer to this grid resolution as high resolution).  We also performed
lower-resolution simulations covering the the semimajor axis by
$\approx 65$ and 80 grid points (we refer to these grid resolutions as
low and middle resolutions).  The accuracy and convergence of the
numerical results for the high grid resolution is found
in~\cite{hotoke2012} and in Appendix A; e.g., the averaged frequency of
gravitational waves emitted from MNSs is determined within $\sim
0.1$\,kHz error.  Numerical simulations with Shen EOS were performed
using a code developed in~\cite{skks2011a,skks2011b}.  For these
simulations, the semi-major diameter of neutron stars is initially 
covered by $\approx 80$ grid points.  The accuracy and convergence of 
the numerical results for the high grid resolution would be slightly 
poorer than those in the piecewise polytropic EOS case.
For comparison, we performed simulations using these two codes with
the same total mass and EOS, H4-135135. Then, we found that
the averaged frequency of gravitational waves emitted by HMNS
agrees with each other within $1\%$ accuracy.   
For all these simulations, the initial data were prepared by a code described
in~\cite{TS2010}, which was developed from LORENE
library~\cite{LORENE}.

\subsection{Hypermassive and supramassive neutron stars}\label{sec:MNS}

Before going ahead, we remind the readers the definitions of the
hypermassive~\cite{BSS00} and supramassive neutron
stars~\cite{CST92,CST94}. We note that in the definitions, we suppose
that neutron stars are cold (i.e., finite-temperature effects are
negligible).

\begin{figure}[t]
\includegraphics[width=84mm,clip]{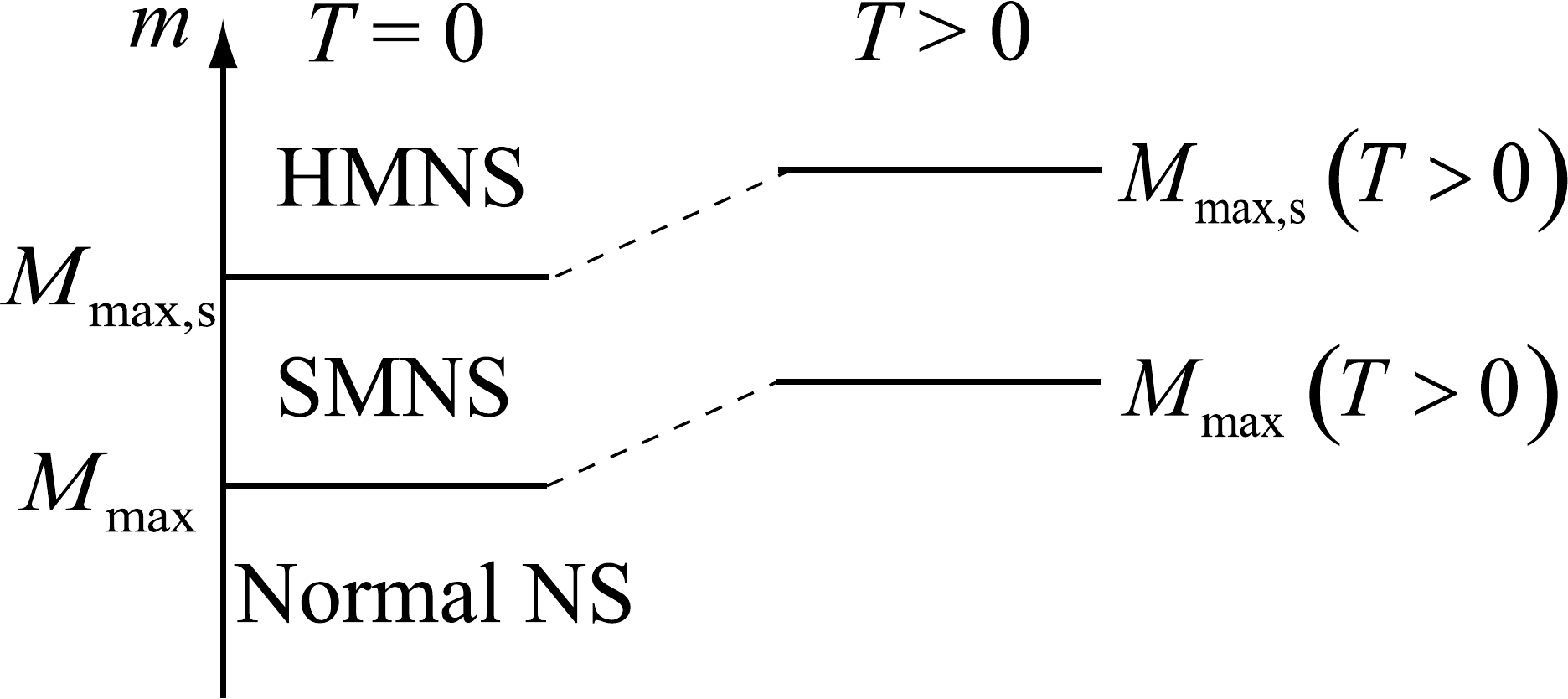}
\caption{A schematic diagram for explaining HMNS and SMNS. 
$M_{\rm max}$ has to be larger than $\sim 2M_{\odot}$ and for such
stiff EOSs, we empirically know that $M_{\rm max,s} \approx 1.2M_{\rm
max}$~\cite{CST94}.  $T$ denotes the typical temperature of MNS.  See
the text for details. }
\label{fig:HMNS0}
\end{figure}

A sequence of spherical neutron stars always has the maximum mass
state.  Rotation can increase the maximum mass of neutron stars. The
maximum mass of a {\em uniformly} rotating neutron star (hereafter
denoted by $M_{\rm max,s}$) is determined by the spin rate at which
the mass element at the equator rotates with the (general
relativistic) Kepler velocity and further speed-up would lead to mass
shedding. This maximum mass for cold neutron stars was determined
numerically for a variety of nuclear-theory-based EOSs
in~\cite{CST94}, which shows that $M_{\rm max,s}$ is by 15\,--\,20\%
larger than $M_{\rm max}$. For stiff EOSs in which $M_{\rm max} >
2M_{\odot}$, it is empirically known that this increase factor is
$\sim 20\%$. Rotating neutron stars with their rest mass exceeding the
maximum rest mass of nonspinning neutron stars for a given EOS are
referred to as supramassive neutron stars (SMNSs)~\cite{CST92}. A
uniformly rotating (cold) neutron star with mass exceeding $\approx
M_{\rm max,s}$ will collapse to a black hole.  However, a uniformly
rotating (cold) neutron star with mass between $M_{\rm max}$ and
$M_{\rm max,s}$ may be alive, and it will collapse to a black hole
only for the case that a process for the dissipation of its angular
momentum is present. The typical dissipation process is an 
electromagnetic emission such as magnetic dipole radiation, for which
the dissipation time scale is much longer than 1\,s for
the typical magnetic-field strength $\sim 10^{12}$\,G and the allowed
spin period ($\agt 1$\,ms) of the neutron stars.

The merger of binary neutron stars does not result in general in a
uniformly rotating remnant, but in a {\em differentially} rotating
one. The inner region of the remnant MNSs often rotates faster than
the envelope: The angular velocity in the inner region can be larger
than the Kepler one of the equator as first found in~\cite{SU00}. This
implies that the centrifugal force, which can have the significant
contribution to supporting the strong self-gravity of the merger
remnant, is enhanced, and thus, the maximum allowed mass of the
remnant MNS can exceed the maximum allowed mass of SMNSs for a
given EOS, $M_{\rm max,s}$.  Such differentially rotating neutron
stars with their mass exceeding $M_{\rm max,s}$ are referred to
as hypermassive neutron stars (HMNSs). Therefore, a differentially
rotating (cold) neutron star with mass exceeding $\approx M_{\rm
max,s}$ does not always have to collapse to a black hole.

The left side of Fig.~\ref{fig:HMNS0} schematically shows the
definition of the HMNS and SMNS. The neutron stars with the mass
smaller than $M_{\rm max}$ are referred to as normal neutron stars in
this figure.

If the degrees of the differential rotation is significantly reduced
by some angular-momentum transport or dissipation processes, HMNSs
will be unstable against gravitational collapse. There are several
possible processes for transporting angular momentum
(e.g.,~\cite{DLSS2004,DLSSS2006a,AEI}). One is the purely hydrodynamics
effect. This becomes an efficient process for HMNSs formed soon after
the merger of binary neutron stars, because such HMNSs usually have a
nonaxisymmetric structure and exert the torque to the matter in the
envelope. Then, the angular momentum in the inner part of the HMNSs is
transported outward, and as the decrease of the angular momentum in
the inner part, the degrees of the differential rotation is
reduced. In the present simulations as well as our simulations
of~\cite{skks2011a,skks2011b,hotoke2012}, only this process is taken
into account, but this is really an efficient process in particular
for the early evolution stage of the HMNSs in which the degrees of
nonaxisymmetry is quite high.

There are two other possible effects for the angular-momentum
transport, both of which are activated in the presence of magnetic
fields.  One is the magnetic winding effect (e.g.,~\cite{BSS00}) for
which the order of the angular-momentum transport time scale is
\beq
\tau_{\rm wind} \sim {R \over v_A} \sim 10^2 \rho_{15}^{1/2}
B_{15}^{-1} R_6\,{\rm ms},
\eeq
where $R$ is the typical radius of the HMNS with $R_6=R/(10^6\,{\rm
cm})$, and $v_A$ is the Alfv{\'e}n velocity
\beqn
v_A \approx {B \over \sqrt{4\pi\rho}}. 
\eeqn
Here, $B$ is the typical magnitude of the radial component of magnetic
fields with $B_{15}=B/(10^{15}\,{\rm G})$, and $\rho$ is the typical
density with $\rho_{15}=\rho/(10^{15}\,{\rm g/cm^3})$. Thus, for the
sufficiently high magnetic-field strength which could be yielded by
the winding itself and compression, the angular-momentum transport is
significantly enhanced.

The other mechanism is the magnetorotational instability
(MRI)~\cite{MRI,MRI2} by which an effective viscosity is likely to be
generated with the effective viscous parameter
\beqn
\nu_{\rm vis} \sim \alpha_{\rm vis} {c_s^2 \over \Omega},
\eeqn
where $\alpha_{\rm vis}$ is the so-called $\alpha$-parameter which
will be 0.01\,--\,0.1~\cite{MRI2}, $c_s$ is the typical sound velocity
of order $\sim 0.1c$, and $\Omega$ is the typical angular velocity
$\sim 10^4$\,rad/s with $\Omega_4=\Omega/(10^4\,{\rm rad/s})$.  Thus,
the viscous angular-momentum transport time scale in the presence of
magnetic fields would be
\beqn
\tau_{\rm mri} &\sim & {R^2 \over \nu_{\rm vis}} \nonumber \\
&\sim & 10^2  R_6^2 \Omega_4  
\biggl({\alpha_{\rm vis} \over 10^{-2}}\biggr)^{-1}
\biggl({c_s \over 0.1c}\biggr)^{-2}\,{\rm ms}, 
\eeqn
and hence, the time scale is as short as $\tau_{\rm wind}$ for the
hypothetical radial field strength $B_{15}\agt 1$. Both effects work
as long as differential rotation is present even in the absence of
nonaxisymmetry. Therefore, unless any process stabilizes them,
HMNSs with sufficiently high mass would collapse to a black hole in the
time scale, $\tau_{\rm wind}$ or $\tau_{\rm mri}$, which is $\sim
10^2$\,ms.

In the definitions of HMNS and SMNS described above, we have not
considered finite-temperature (thermal) effects. This effect could be
important for the remnant MNS of binary neutron stars. The reason for
this is that during the merger process, strong shocks are often formed
and the maximum temperature of MNSs is increased up to
30\,--\,50\,MeV~\cite{skks2011a,skks2011b}. The thermal pressure
associated with this high temperature is several 10\% of the cold-part
pressure caused by the repulsive nuclear force, and hence, it is never
negligible. Although it is not easy to strictly determine their
values, it is reasonable to consider that the finite-temperature
effect could increase the values of $M_{\rm max}$ and $M_{\rm max,s}$
by $\sim 0.1M_{\odot}$. The right-hand side of Fig.~\ref{fig:HMNS0}
schematically shows the possible increase of these values.  Hereafter
we refer to these values as $M_{\rm max}(T>0)$ and $M_{\rm
max,s}(T>0)$, supposing that these are larger than $M_{\rm max}$ and
$M_{\rm max,s}$ by $\sim 0.1M_{\odot}$.

The finite-temperature neutron stars will eventually dissipate the
thermal energy in a short time scale $\sim 1$\,--\,$10$\,s because of
the neutrino cooling~\cite{ST83}.
References~\cite{skks2011a,skks2011b} indeed show that the time scale
of the neutrino emission is of order seconds.  Here, when we consider
the possible evolution processes of HMNS and SMNS, we have to keep in
mind that in this time scale, the values of $M_{\rm max}(T>0)$ and
$M_{\rm max,s}(T>0)$ could be by $\sim 0.1M_{\odot}$ larger than
$M_{\rm max}$ and $M_{\rm max,s}$.  For example, consider a
differentially rotating and hot remnant MNS for which the mass is
larger than $M_{\rm max,s}$ and smaller than $M_{\rm max,s}(T>0)$.  If
a process of angular-momentum transport works and the degrees of
differential rotation is significantly reduced, such a remnant will be
unstable against gravitational collapse, for the case that the thermal
effect is negligible. However, if the thermal effect is important even
after the angular-momentum transport process works, it will be stable
in the cooling time scale.  It will eventually collapse to a black
hole after the neutrino cooling. However, its lifetime $\agt 1$\,s
could be much longer than the angular-momentum transport time scale 
$\alt 100$\,ms. 

\section{Properties and evolution process of MNS}\label{sec3}

\begin{figure*}[t]
\includegraphics[width=82mm,clip]{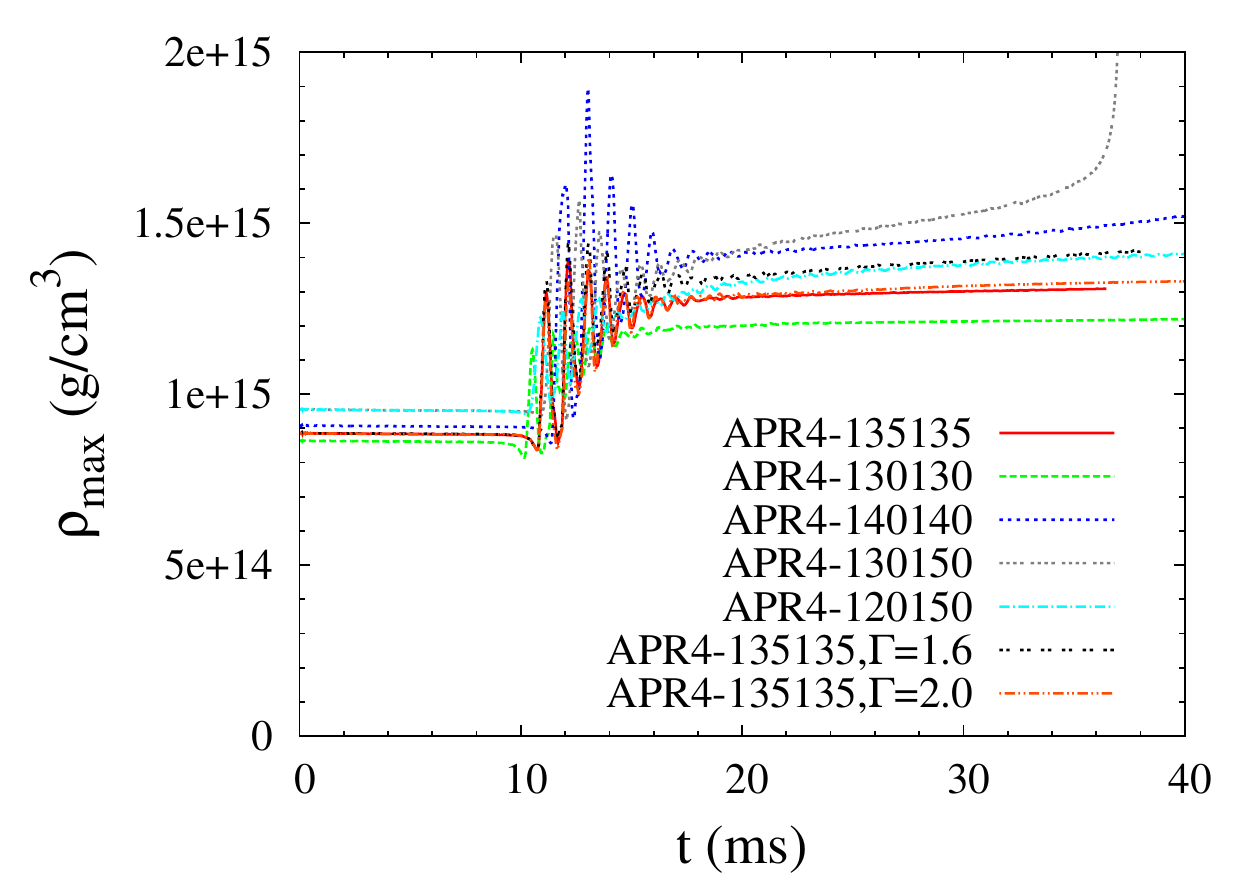}~~
\includegraphics[width=82mm,clip]{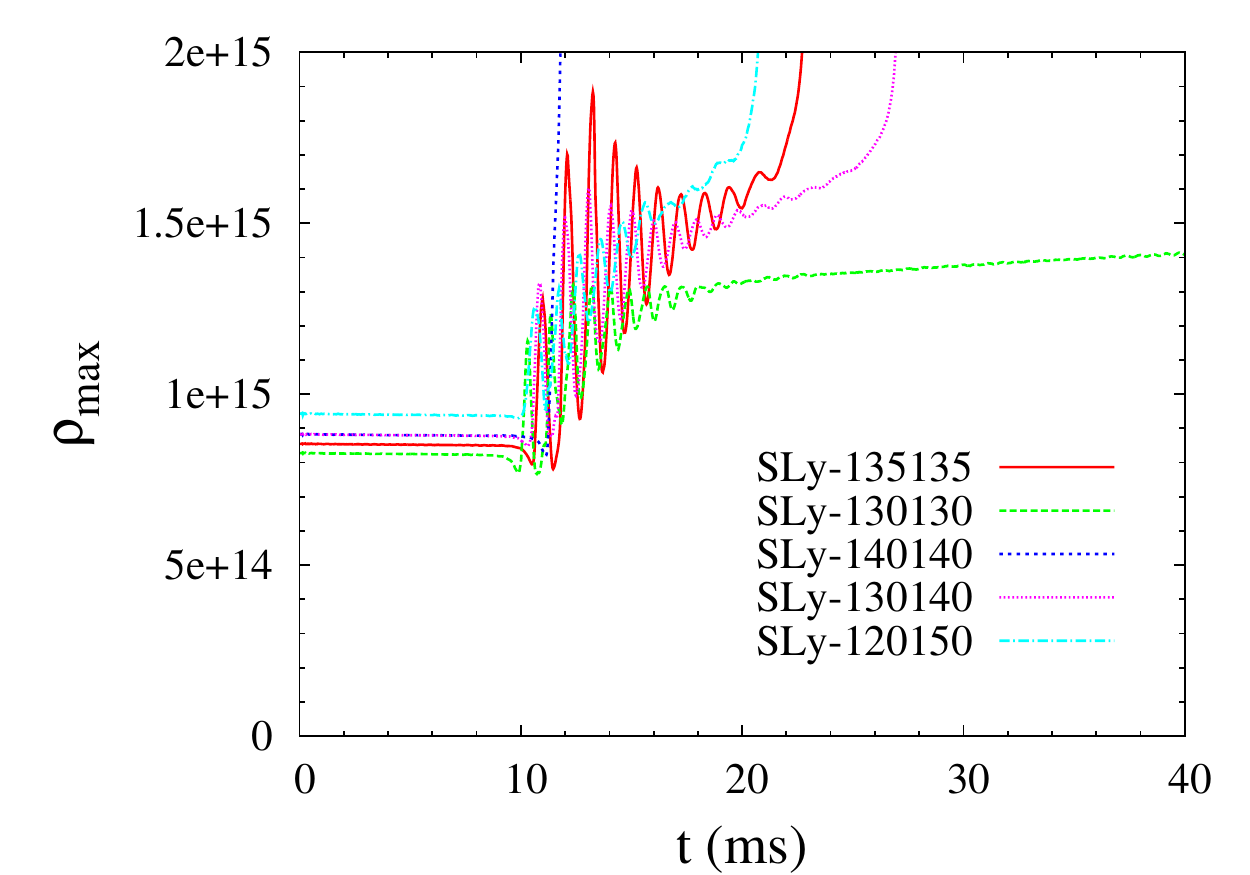}\\
\includegraphics[width=82mm,clip]{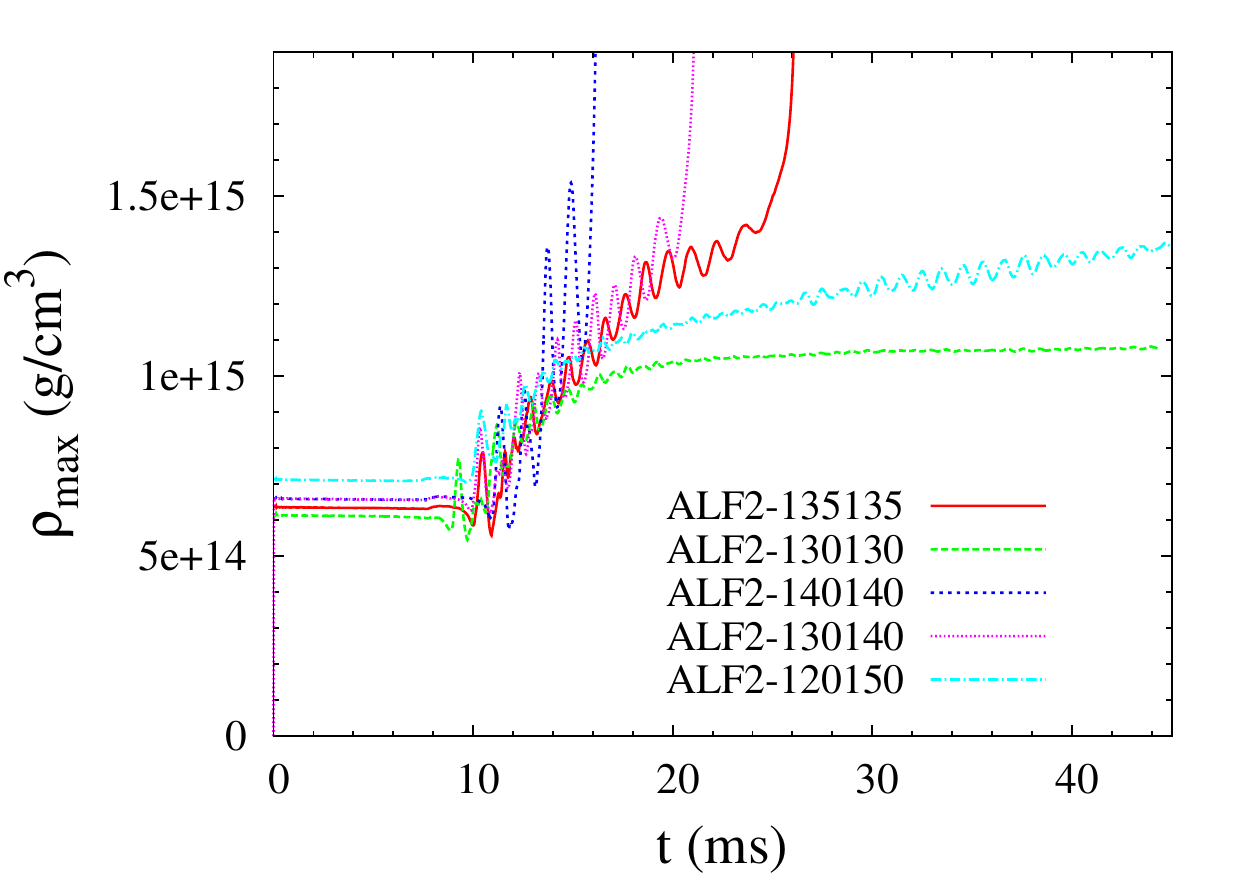}~~
\includegraphics[width=82mm,clip]{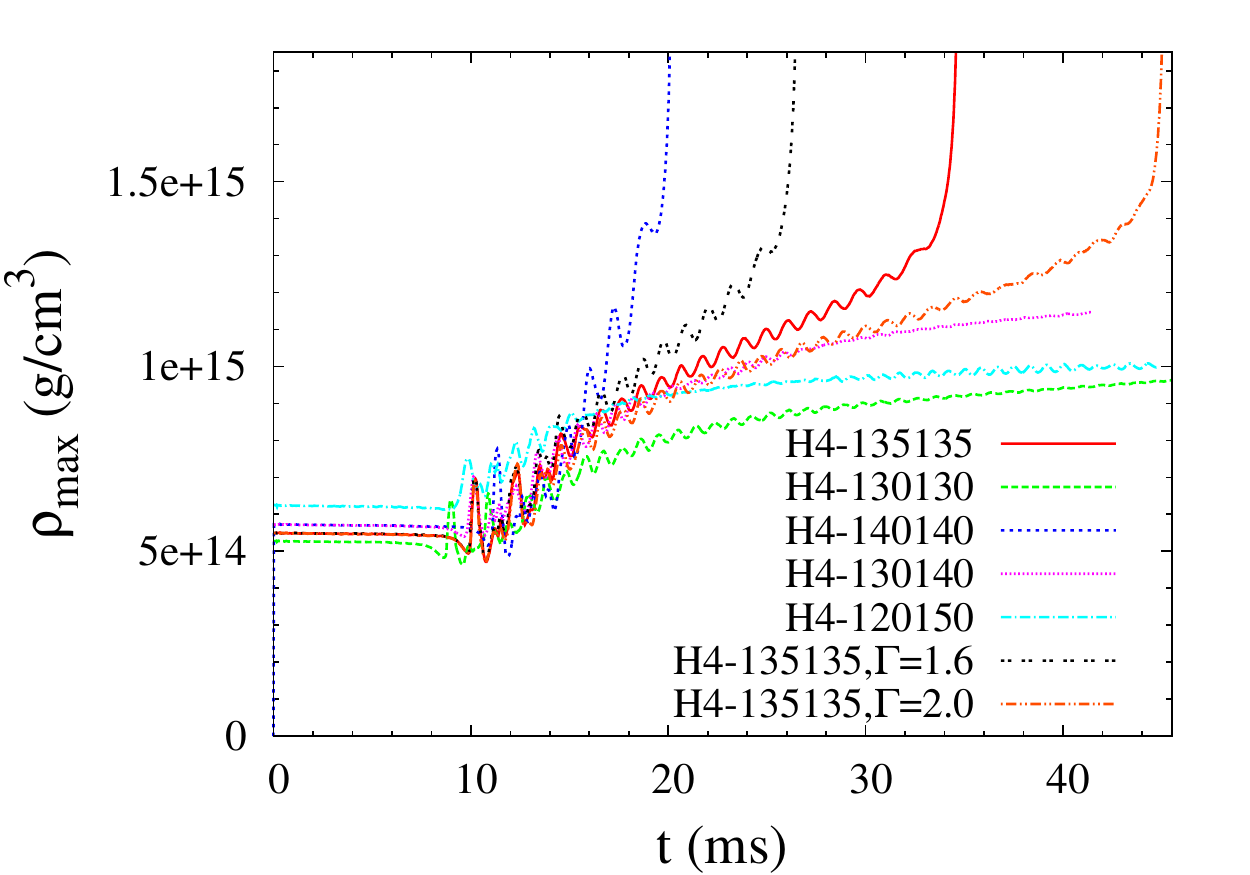}\\
\includegraphics[width=82mm,clip]{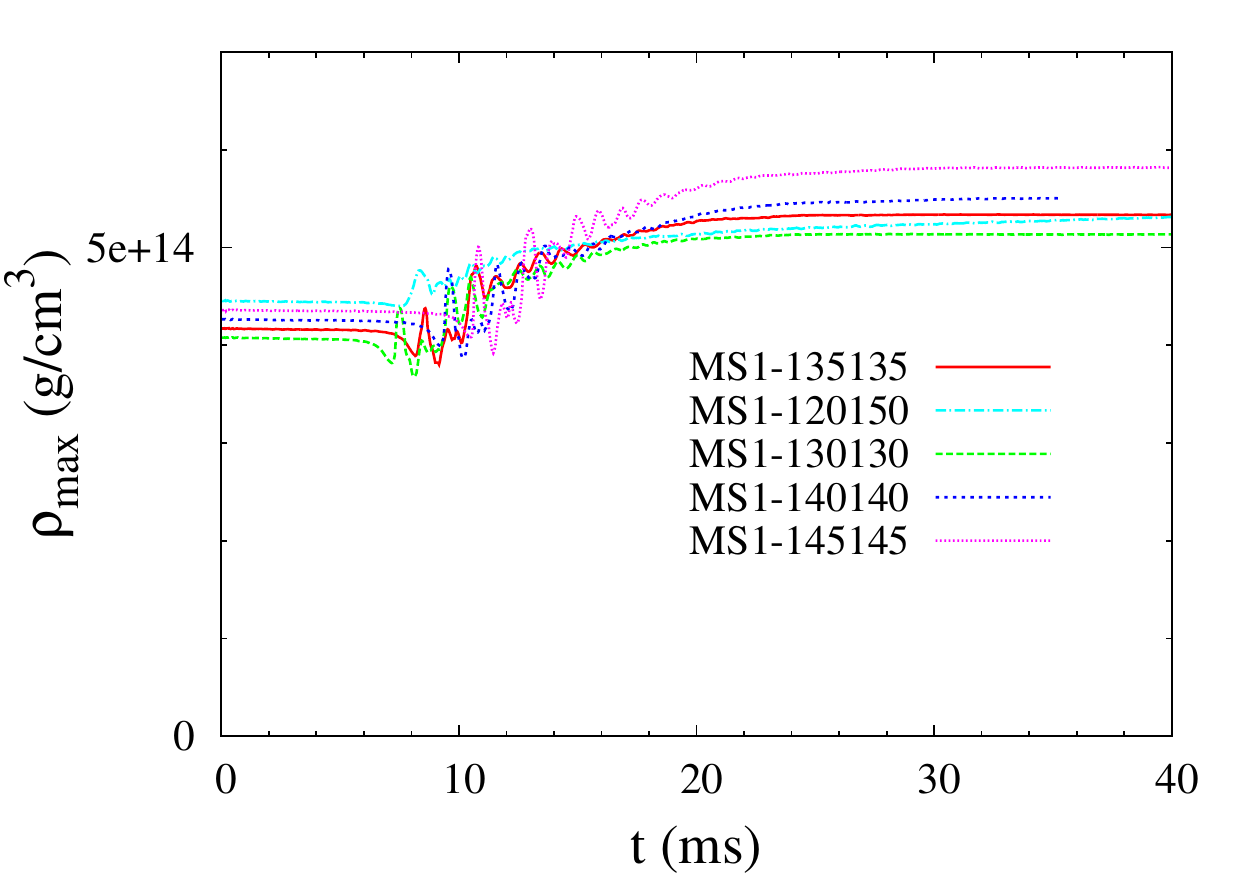}~~
\includegraphics[width=82mm,clip]{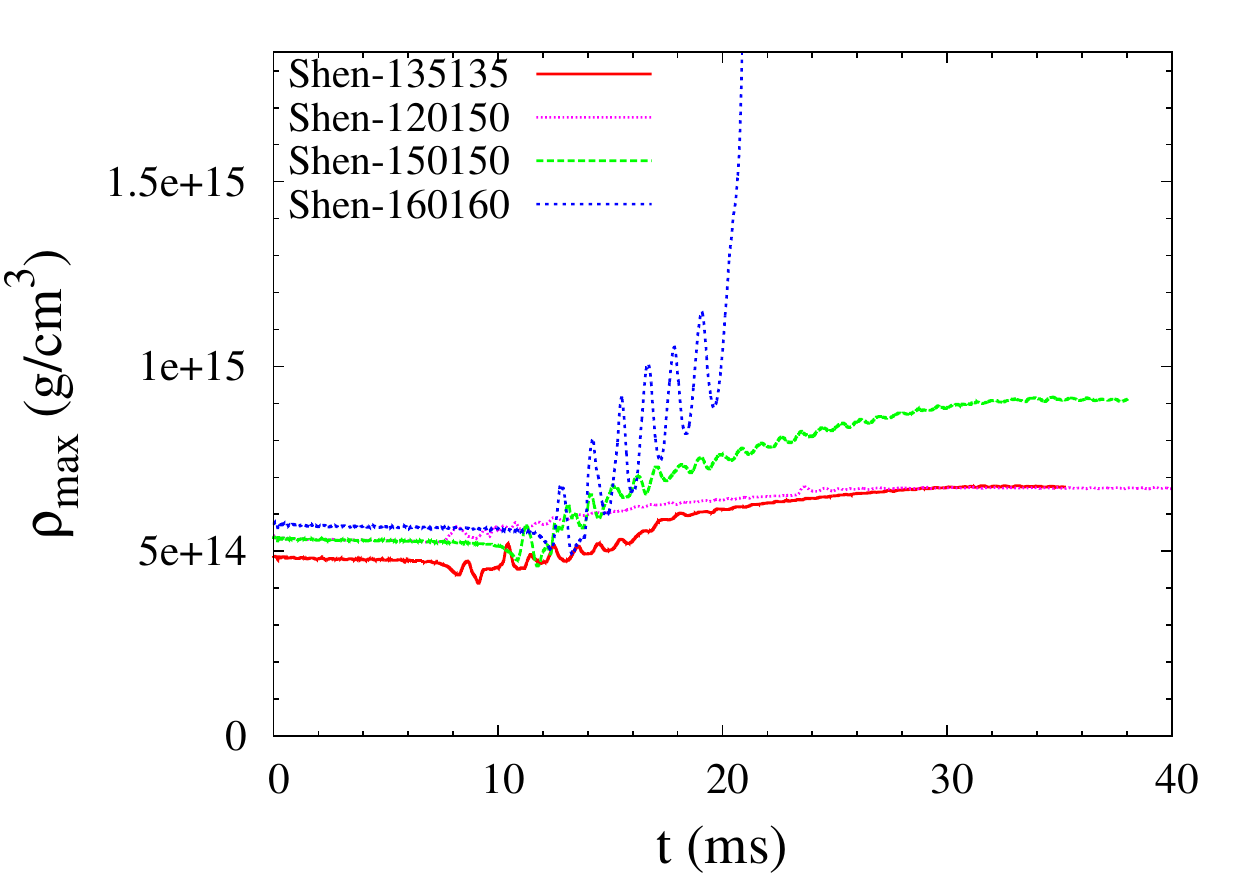}
\caption{Maximum density as a function of time for APR4 (top left),
SLy (top right), ALF2 (middle left), H4 (middle right), MS1 (bottom
left), and Shen (bottom right) EOSs with several values of binary
mass. Note that $\Gamma$ for the panel of APR4 and ALF2 denotes
$\Gamma_{\rm th}$ and the absence of $\Gamma$ value means that
$\Gamma_{\rm th}=1.8$. For APR4-140140 and ALF2-120150, we
confirmed that a black hole was eventually formed at $t \sim 47$\,ms
and 52\,ms.}
\label{figrho}
\end{figure*}

\begin{figure*}[t]
\includegraphics[width=160mm,clip]{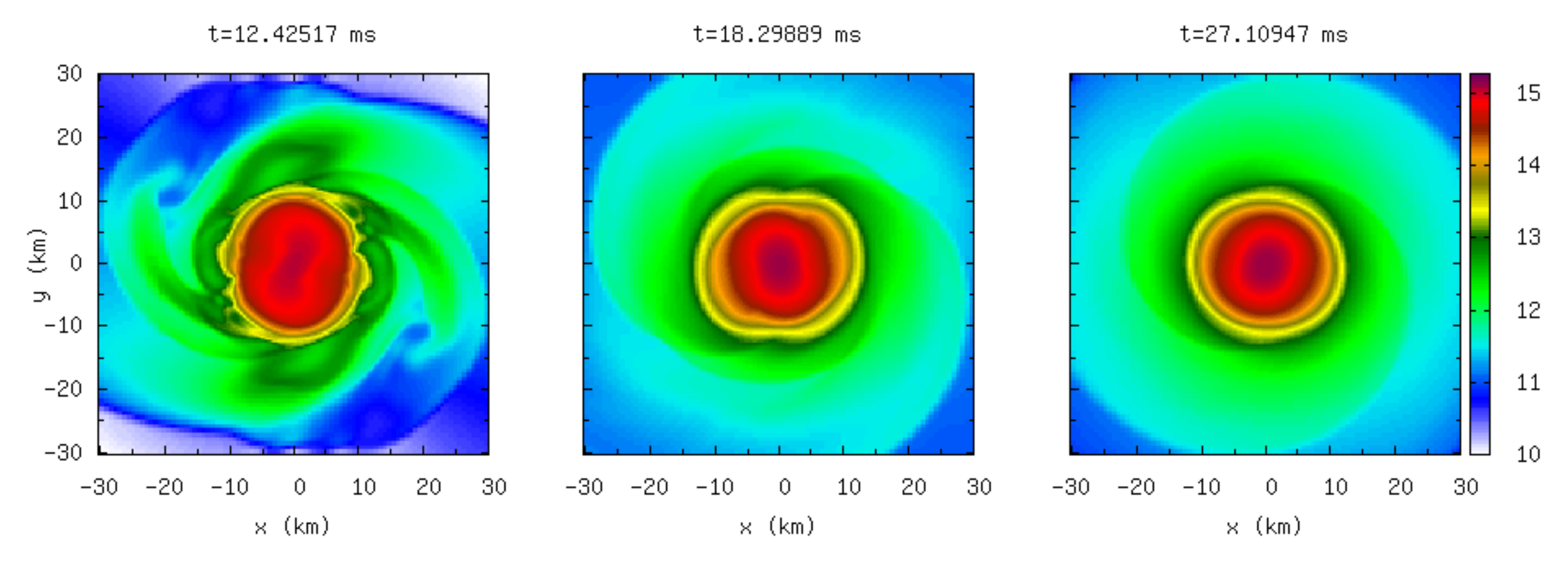}
\includegraphics[width=160mm,clip]{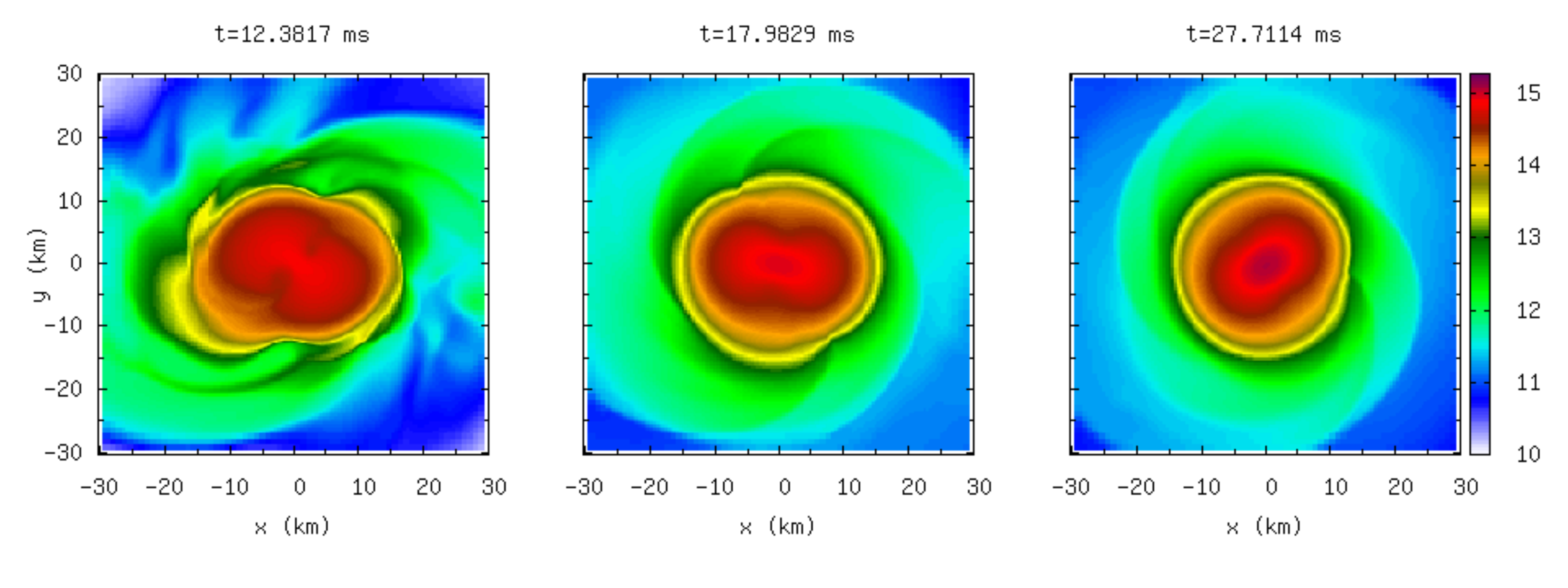}
\vspace{-5mm}
\caption{Snapshots of the density profile in the equatorial plane at
selected time slices for equal-mass models APR4-135135 (upper panel) 
and H4-135135 (lower panel) with $\Gamma_{\rm th}=1.8$. 
}
\label{figcont1}
\end{figure*}

\begin{figure*}[t]
\includegraphics[width=160mm,clip]{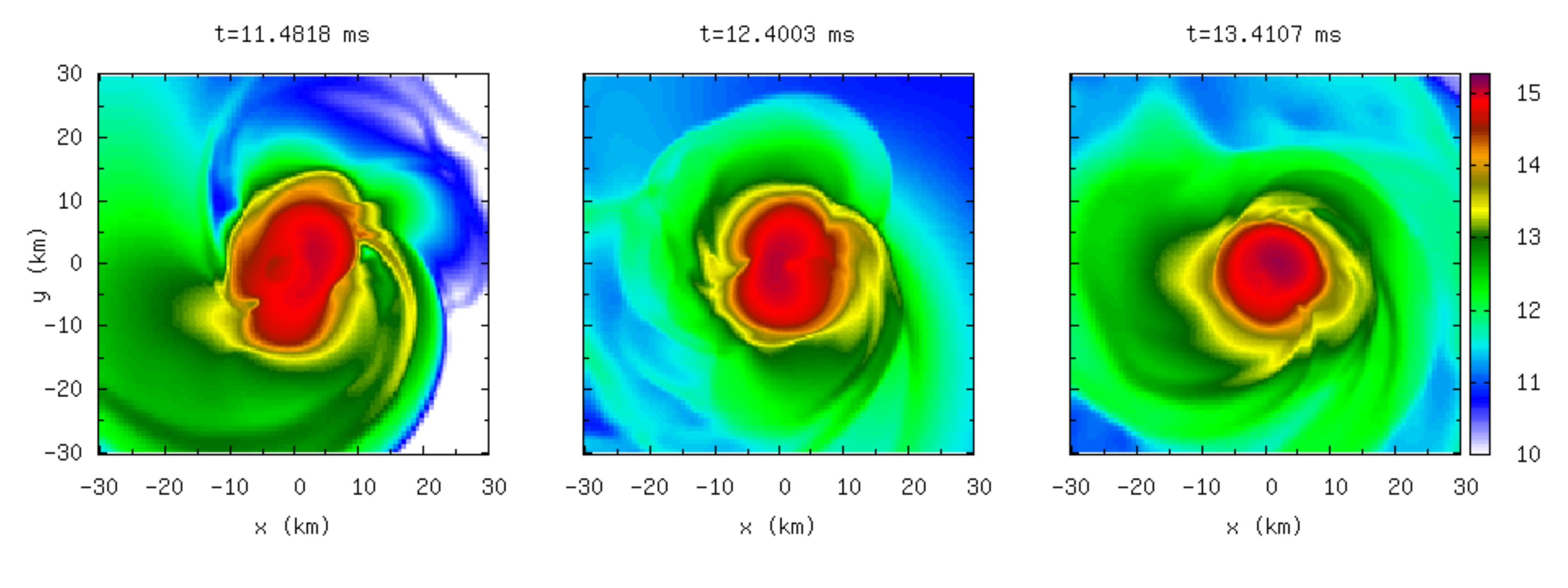} \\
\vspace{-1cm}
\includegraphics[width=160mm,clip]{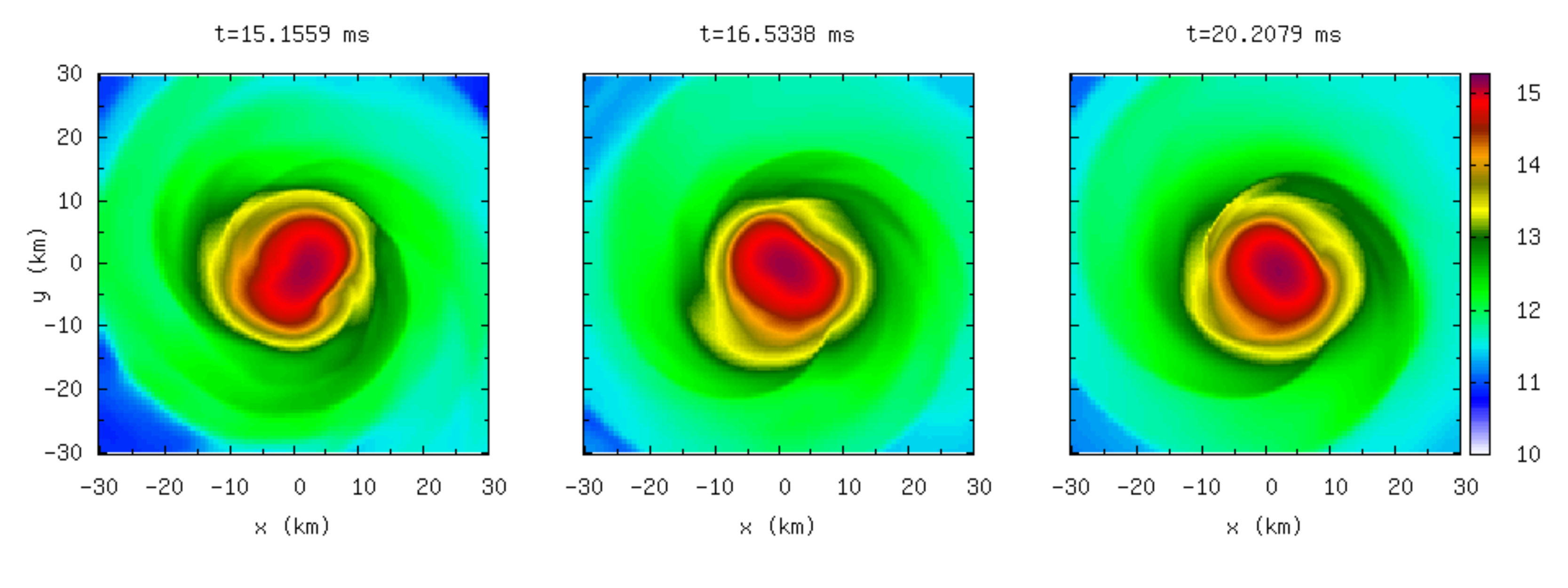} \\
\includegraphics[width=160mm,clip]{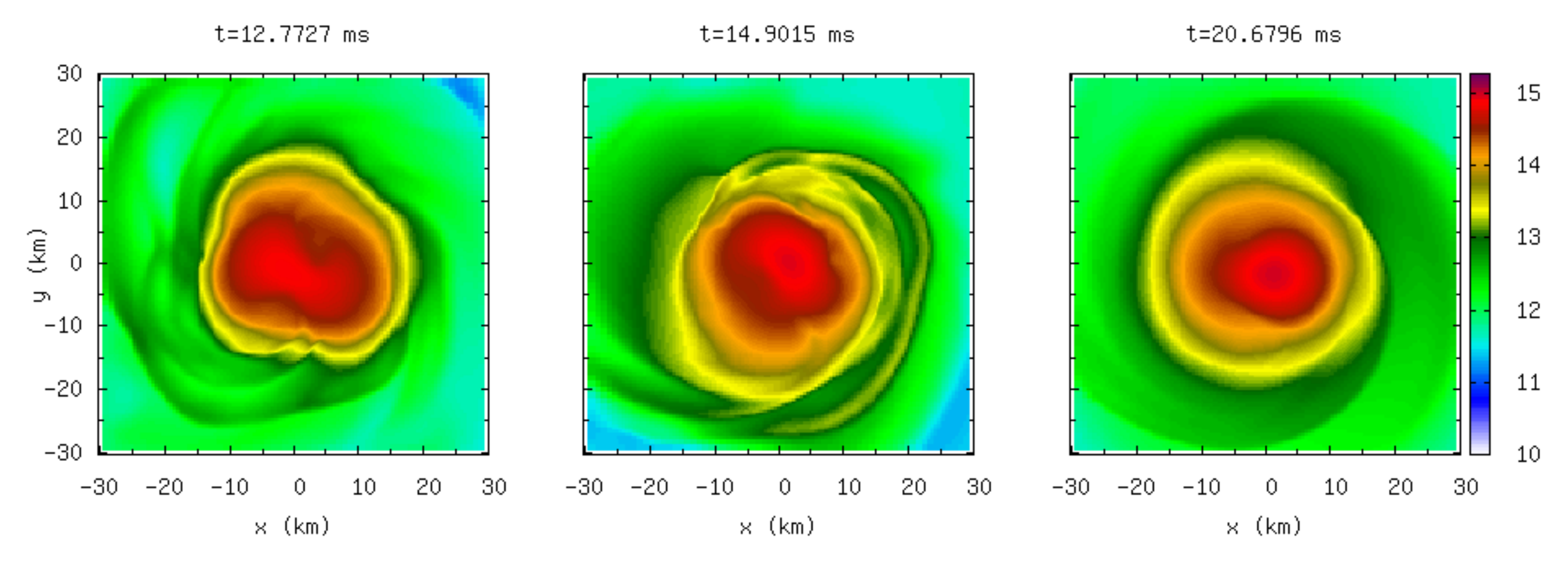}
\vspace{-0.5cm}
\caption{The same as Fig.~\ref{figcont1} but for unequal-mass models 
APR4-120150 (upper and middle panels) and H4-120150 (bottom panel). }
\label{figcont2}
\end{figure*}

Previous studies (e.g.,~\cite{skks2011a,hotoke,hotoke2012})
clarified that soon after the onset of the merger, either a long-lived
MNS (HMNS or SMNS or normal neutron star) or a black hole is formed.
For most of the simulations in this paper performed with stiff EOSs
and with the canonical total mass $2.6$\,--\,$2.8M_{\odot}$, we found
that a long-lived MNS is formed with its lifetime much longer than its
dynamical time scale $\sim 0.1$\,ms and its rotation period $\sim 1$\,ms
(cf. Figs.~\ref{figrho}\,--\,\ref{figcont2}).  The unique properties to
be particularly noticed are that the MNSs are rapidly and differentially
rotating, and nonaxisymmetric (cf. Figs.~\ref{figcont1} and
\ref{figcont2}): Thus, they could be temporarily stable even if they
are very massive, and in addition, they could be strong sources of
gravitational waves.  The purpose of this section is to explore the
properties and evolution processes of such MNSs. In the subsequent
sections, we will clarify the properties of gravitational waves. 


\subsection{Dependence on EOS}\label{sec:depEOS}

Figure~\ref{figrho} plots the evolution of the maximum density for
five piecewise polytropic EOSs and Shen EOS with several binary masses
in the range $m=2.6M_{\odot}$\,--\,$2.8M_{\odot}$.  For Shen and MS1
EOSs, the results with more massive cases are also plotted.  This
figure shows that the evolution process of MNSs depends strongly on
the EOSs and total mass as described in the following. 

{\bf APR4}: For this EOS, the pressure at $\rho=\rho_2$ (i.e., the
value of $P_2$) is lowest among all the EOSs employed. However, the
pressure for $\rho \geq \rho_3$ is rather high because the adiabatic
index for this density range (i.e., $\Gamma_3$) is highest. Reflecting
the small pressure for $\rho < \rho_3$, the maximum density increases
steeply during the early stage of the merger due to the sudden
increasing strength of the self-gravity.  However, also, reflecting
the high pressure for $\rho > \rho_3$ due to the high value of
$\Gamma_3$, the steep increase of the density is hung up and
subsequently the maximum density oscillates with high amplitude for
several oscillation periods (for $\sim 5$\,ms). This is the unique
feature for this type of EOS (i.e., APR4 and SLy).  After a subsequent
relaxation process through the interaction with the envelope
surrounding the central core, the maximum density eventually relaxes
approximately to a constant.

In this relaxation process, the angular momentum is transported
substantially from the inner region to the outer region via the
hydrodynamical angular-momentum transport process, because the MNS has
a highly nonaxisymmetric structure and can exert the torque to the
surrounding matter in its early evolution stage. It is worthy to note
that the high-amplitude oscillation also plays an important role for
enhancing the angular-momentum transport because the MNS interacts
directly with the envelope during this oscillation.

The resulting MNSs formed after the relaxation evolve in a
quasistationary manner. For a relatively small total mass with $m \alt
2.7M_{\odot}$, the maximum density in the quasistationary stage
remains approximately constant for a sufficiently long time $\gg
10$\,ms. This is due to the facts that the gravitational-wave emission
does not yield significant dissipation, i.e., the system relaxes to a
quasistationary state for this stage, and that the final mass of the
MNS is likely to be smaller than $M_{\rm max,s}$ or $M_{\rm
max,s}(T>0)$. The first fact can be found by estimating the
dissipation time scale by the gravitational-wave emission which is
much longer than the dynamical time scale (see also
Figs.~\ref{figGW1}\,--\,\ref{figGW3} from which we find that
gravitational-wave amplitude for the $t-t_{\rm merge} > 10$\,ms is
much smaller than that for $t -t_{\rm merge} \alt 5$\,ms). To confirm
the second fact, we calculated the final gravitational mass of the
system applying the formula of the ADM mass for a finite sphere of
radius $\approx 300$\,km (see Table~\ref{table:ID}) which is
approximately equal to the initial ADM mass minus energy carried out
by gravitational waves, and found that for $m=2.6M_{\odot}$ and
$2.7M_{\odot}$, the final mass is $\approx 2.53M_{\odot}$ and
$2.60M_{\odot}$, respectively. The value of $M_{\rm max}$ for this EOS
is $\approx 2.20M_{\odot}$, and thus, that of $M_{\rm max,s}$ should
be $\sim 2.6M_{\odot}$ according to the numerical results
of~\cite{CST94}. Then, it is reasonable to consider that the value of
$M_{\rm max,s}(T>0)$ is larger than $2.6M_{\odot}$. Therefore, for $m
= 2.6M_{\odot}$, the remnant is a SMNS, and hence, it will be alive
for a long time $\gg 1$\,s, even in the presence of a realistic
process of the angular-momentum transport and dissipation (see
Sec.~\ref{sec:MNS}), and for $m=2.7M_{\odot}$, the remnants may be
HMNSs. However, the mass would be smaller than $M_{\rm max,s}(T>0)$.
For this system, the angular-momentum transport alone may not trigger
the gravitational collapse, and the lifetime of the HMNS would be
determined by the neutrino cooling in reality.

For $m \agt 2.8M_{\odot}$, on the other hand, the remnants are HMNSs
which evolve due to the gravitational-wave emission and hydrodynamical
angular-momentum transport, resulting in the slow but monotonic
increase of the maximum density with time for the quasistationary
stage. For example, for APR4-130150 and APR4-140140, a black hole is
formed at $\sim 30$ and $35$\,ms after the onset of the merger in
our simulations. The resulting object is a black hole surrounded by a
massive disk (or torus) of mass $\sim 0.1M_{\odot}$. Here, the high
mass of the disk is a result of the long-term angular momentum transport
process in the HMNS stage.

{\bf SLy}: For this EOS, the evolution process agrees qualitatively
with that for APR4. However, the value of $M_{\rm max}$ for this EOS
is slightly (by $0.14M_{\odot}$) smaller than that for APR4. Thus, the
values of $M_{\rm max,s}$ and $M_{\rm max,s}(T>0)$ would be also
smaller by $\sim 0.15M_{\odot}$; the plausible value of $M_{\rm
max,s}$ would be $\sim 2.45M_{\odot}$. Reflecting this fact, the
threshold mass for the prompt formation of the black hole becomes $m
\approx 2.8M_{\odot}$ for this EOS. For $m=2.7M_{\odot}$ with which the mass
of the remnant MNS is $\sim 2.6M_{\odot}> M_{\rm max,s}$ and thus the
MNS is hypermassive, a black hole is formed in $\sim 10$\,--\,15\,ms
after the onset of the merger irrespective of the mass ratio; the
lifetime of the HMNS is not very long. For $m=2.6M_{\odot}$, the mass
of the remnant MNS is $\agt 2.5M_{\odot}$, and thus, the MNS is
hypermassive as well. However, for this mass, the lifetime is $\gg
10$\,ms; the gravitational-wave emission and hydrodynamical
angular-momentum transport process also are not sufficient for
inducing the collapse.  Subsequent evolution of such HMNS will be
determined by angular-momentum transport processes or cooling in
reality.  If the thermal pressure plays a sufficiently important role,
the HMNS collapses to a black hole after the neutrino cooling with the
time scale of seconds, and if it does not, the collapse to a black
hole occurs in some angular-momentum transport time scale $\sim
100$\,ms.

As argued in the following, binary neutron stars with $m=2.8M_{\odot}$
do not result in a black hole formation promptly after the onset of
the merger for ALF2 and H4 EOSs for which the value of $M_{\rm max}
\sim 2M_{\odot}$, by contrast to the case of SLy. This suggests that
for a given value of $M_{\rm max}$, the black hole formation is more
subject to EOSs with smaller values of $P_2$, or in other words, with
smaller radii of canonical-mass neutron stars. As shown in
Sec.~\ref{sec:GW}, a characteristic peak in the Fourier spectrum of
gravitational waves for a high frequency band $\sim 2$\,--\,4\,kHz is
present for the case that a MNS is formed after the merger. This
implies that if high-frequency gravitational waves from the merger of
binary neutron stars with particular total mass, say $2.8M_{\odot}$,
are observed, we will be able to constrain the EOS of neutron stars
only by determining whether the peak is present or
not~\cite{shibata2005}.

{\bf ALF2}: For this EOS, not only neutron stars of mass
1.2\,--\,$1.5M_{\odot}$ but also MNSs just after the formation with
$m=2.6$\,--\,$2.8M_{\odot}$ have the maximum density between $\rho_2$
and $\rho_3$. Thus, although the oscillation of the maximum density is
observed, its amplitude is not as high as for APR4 and SLy, and hence,
the angular-momentum transport process does not seem to be as
efficient as for APR4 and SLy as well.  For ALF2, however, the
adiabatic index for this density range is small ($\Gamma_2 \sim 2.4$),
although the pressure for $\rho_2 \leq \rho \leq \rho_3$ is relatively
high. Because of this property, the maximum density of the MNSs
increases as a result of the gravitational radiation reaction and
hydrodynamical angular-momentum transport with a relatively short time
scale.  For the models with $m \agt 2.7M_{\odot}$, the maximum density
becomes eventually larger than $\rho_3$. For $\rho > \rho_3$, the
adiabatic index is quite small $(\Gamma_3 \sim 1.9$), and hence, the
increase of the maximum density is enhanced, leading to the eventual
gravitational collapse to a black hole. For this evolution process,
the formation time scale of the black hole is determined by the time
scale of gravitational-wave emission or hydrodynamical
angular-momentum transport.

For $m = 2.7M_{\odot}$ and $2.8M_{\odot}$, the remnant mass is
$\approx 2.63M_{\odot}$ and $2.72M_{\odot}$, and thus, the remnants
are very hypermassive, because for this EOS, $M_{\rm max} \approx
2.0M_{\odot}$ and $M_{\rm max,s}$ would be $\alt 2.4M_{\odot}$. Since
the black hole is formed for $m \geq 2.7M_{\odot}$, the thermal
pressure is not sufficient for sustaining the additional self-gravity
of the HMNSs;  $M_{\rm max,s}(T>0)-M_{\rm max,s}$ would be smaller than
$\sim 0.2M_{\odot}$. 

For relatively small mass $m = 2.6M_{\odot}$, on the other hand, the
emission of gravitational waves and hydrodynamical angular-momentum
transport become inactive before the maximum density significantly
exceeds $\rho_3$. In this case, the increase of the maximum density is
stopped and the HMNS relaxes to a quasistationary state.
The resulting remnant mass is $\approx 2.54M_{\odot}$ which is likely to
be larger than $M_{\rm max,s}$.  Thus, the remnant is likely to be a
HMNS. Subsequent evolution of such HMNS will be determined by
angular-momentum transport processes or cooling in reality, as in the
case of model SLy-130130.

One point worthy to be noted is that the evolution process for
$m=2.7M_{\odot}$ depends on the mass ratio (compare the plots for
ALF2-135135 and ALF2-120150). For the sufficiently large asymmetry
(i.e., $q=0.8$), the lifetime of the HMNS becomes much longer than
that of the equal-mass model for this EOS.  The reason is that for the
asymmetric case, the merger occurs at a larger orbital separation than
for the equal-mass case; i.e., before a sufficient amount of angular
momentum is dissipated by the gravitational-wave emission, the merger
sets in. In addition, a large fraction of the materials (in particular
the materials of the less-massive neutron star) obtain a sufficient
angular momentum during the merger process resulting in a disk or a
material ejected from the system~\cite{hotoke2012}. This reduces the
mass of the HMNS and the collapse to a black hole is delayed
\footnote{For SLy, this dependence of the lifetime on the 
mass ratio is not found. The reason is that for SLy (and APR4), a
large fraction of mass is ejected from the system even for the
equal-mass binary. For SLy-135135, $\sim 0.02M_{\odot}$ is ejected
while for SLy-120150, $\sim 0.01M_{\odot}$ is ejected. }.

{\bf H4}: The evolution process of MNSs in this EOS is similar to that
in ALF2. As the value of $M_{\rm max}$ is approximately equal to that
in ALF2, the criterion for the MNS formation is also very similar. For
this EOS, however, the adiabatic index does not decrease with the
increase of the density as drastically as for ALF2. Thus, the increase
rate of the maximum density with time is relatively slow, and
reflecting this fact, the configuration of the MNS relaxes to a
quasistationary one in a short time scale after its formation. The
resulting quasistationary MNS evolves through the hydrodynamical
angular-momentum transport process and gravitational-wave emission
subsequently. However the evolution time scale is much longer than
10\,ms.

A point clearly seen for the density-evolution plot of this EOS is
that the difference in the shock-heating efficiency (i.e., the value
of $\Gamma_{\rm th}$) is reflected in the change of the lifetime of
the HMNSs: By the increasing efficiency of the shock heating (for the
larger values of $\Gamma_{\rm th}$), the lifetime of the HMNSs becomes
longer (this effect should be universally found irrespective of the
EOSs).  It is also found that the presence of the mass asymmetry
increases the lifetime of the HMNS because a large fraction of the
materials escapes from the HMNS during the early stage of the
merger for this EOS.

{\bf MS1}: For this EOS, the maximum mass of spherical neutron stars
is too high ($M_{\rm max}\approx 2.77M_{\odot}$) to form SMNSs or
HMNSs for $m \leq 2.8M_{\odot}$ because the remnant mass for such
initial mass range is smaller than $2.75M_{\odot}$ (see
Table~\ref{table:ID}).  For this case, the merger remnant relaxes to a
quasistationary MNS in a time scale $\sim 10$\,--\,15\,ms. In a real
MNS, a dissipation or a transport process of the angular momentum
plays a role for the subsequent evolution for it.  However, a black
hole will not be formed for $m \leq 2.8M_{\odot}$ for which the
remnant mass is smaller than $M_{\rm max} \approx 2.77M_{\odot}$.  For $m =
2.9M_{\odot}$, a quasistationary MNS is also formed. For this case,
the MNS is likely to be supramassive, but not hypermassive. Thus, this
MNS will be also alive for a long time scale $\gg 1$\,s. 

{\bf Shen}: The evolution process of MNSs in this EOS is similar to
that in H4, although the threshold mass for the eventual formation of
a black hole is much higher than that for H4 ($m > 3.0M_{\odot}$ while
$m \agt 2.7M_{\odot}$ for H4). For this EOS, a long-lived HMNS is
formed even for $m=3.0M_{\odot}$ which is by 36\% larger than the value
of $M_{\rm max} \approx 2.2M_{\odot}$. By contrast, for H4, a black
hole is eventually formed if $m \agt 2.7M_{\odot} \approx 1.33M_{\rm
max}$. This suggests that in the tabulated EOS in which the heating
effects are taken into account in a more strict way, the shock
heating may play a more important role for sustaining the self-gravity
of the HMNS.

Before closing this subsection, it is worthy to summarize the
dependence of the evolution process for MNSs of canonical mass $m
\approx 2.7M_{\odot}$ on EOSs as follows: 
\begin{itemize}
\item For the EOSs such as APR4 and SLy for which $P_2$ has a
relatively small value, the evolution process of the MNSs depends
primarily on the adiabatic index for $\rho > 10^{15}\,{\rm g/cm^3}$
(i.e., $\Gamma_3$). 
\item For the EOSs such as ALF2 and H4 for which $P_2$ has a fairly
large value, the evolution process of the MNSs depends on the
adiabatic index for $\rho \agt 5 \times 10^{14}\,{\rm g/cm^3}$ (i.e.,
both on $\Gamma_2$ and $\Gamma_3$).
\item For the stiff EOSs such as MS1 and Shen for which $P_2$ has a
large value, the evolution process of the MNSs depends only on the
adiabatic index for $\rho \alt 10^{15}\,{\rm g/cm^3}$ (i.e.,
$\Gamma_2$).
\end{itemize}
Therefore, future gravitational-wave observation for MNSs will be used
for exploring the properties of the EOS in a specific density range.

\subsection{Characteristic time scales}\label{sec:timescale}

\begin{figure*}[t]
\includegraphics[width=100mm,clip]{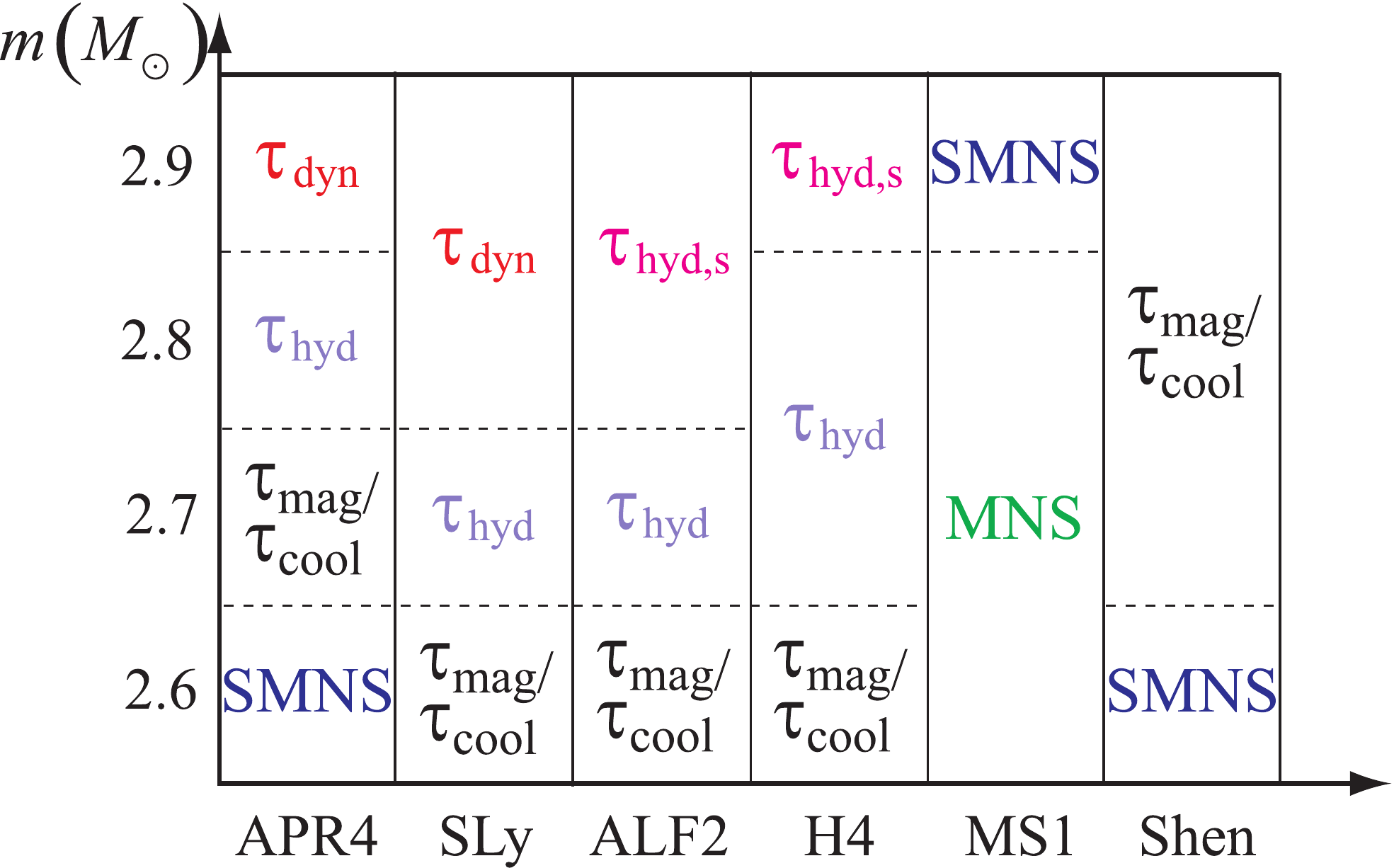}
\caption{The evolution time scale of the system in the plane composed
of EOSs and total mass.  $\tau_{\rm dyn}$: A black hole is formed in
the dynamical time scale after the onset of the merger.  $\tau_{\rm
hyd}$: A HMNS is formed and its lifetime is determined by the
hydrodynamical angular-momentum transport time scale.  $\tau_{\rm
hyd,s}$: The same as for $\tau_{\rm hyd}$ but the lifetime is shorter
than $\sim 10$\,ms.  $\tau_{\rm mag}/\tau_{\rm cool}$: A HMNS is
formed and its lifetime would be determined by the time scale of
angular-momentum transport by some magnetohydrodynamics effects or by
the neutrino cooling time scale. The evolution time scale for a given
total mass depends weakly on the mass ratio. For MS1, only the MNS or
SMNS is formed for $m
\leq 2.9M_{\odot}$.  For APR4 and Shen, the remnant for the $m \alt
2.6M_{\odot}$ case is likely to be a SMNS (not HMNS).}
\label{fig:HMNS}
\end{figure*}

As their lifetime is tabulated in one of the columns of
Table~\ref{table:ID}, HMNSs collapse to a black hole for several
relatively massive models. This collapse is triggered by the
angular-momentum loss by the gravitational-wave emission and by the
angular-momentum transport process from the inner region of the HMNS
to its outer envelope.  The transport process can work because the
HMNS formed has a nonaxisymmetric structure and exerts the torque to
the envelope surrounding it, as already mentioned in
Sec.~\ref{sec:MNS}.  We note that the mass of the disk surrounding the
remnant black hole formed after the collapse of the HMNS is in general
larger for the longer lifetime of the HMNS for a given EOS (see Table
\ref{table:ID}).  In addition, the emissivity of gravitational waves
is quite low for not-young HMNS as shown in Sec.~\ref{sec:GW}: This is
because the degree of the nonaxisymmetry for the HMNS decreases with
time.  These facts obviously show that the hydrodynamical
angular-momentum transport process plays an essential role for the
black hole formation. Therefore, for the HMNS of lifetime $\sim
10$\,--\,50\,ms, we conclude that the black hole formation is
determined primarily by the hydrodynamical angular-momentum transport
process, and write the time scale as $\tau_{\rm hyd}$.

On the other hand, for less-massive HMNSs and SMNSs, neither the
emission of gravitational waves nor the hydrodynamical effect are
likely to determine their lifetime. For such systems, other
dissipation processes (which are not taken into account in our
numerical simulations) will play an important role, and the evolution
proceeds with the dissipation time scale.  If the system is
hypermassive and its degree of differential rotation is sufficiently
high, the angular-momentum transport process via magnetohydrodynamics
effects could trigger the eventual collapse of the HMNS to a black
hole (e.g.,~\cite{DLSSS2006a}) with a relatively short time
scale $\tau_{\rm wind}$ or $\tau_{\rm mri}\sim 100$\,ms or less, which
is comparable to $\tau_{\rm hyd}$.  If the degree of differential
rotation is not high and the thermal effect plays an important role
for sustaining the self-gravity of the HMNS, neutrino cooling will
play a dominant role for determining the process toward the black-hole
formation.  According to~\cite{skks2011a,skks2011b}, the cooling
time scale via the neutrino emission is of order seconds (hereafter
denoted by $\tau_{\rm cool}$), and hence, it is much longer than
$\tau_{\rm hyd}$. However, if the degree of differential rotation is
not high, $\tau_{\rm cool}$ could be shorter than $\tau_{\rm wind}$
and $\tau_{\rm mri}$. Furthermore, if the remnant mass is smaller than
$M_{\rm max,s}(T>0)$, the magnetic winding and MRI would not trigger
the collapse to a black hole. For such a system, the neutrino cooling
will trigger the collapse eventually. Our previous
work~\cite{skks2011a} suggests that this is likely to be the case.

For a smaller-mass system with $M_{\rm max} \alt m \alt M_{\rm
max,s}$, the remnant neutron star is not hypermassive, and it evolves
simply to a cold SMNS in $\tau_{\rm cool}$. The cold SMNS will
collapse eventually to a black hole after its angular momentum is
dissipated by some process such as magnetic dipole radiation.  For an
even smaller-mass system with $m \alt M_{\rm max}$, the remnant
neutron star is not supramassive , and it evolves simply to a cold
neutron star in $\tau_{\rm cool}$. This is the case for MS1 with $m
\alt 2.8M_{\odot}$.

We may classify the remnant MNSs by its evolution time scale.
Figure~\ref{fig:HMNS} shows such a classification.  In this figure,
$\tau_{\rm dyn}$ shows that a black hole is formed in the dynamical
time scale after the onset of the merger; $\tau_{\rm hyd}$ shows that
a HMNS is formed and its lifetime is determined by the time scale of
the hydrodynamical angular-momentum transport (and partly
gravitational-wave emission).  $\tau_{\rm hyd,s}$ implies that the
evolution process is the same as for $\tau_{\rm hyd}$ but the lifetime
is shorter than $\sim 10$\,ms (for this case, the gravitational-wave
emission could play an important role for inducing gravitational
collapse to a black hole); $\tau_{\rm mag}/\tau_{\rm cool}$ shows that
a HMNS is formed and its lifetime, which is longer than $\tau_{\rm
hyd}$, would be determined by the time scale of angular-momentum
transport by some magnetohydrodynamics effects or of the neutrino
cooling; ``SMNS'' shows that a SMNS is formed and its lifetime would
be much longer than $\tau_{\rm mag}$ and $\tau_{\rm cool}$.

Figure~\ref{fig:HMNS} clearly shows that the evolution process and its
lifetime of a HMNS depend strongly on its EOS and binary initial mass
$m$.  Furthermore, the dependence of the lifetime of a HMNS on the
initial mass depends strongly on the EOS. This property is well
reflected in the gravitational waveforms, as shown in
Sec.~\ref{sec:GW}.

We note that for a given EOS, a disk surrounding a black hole which is
formed after the evolution of a HMNS is larger for the {\em smaller}
total mass because of the longer lifetime and the longer
angular-momentum transport process. The most popular scenario for the
generation of short-hard gamma-ray bursts is that the merger of binary
neutron stars produces a system composed of a black hole and a massive
disk surrounding it, and the massive disk of high temperature or high
magnetic fields subsequently becomes the engine of a gamma-ray burst
jet~\cite{GRB-BNS}.  For more massive disks, the total generated
energy of the gamma-ray bursts would be higher. Thus, the total mass
of the binary system may be well reflected in the total power of the
short-hard gamma-ray bursts. The gravitational-wave observation
together with the observation of short gamma-ray bursts could test
this hypothesis~\cite{GRB-GW}.

Before closing this section, we give a comment on the convergence.  In
general, for lower grid resolutions, the lifetime of the HMNSs is
shorter.  The reason inferred is that the lower resolution results in
higher numerical dissipation. Hence, the lifetime of the HMNSs found
in the numerical result should be considered as the lower limit.  For
the case that a black hole is formed in a few ms after the onset of
the merger, by contrast, the dependence of the lifetime on the grid
resolution is quite weak (less than dynamical time scale $<$ 1 ms). 

\subsection{Dependence of the MNS evolution on binary mass ratio}
\label{sec:depQ}

A MNS formed after the merger is rapidly rotating and nonaxisymmetric
(cf.~Figs.~\ref{figcont1} and \ref{figcont2}).  Due to this fact, it
becomes a strong emitter of gravitational waves.  Here, the detailed
property of the gravitational waveform depends on the density and
velocity profiles of the MNS. The EOS determines the characteristic
radius of the MNS, and hence, the frequency of gravitational waves
depends strongly on the EOS (see Sec.~\ref{sec:GW}). The merger
process depends not only on the EOS but also on the mass ratio and
total mass. The mass ratio in particular becomes a key ingredient for
determining the evolution process of the density profile and the
configuration of a MNS in a quasistationary state.  Through this
fact, the mass ratio gives an impact on the gravitational waveforms.
In this section, we pay special attention to the dependence for the
evolution of the MNS configuration on the binary mass ratio.

First, we summarize the evolution process of MNSs for the equal-mass
case (see Fig.~\ref{figcont1}).  For this case, a dumbbell-shaped MNS
composed of two cores is formed soon after the onset of the merger
irrespective of the EOSs employed. Then, due to the loss of their
angular momentum by the hydrodynamical angular-momentum transport and
gravitational-wave emission, the shape changes gradually to an
ellipsoidal one, and the ellipticity decreases with time. Here, the
time scale of the angular-momentum loss depends on the EOS. For APR4
and SLy for which a quasiradial oscillation violently occurs in the
early evolution stage of the MNS (see Fig.~\ref{figrho}), the time
scale of the angular-momentum loss is short ($\sim 10$\,ms), while for
H4, MS1, and Shen, the time scale is rather long. For these EOSs, the
evolution time scale of MNSs from the dumbbell-like to the spheroidal shape
is relatively long $>10$\,ms.  These facts can be found from
Fig.~\ref{figcont1}, and also Fig.~4 of~\cite{skks2011a}.

For the unequal-mass case, the evolution process of the MNS
configuration is different from that for the equal-mass case.  To make
the difference clear, we focus here on the case of $m_1=1.2M_{\odot}$
and $m_2=1.5M_{\odot}$ (see Fig.~\ref{figcont2}). For this case, the
configuration of the MNS changes with the dynamical time scale in the
early evolution stage. The reason is as follows: In the merger stage,
the less-massive neutron star is tidally deformed and its outer part
is stripped during the merger. Then, the stripped material forms an
envelope of the remnant MNS while the core of the less-massive neutron
star interacts with the core of the massive companion, and the MNS is
composed of two asymmetric cores (see the first two panels of
Fig.~\ref{figcont2}).  Because the gravity of the less-massive core is
much weaker than that of the massive one, it behaves as a satellite
that is significantly and dynamically deformed by the main core,
varying its configuration with time like an amoeba. During its
evolution, the satellite is significantly elongated, encompassing the
main core. For such a case, the shape of the MNS (composed of the main
core and elongated satellite) becomes approximately spheroidal at a
moment (see the third panel of Fig.~\ref{figcont2}). For such a
moment, the emission of gravitational waves is suppressed transiently
(see Sec.~\ref{sec:GW}). 

However, after a substantial hydrodynamical angular-momentum transport
process which occurs via the interaction with the envelope, the MNS
relaxes to a quasistationary state irrespective of the EOSs
employed. The quasistationary MNS appears to be composed of major and
minor cores which are rotating in a quasistationary manner (see the
late-time snapshots of Fig.~\ref{figcont2}).  This system looks like a
hammer thrower rotating with a hammer (here the thrower is the major
core and the hammer is the minor core).  This system subsequently
loses the angular momentum primarily through the hydrodynamical
angular-momentum transport process, and thus, the degrees of asymmetry
decreases gradually, although the time scale of this change is much
longer than the dynamical time scale.

All these evolution processes of the MNSs are well reflected in their
gravitational waveforms. In the next section, we will summarize the
properties of the gravitational waveforms.


\section{Gravitational waves from MNS}\label{sec:GW}

Gravitational waves are extracted by calculating the outgoing part of
the complex Weyl scalar $\Psi_4$ at finite coordinate radii $r =
200$\,--\,$400 M_\odot$ and by integrating $\Psi_4$ twice in time (see,
e.g., \cite{kst2011} for our method). In this work, we focus only
on $(l,m)=(2,2)$ modes, which dominate over the gravitational-wave
amplitude during the MNS phase.  

\begin{figure*}[p]
\includegraphics[width=80mm,clip]{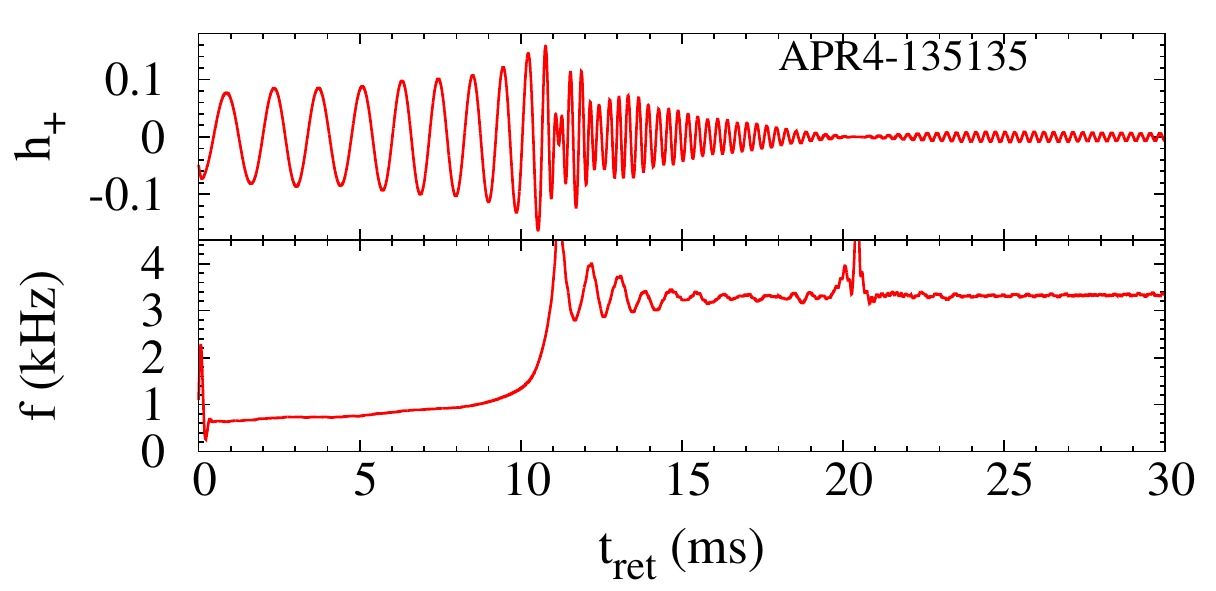}~~
\includegraphics[width=80mm,clip]{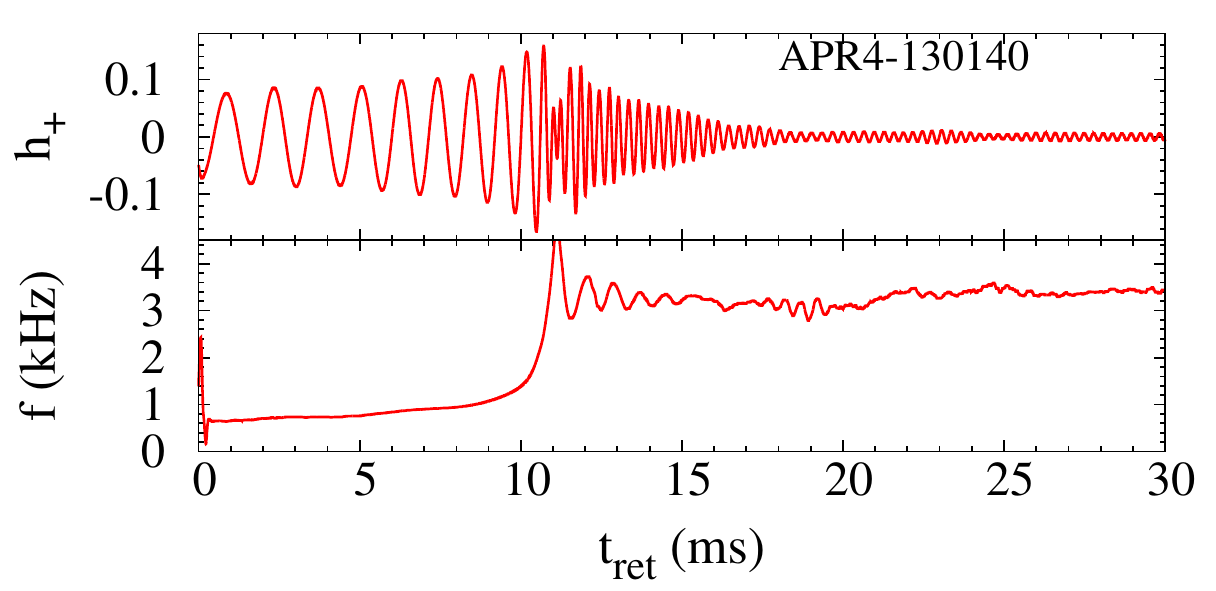}\\
\includegraphics[width=80mm,clip]{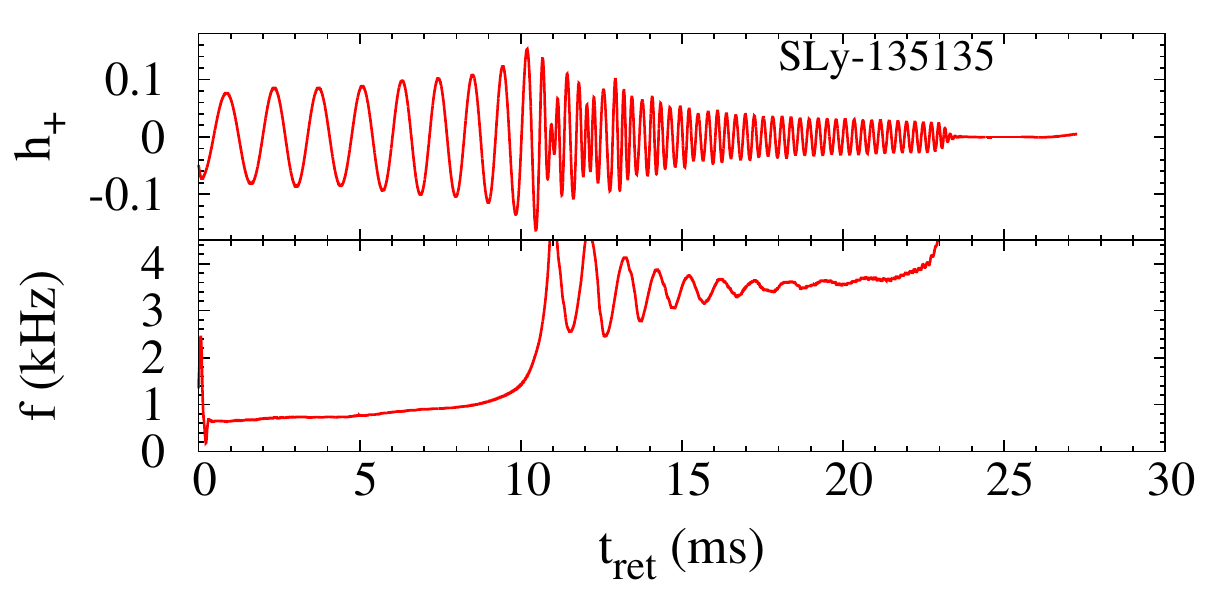}~~
\includegraphics[width=80mm,clip]{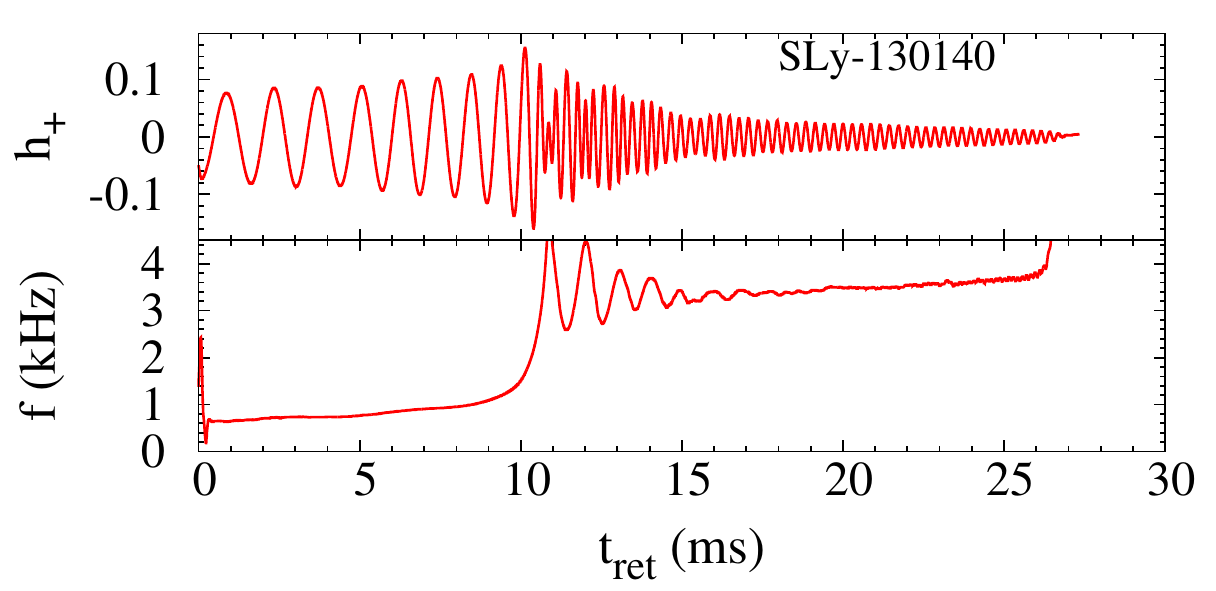}\\
\includegraphics[width=80mm,clip]{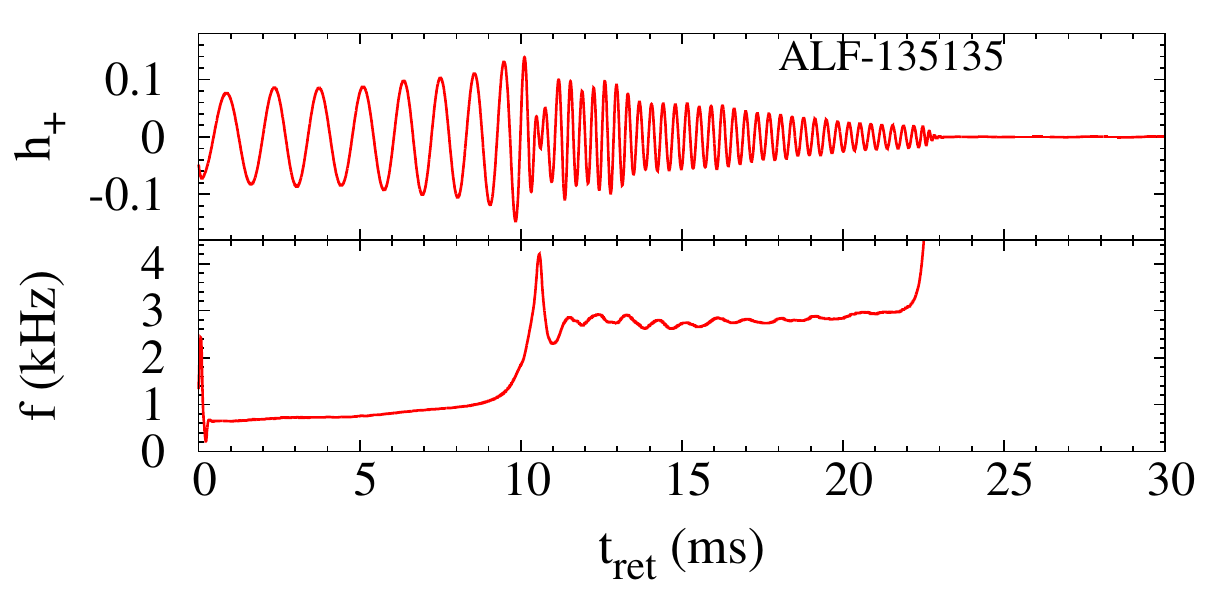}~~
\includegraphics[width=80mm,clip]{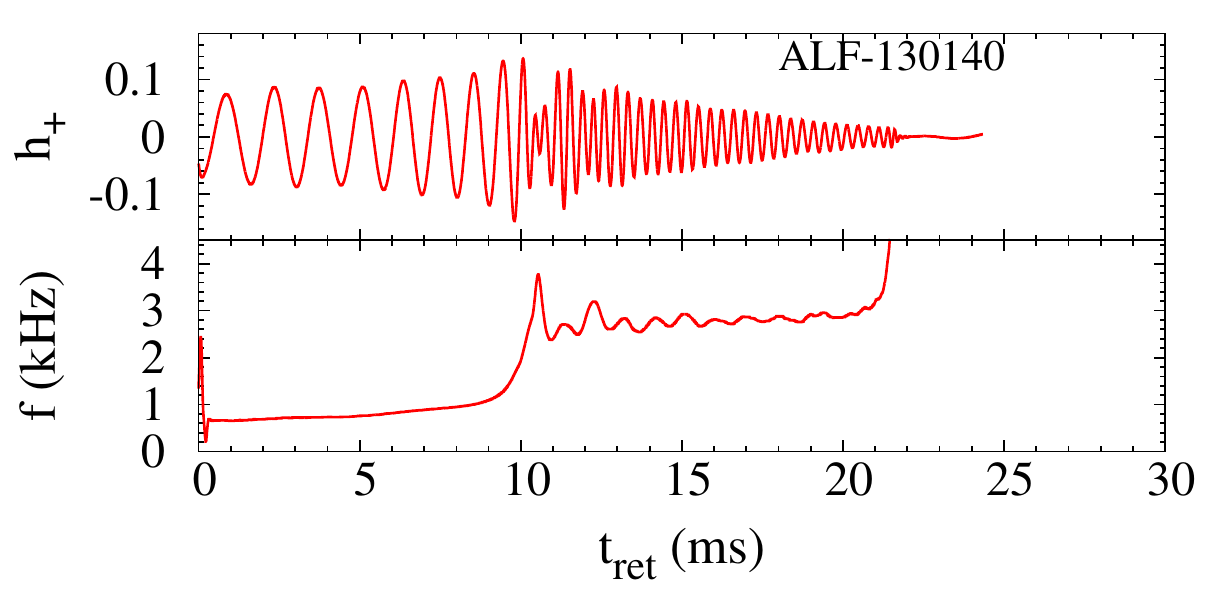}\\
\includegraphics[width=80mm,clip]{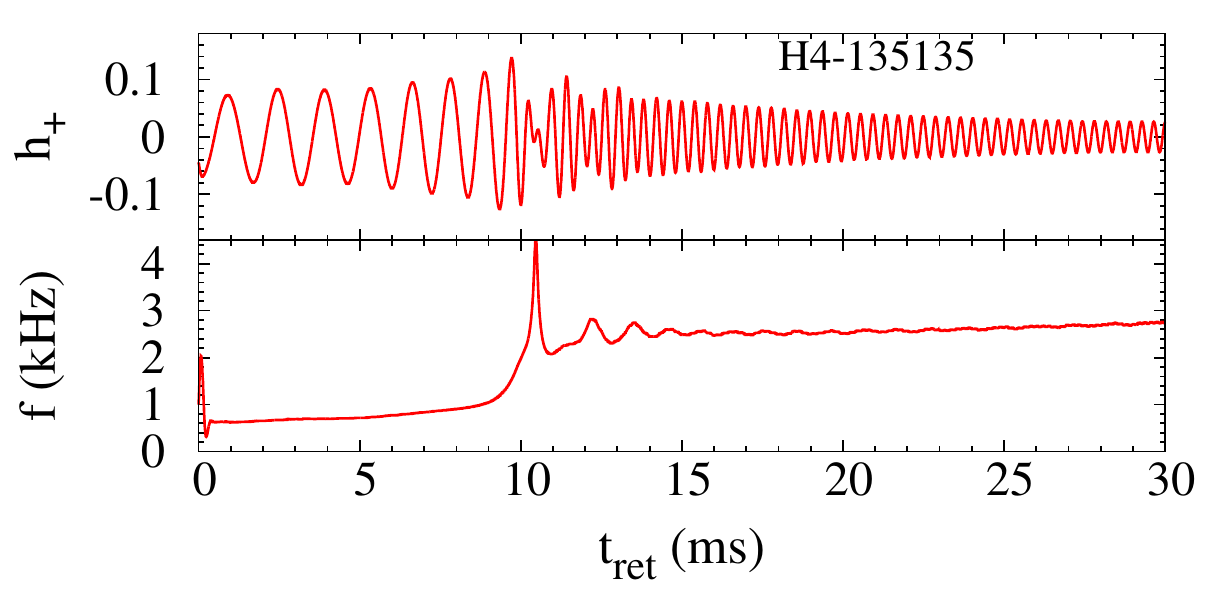}~~
\includegraphics[width=80mm,clip]{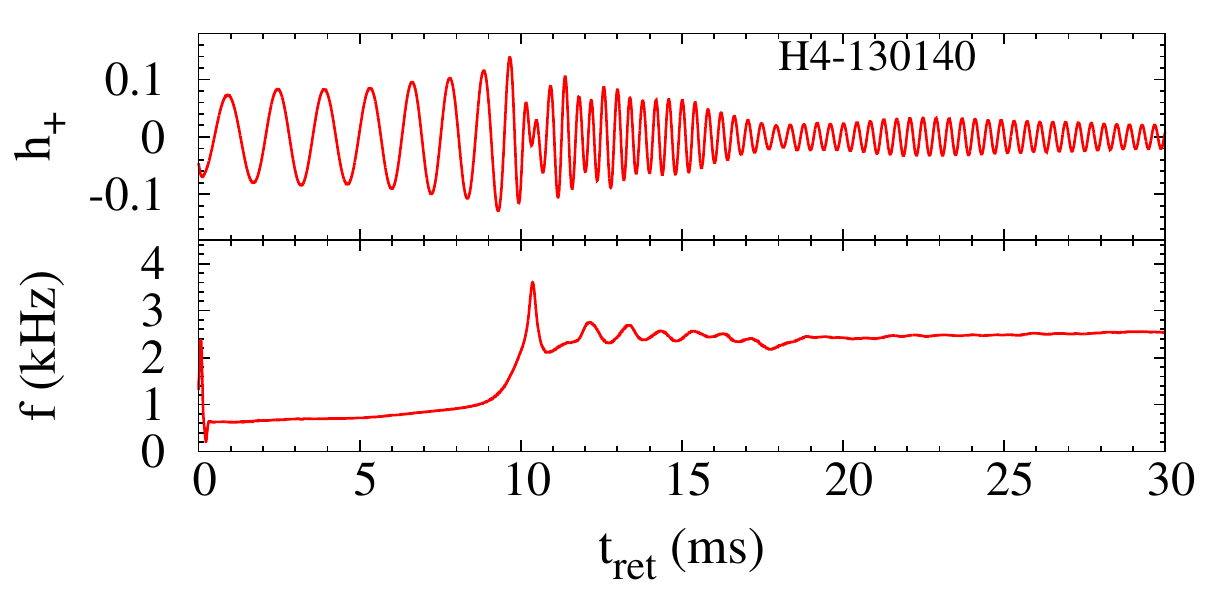}\\
\includegraphics[width=80mm,clip]{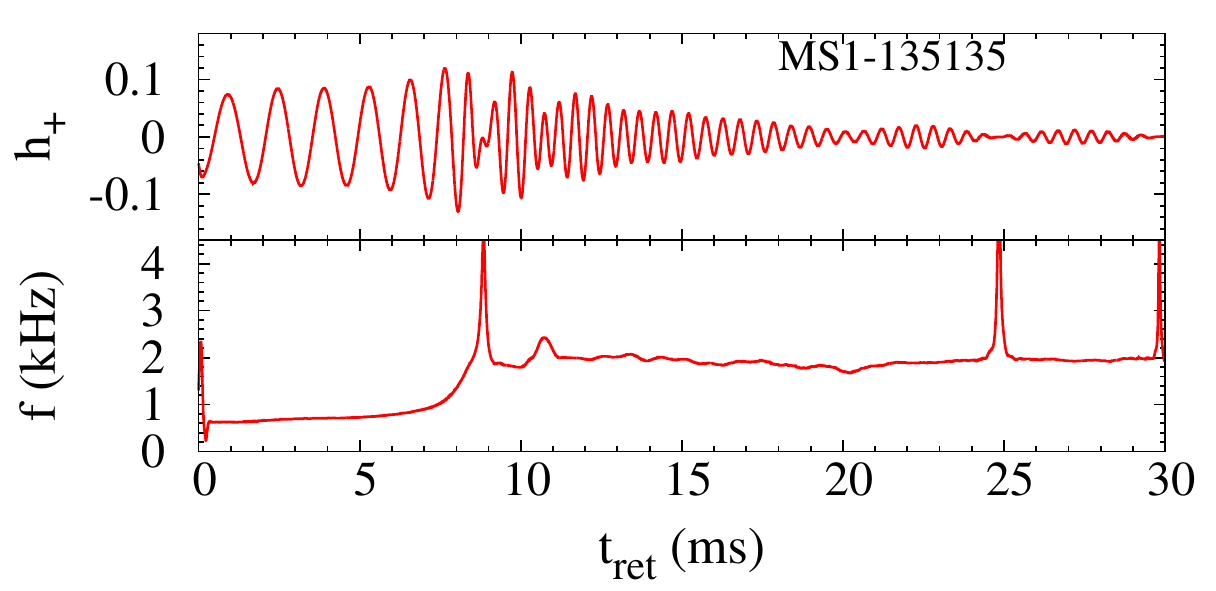}~~
\includegraphics[width=80mm,clip]{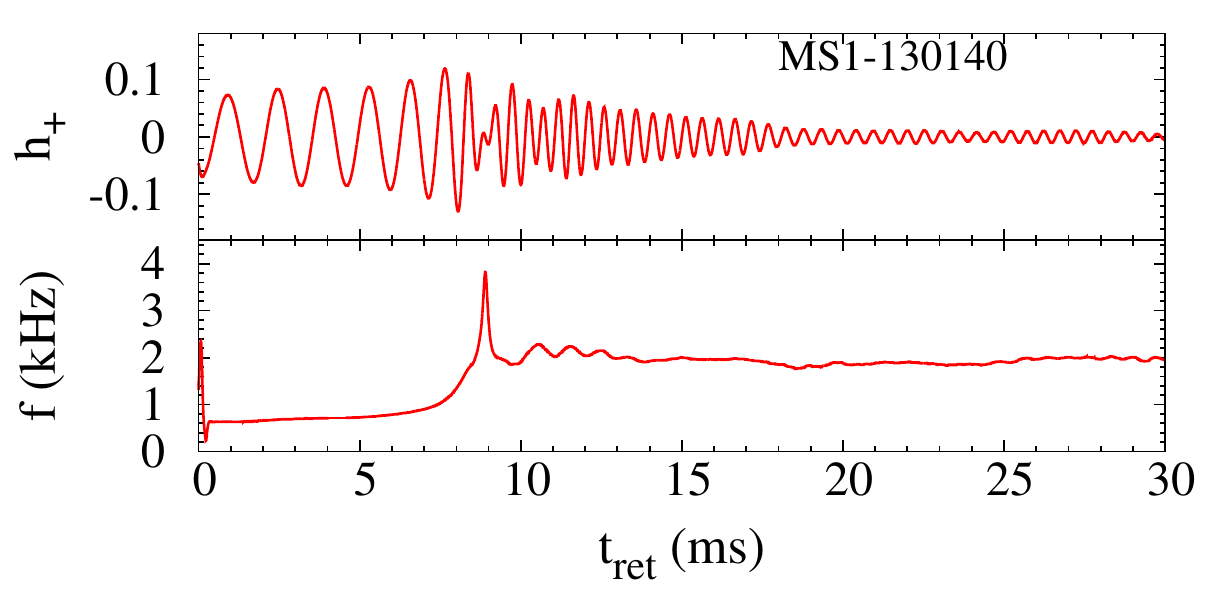}
\caption{Gravitational waves ($h_+ D/m$) and the frequency of
gravitational waves $f$\,(kHz) as functions of retarded time for
models $m_1=m_2=1.35M_{\odot}$ and $(m_1, m_2)=(1.30M_{\odot},
1.40M_{\odot})$ with APR4 (top row), SLy (second row), ALF2 (third
row), H4 (fourth row), and MS1 (bottom row).  For SLy and ALF2, a
black hole is eventually formed for $t_{\rm ret} < 30$\,ms.  For all
the models, $\Gamma_{\rm th}=1.8$.  The vertical axis of the
gravitational waveforms shows the non-dimensional amplitude, $h_+
D/m$, with $D$ being the distance to the source.  Spikes in the curves
of $f(t)$ (for the plot of APR4-135135 and MS1-135135) are not
physical; these are generated when the gravitational-wave amplitude is
too low to determine the frequency accurately. }
\label{figGW1}
\end{figure*}

\begin{figure*}[p]
\includegraphics[width=80mm,clip]{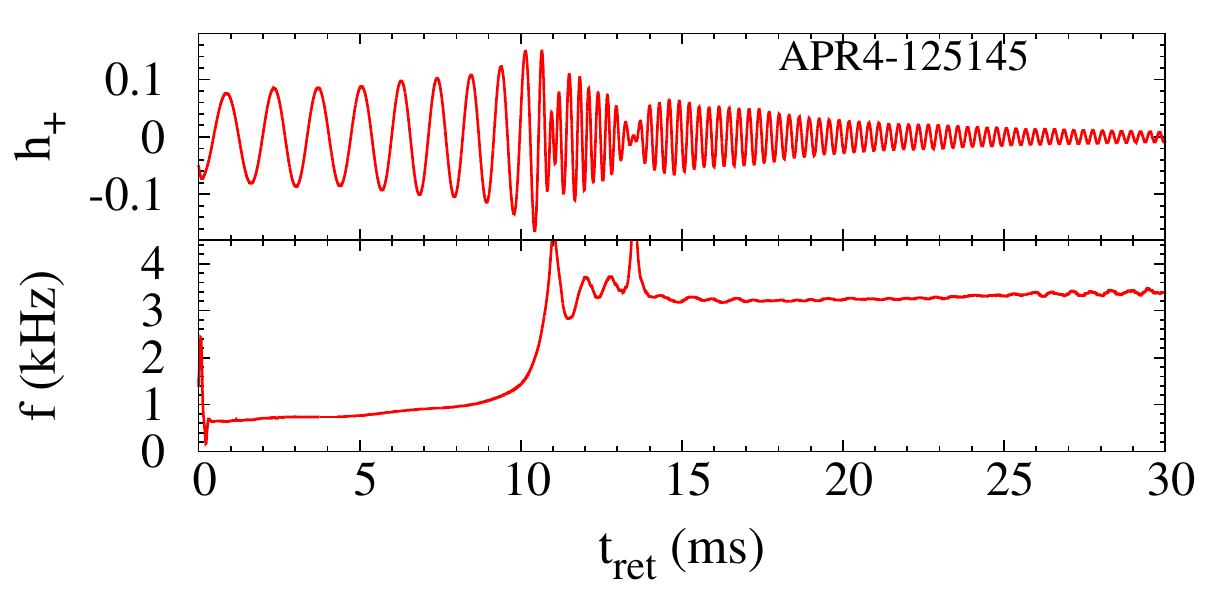}~~
\includegraphics[width=80mm,clip]{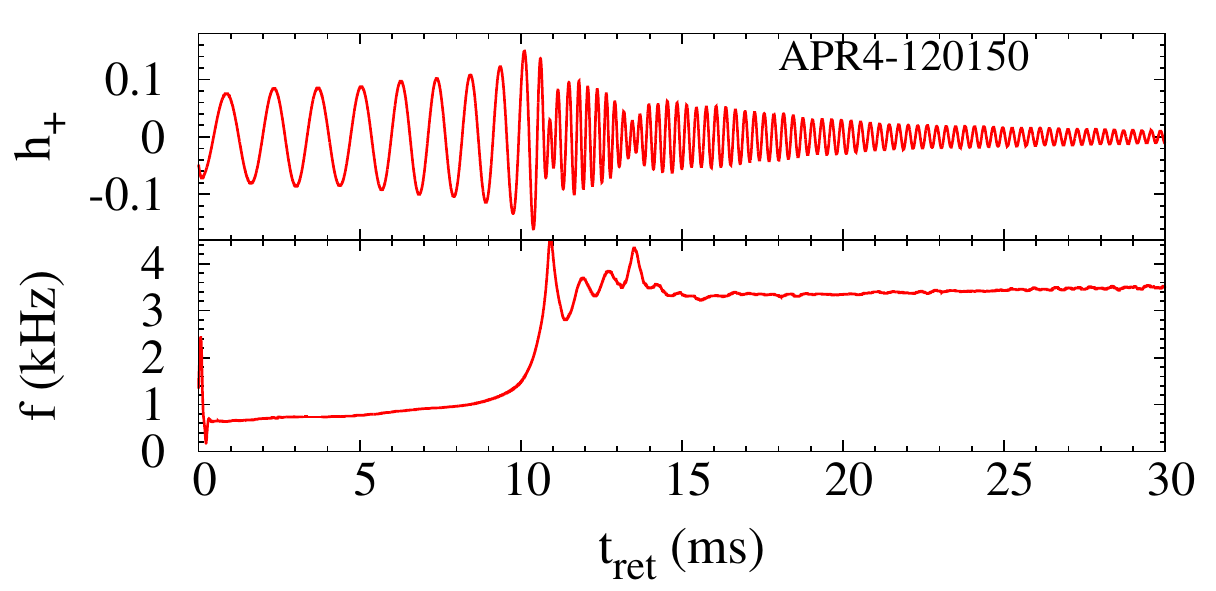}\\
\includegraphics[width=80mm,clip]{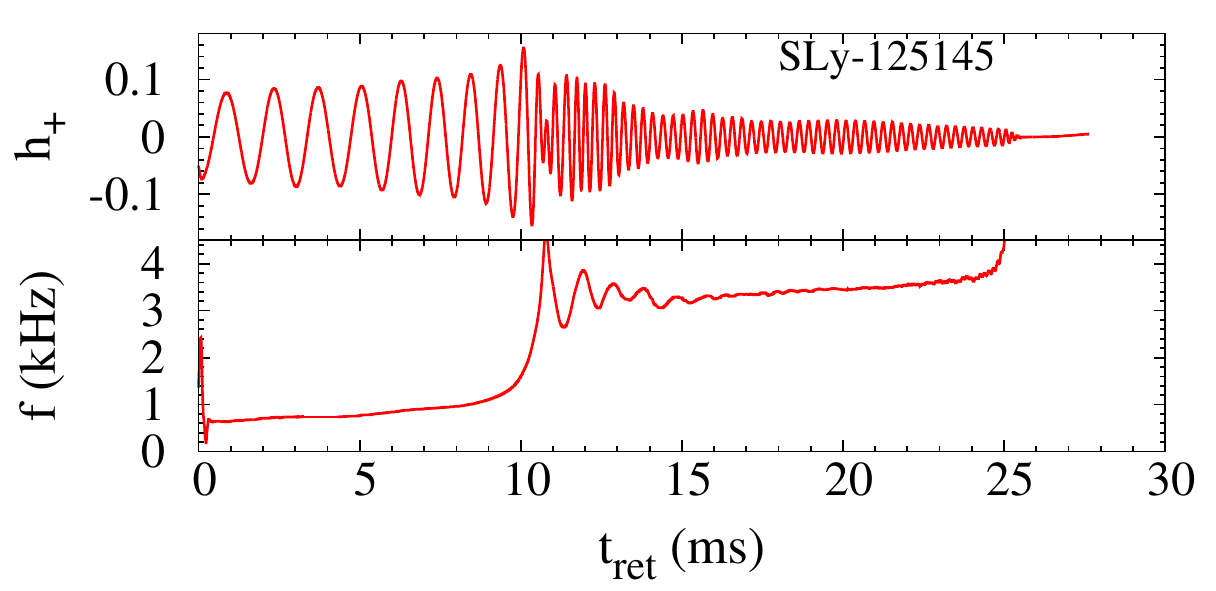}~~
\includegraphics[width=80mm,clip]{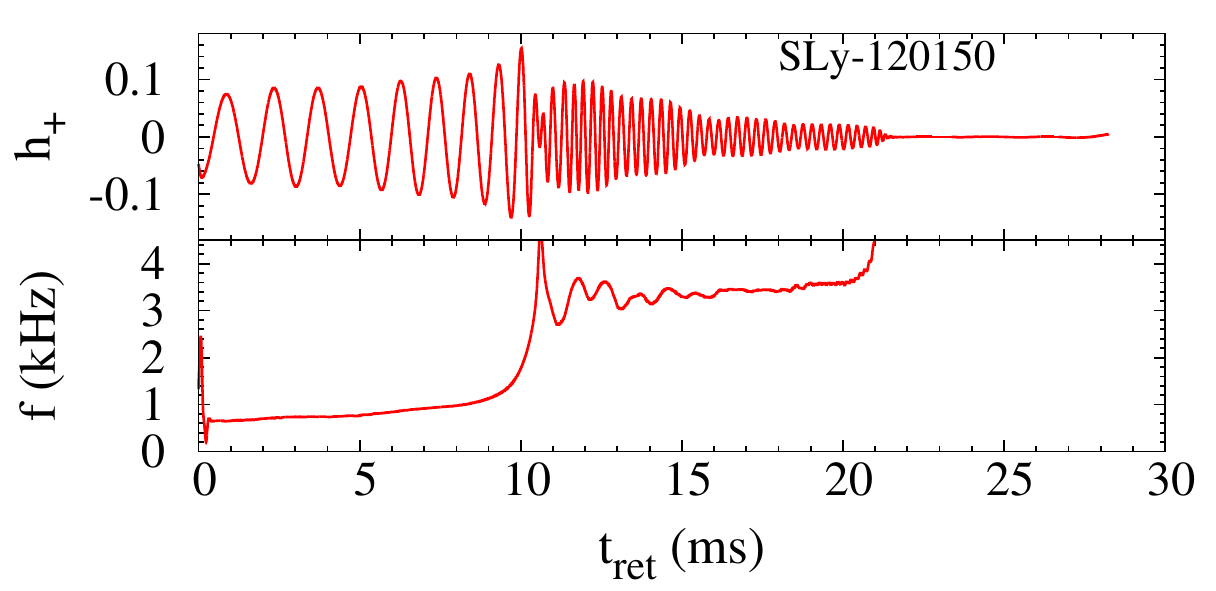}\\
\includegraphics[width=80mm,clip]{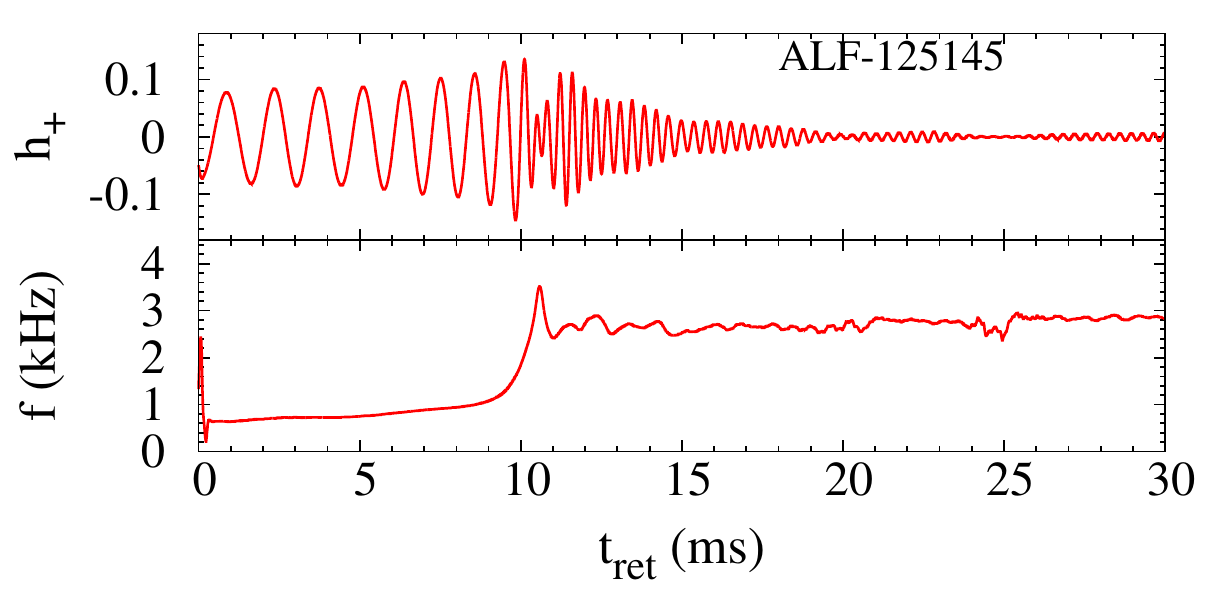}~~
\includegraphics[width=80mm,clip]{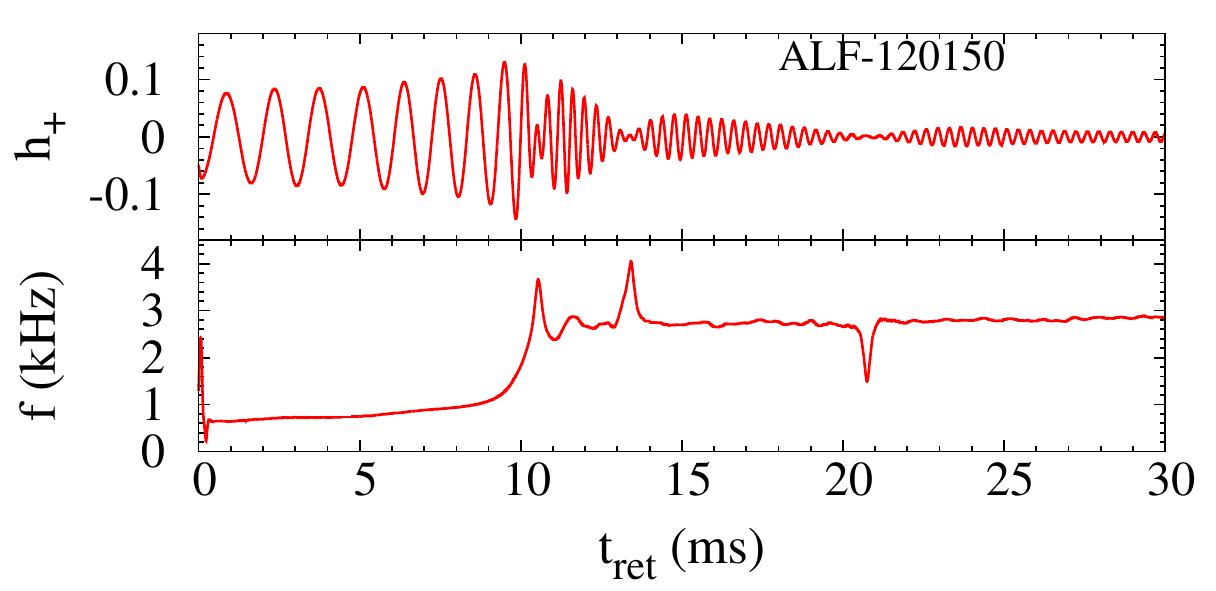}\\
\includegraphics[width=80mm,clip]{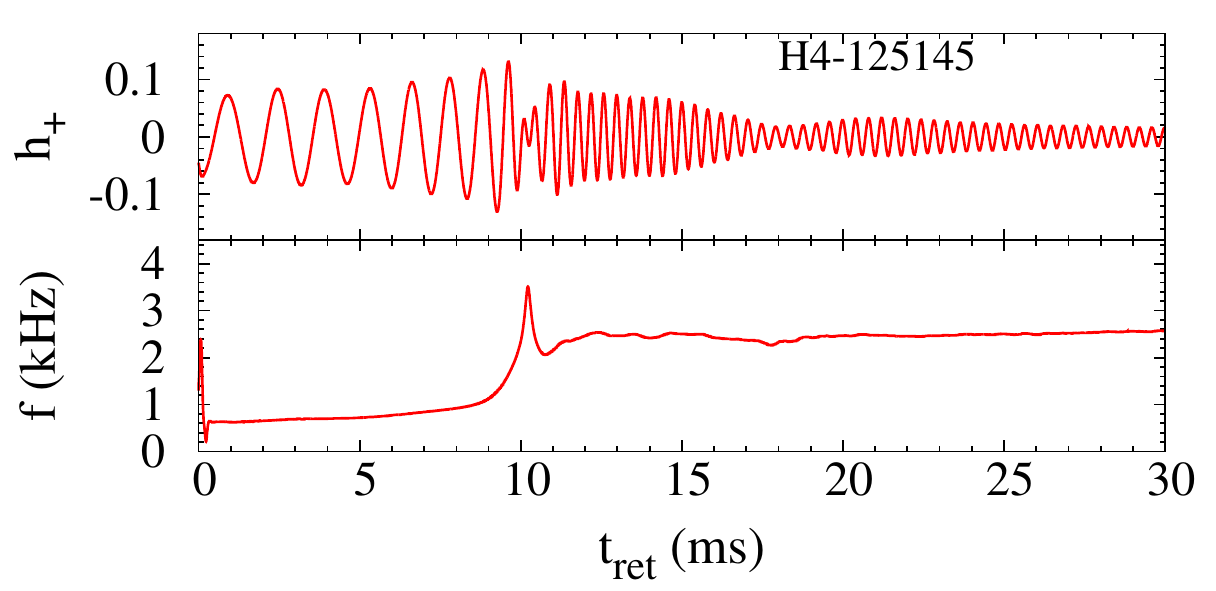}~~
\includegraphics[width=80mm,clip]{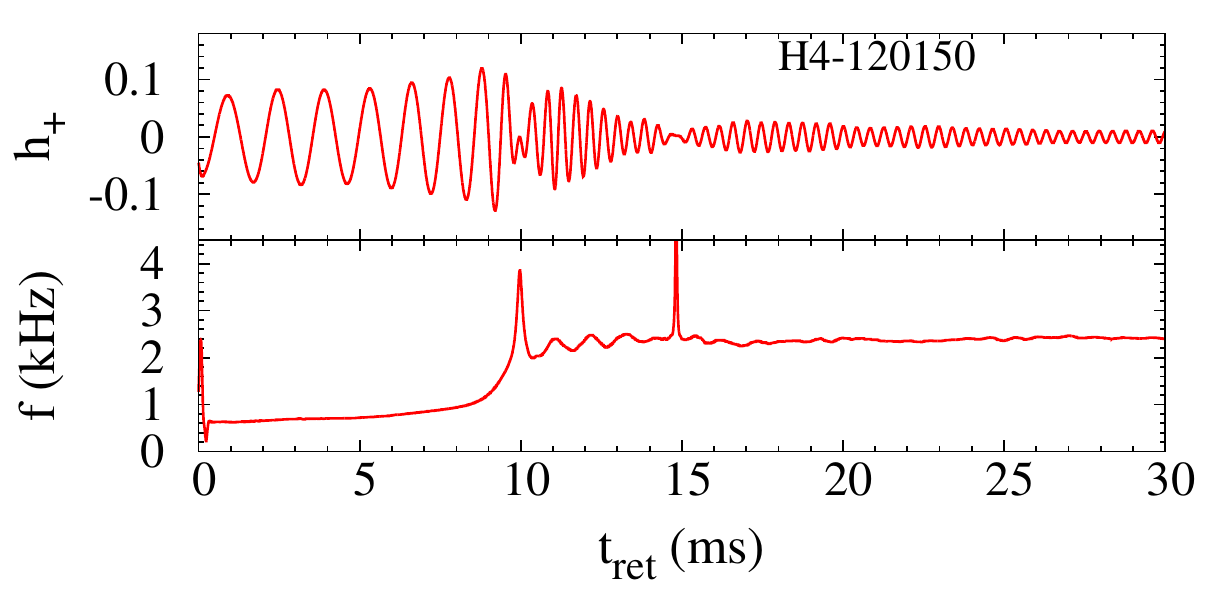}\\
\includegraphics[width=80mm,clip]{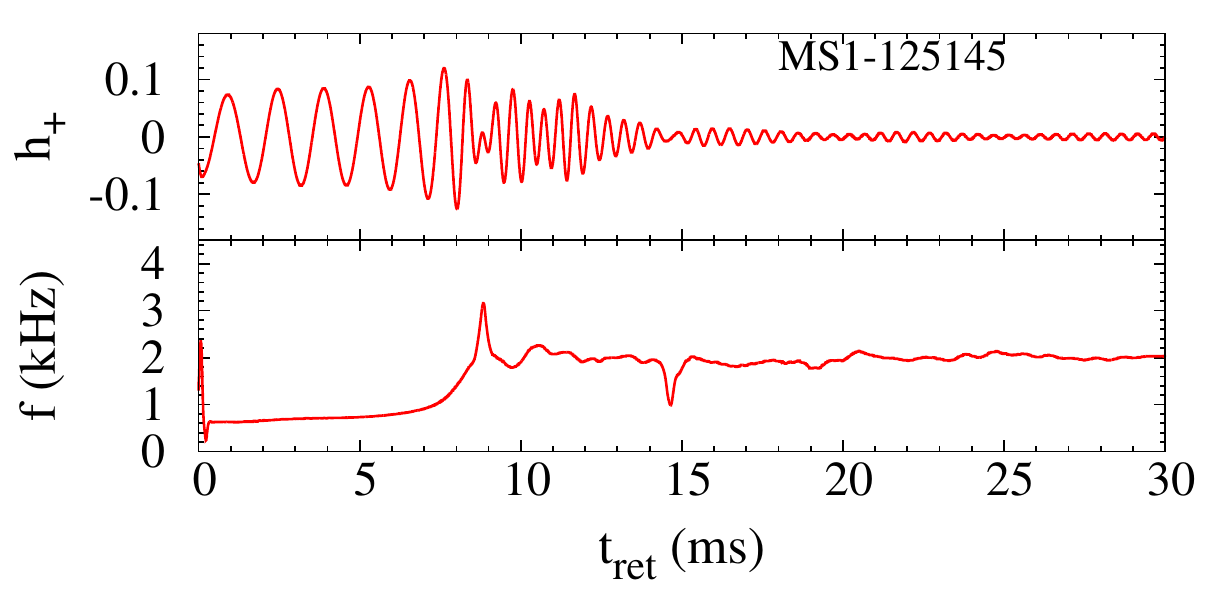}~~
\includegraphics[width=80mm,clip]{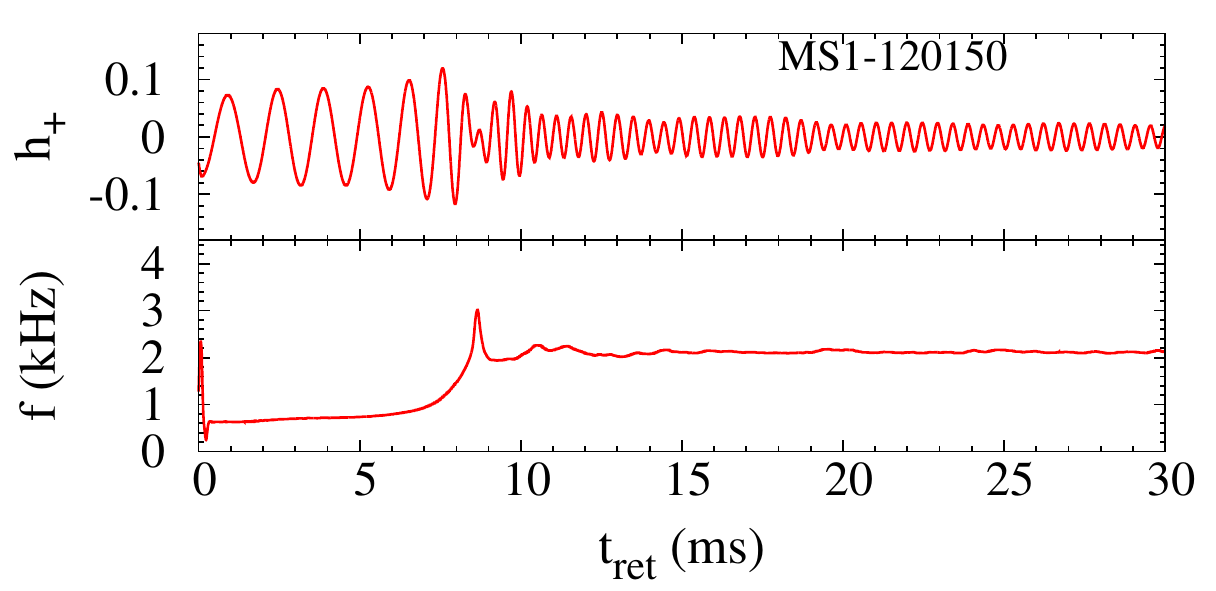}
\caption{The same as Fig.~\ref{figGW1} but for $(m_1,
m_2)=(1.25M_{\odot}, 1.45M_{\odot})$ (left) and $(1.20M_{\odot},
1.50M_{\odot})$ (right).  }
\label{figGW2}
\end{figure*}

\begin{figure*}[p]
\includegraphics[width=80mm,clip]{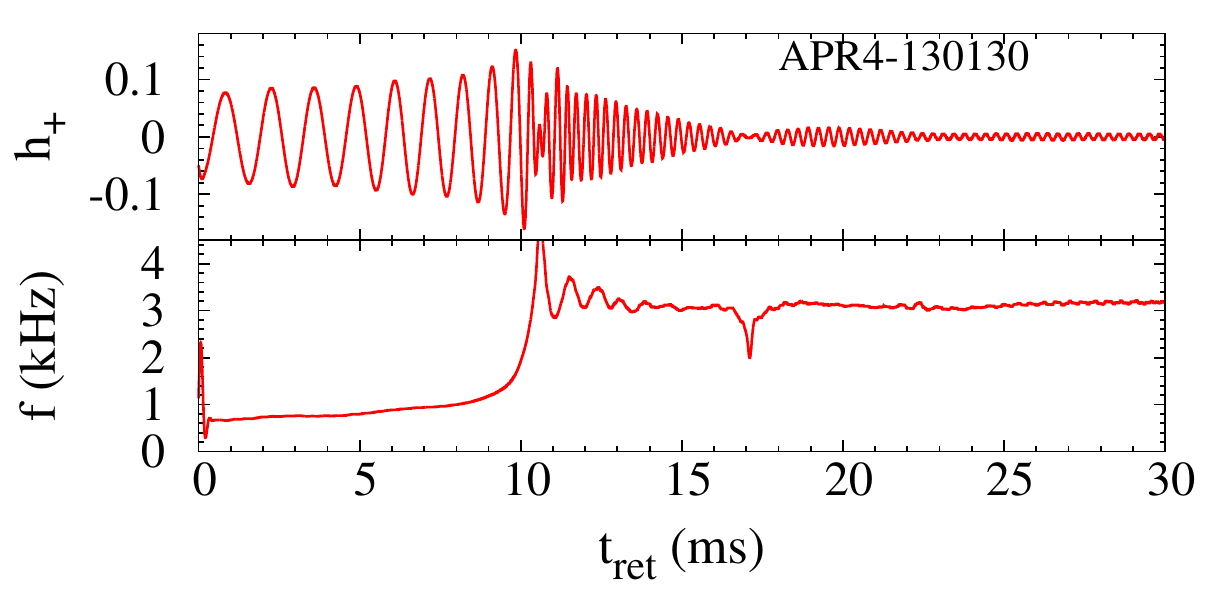}~~
\includegraphics[width=80mm,clip]{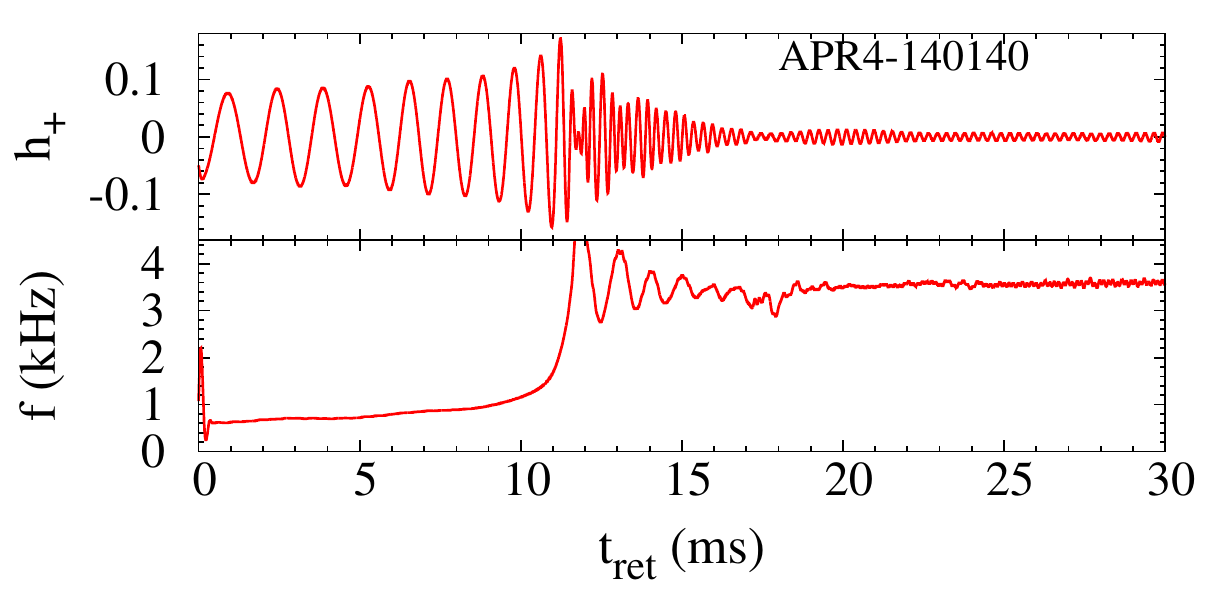} \\
\includegraphics[width=80mm,clip]{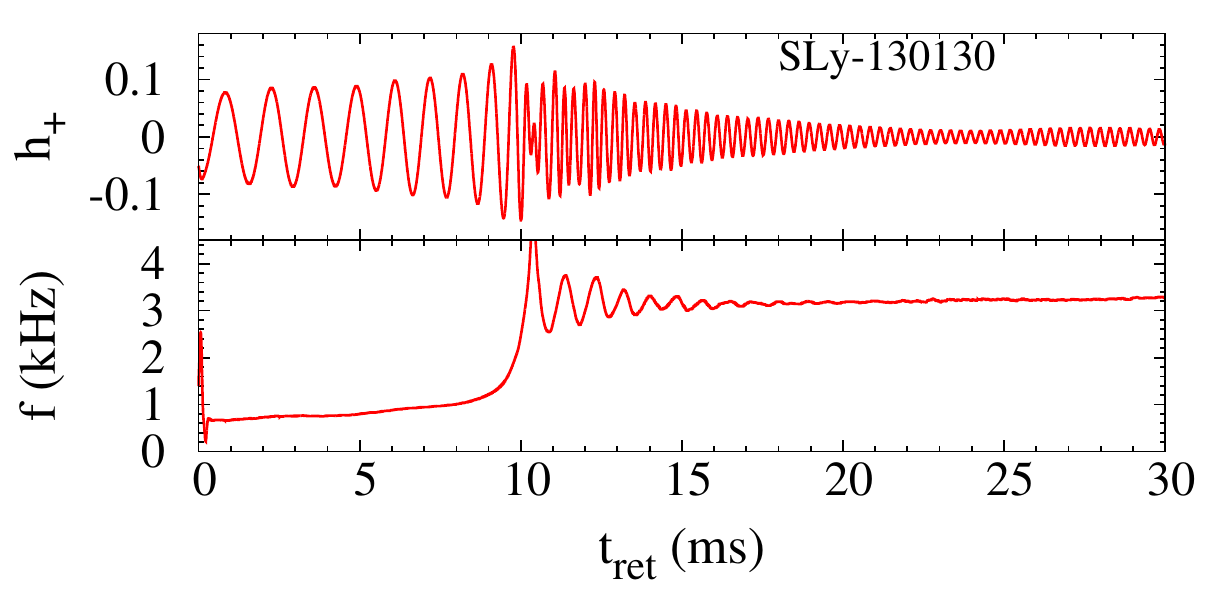}~~ 
\includegraphics[width=80mm,clip]{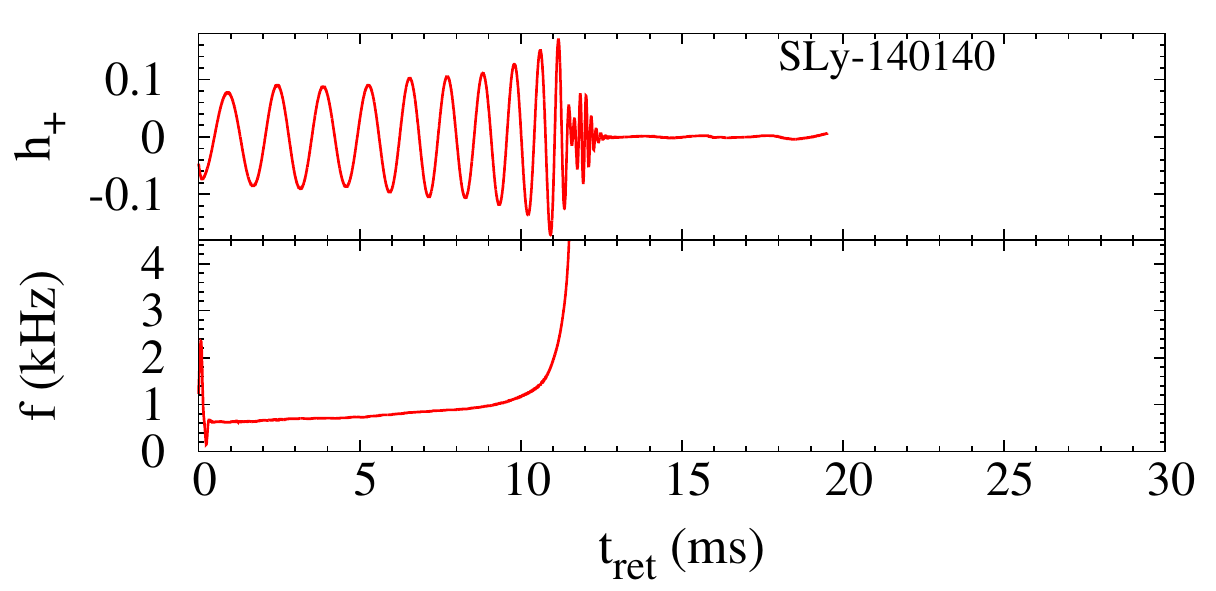}\\
\includegraphics[width=80mm,clip]{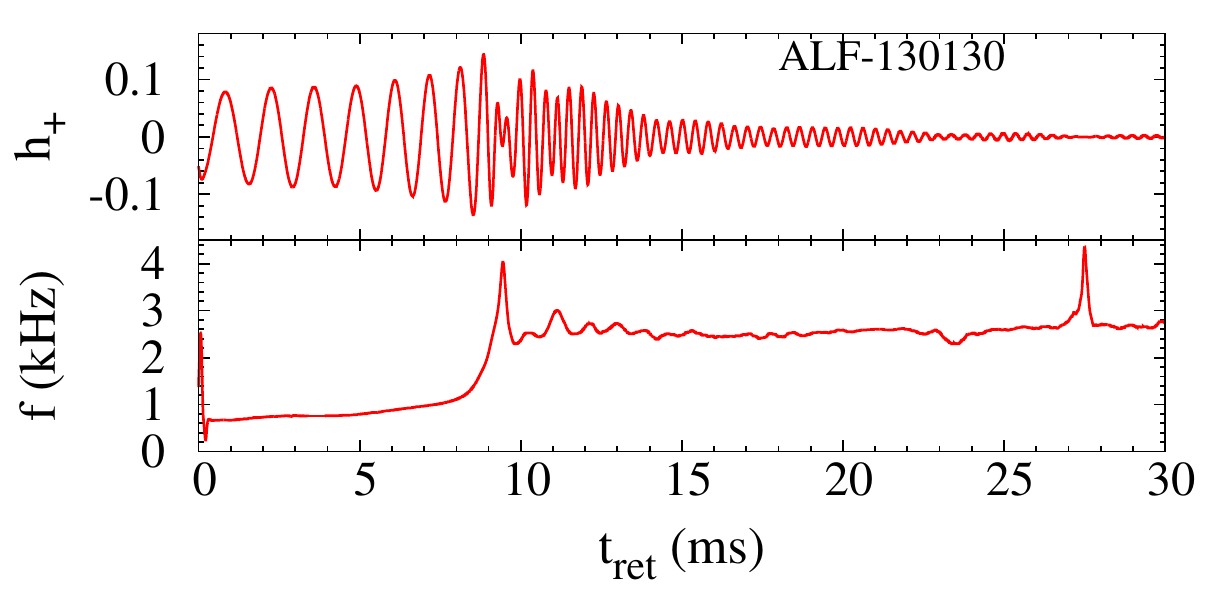}~~ 
\includegraphics[width=80mm,clip]{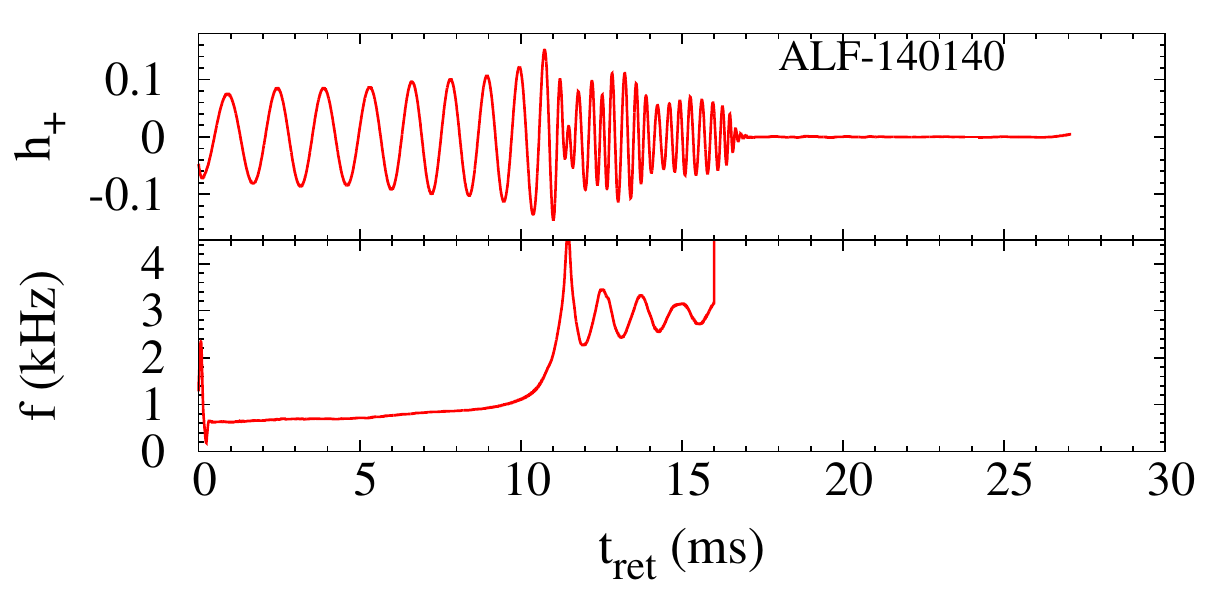}\\
\includegraphics[width=80mm,clip]{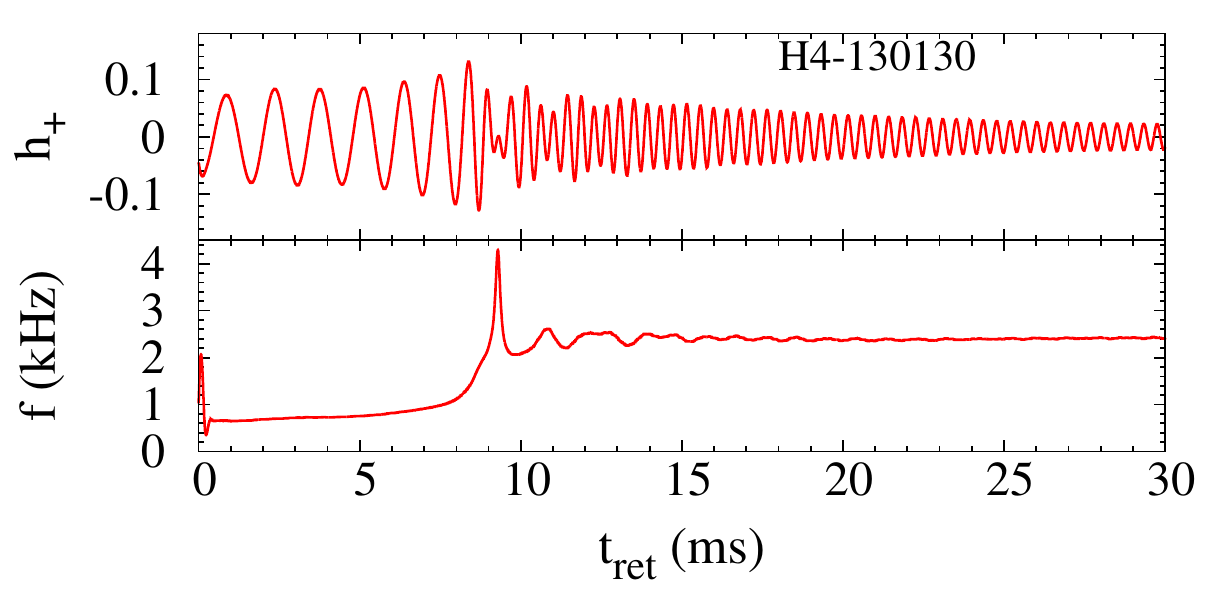}~~
\includegraphics[width=80mm,clip]{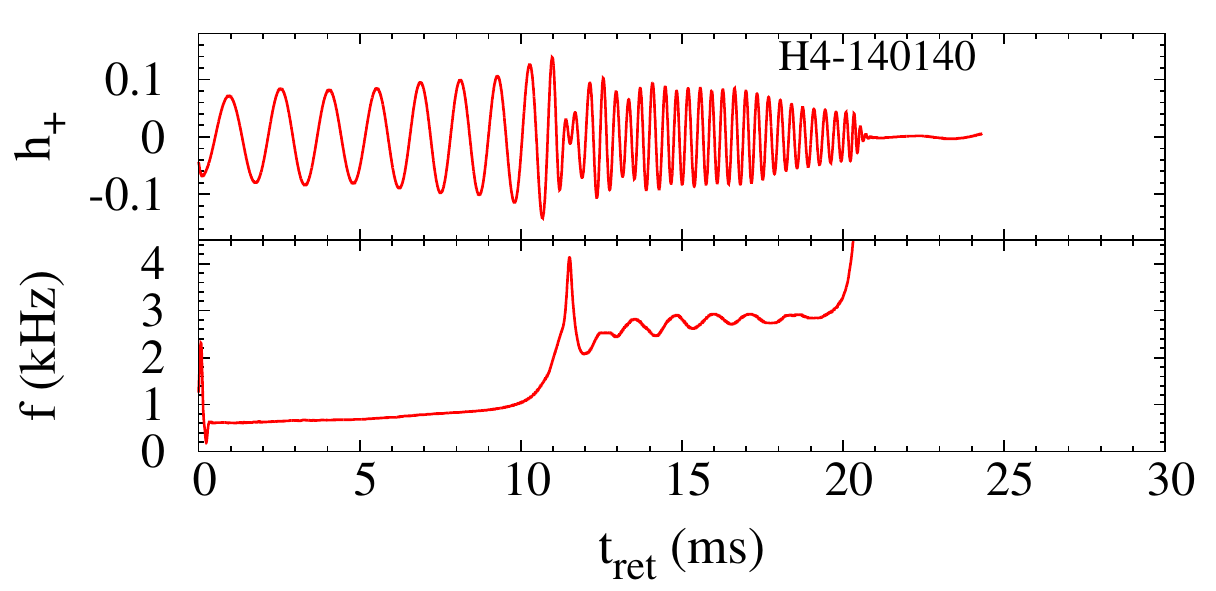} \\
\includegraphics[width=80mm,clip]{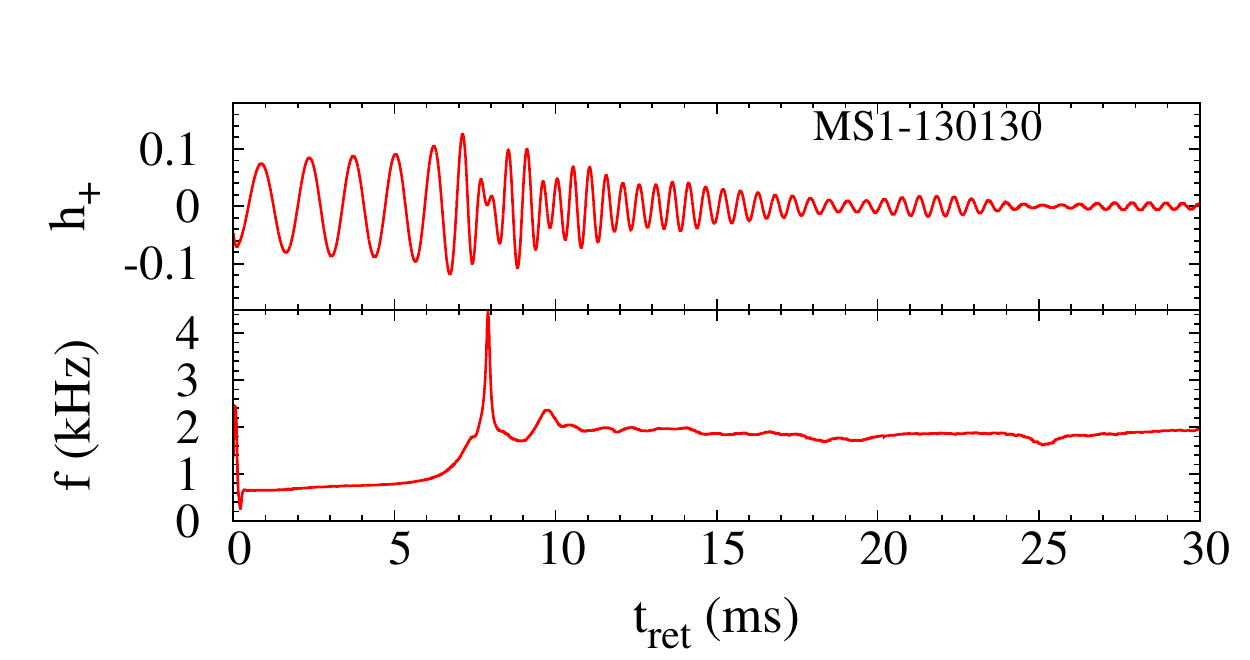}~~
\includegraphics[width=80mm,clip]{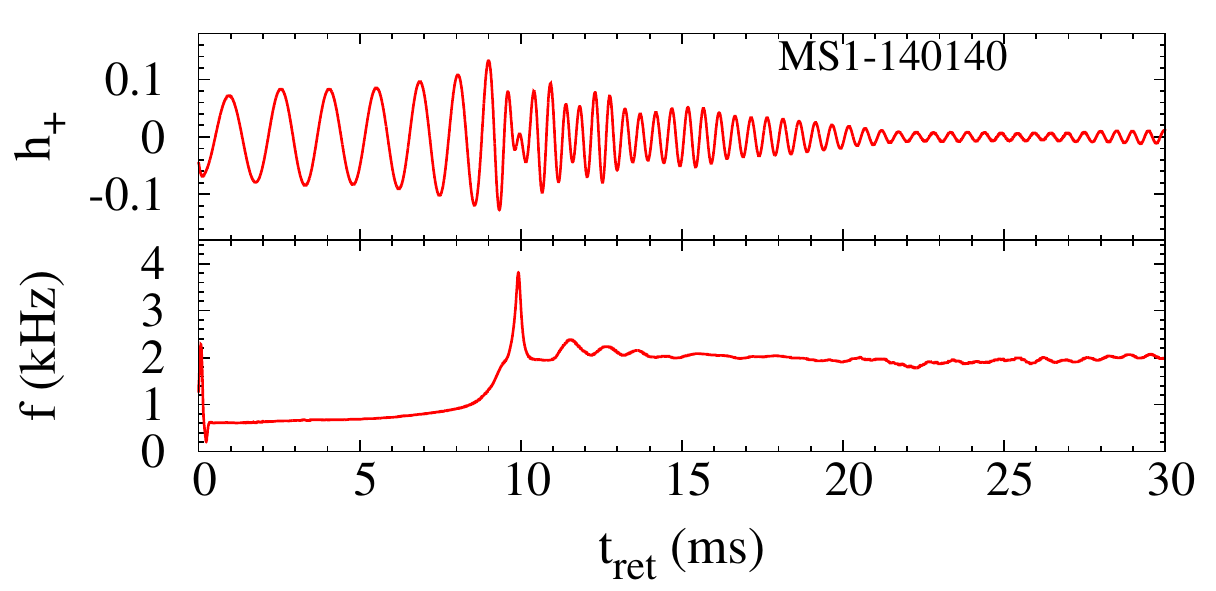} 
\caption{The same as Fig.~\ref{figGW1} but for $m_1=m_2=1.3M_{\odot}$
(left) and $m_1=m_2=1.4M_{\odot}$ (right). }
\label{figGW3}
\end{figure*}

\begin{figure*}[t]
\includegraphics[width=80mm,clip]{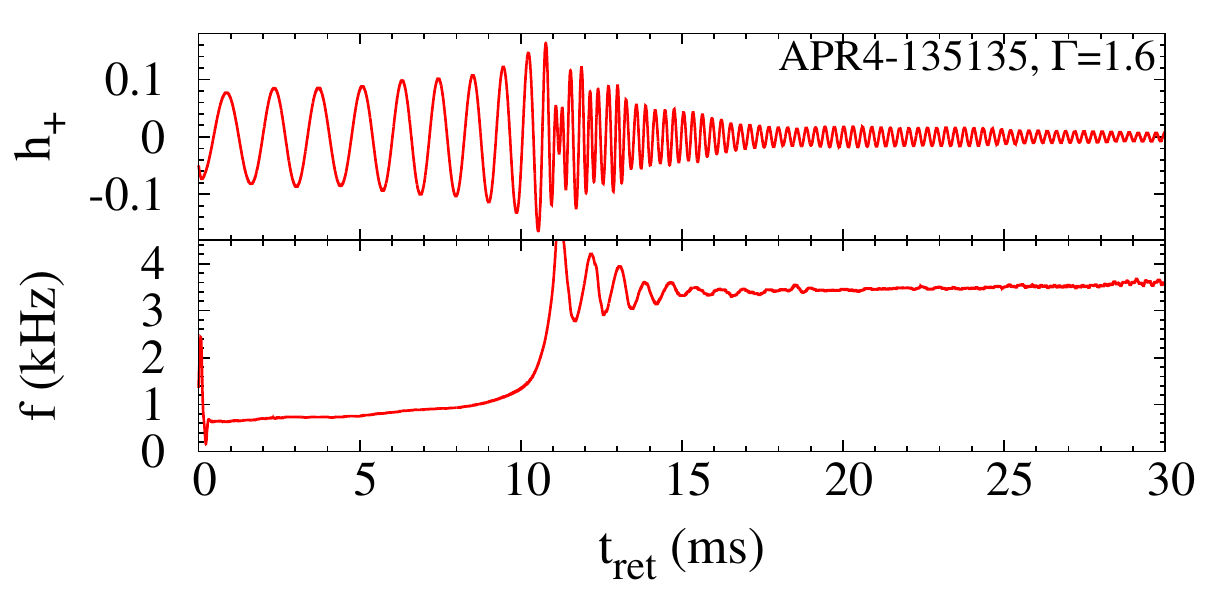}~~
\includegraphics[width=80mm,clip]{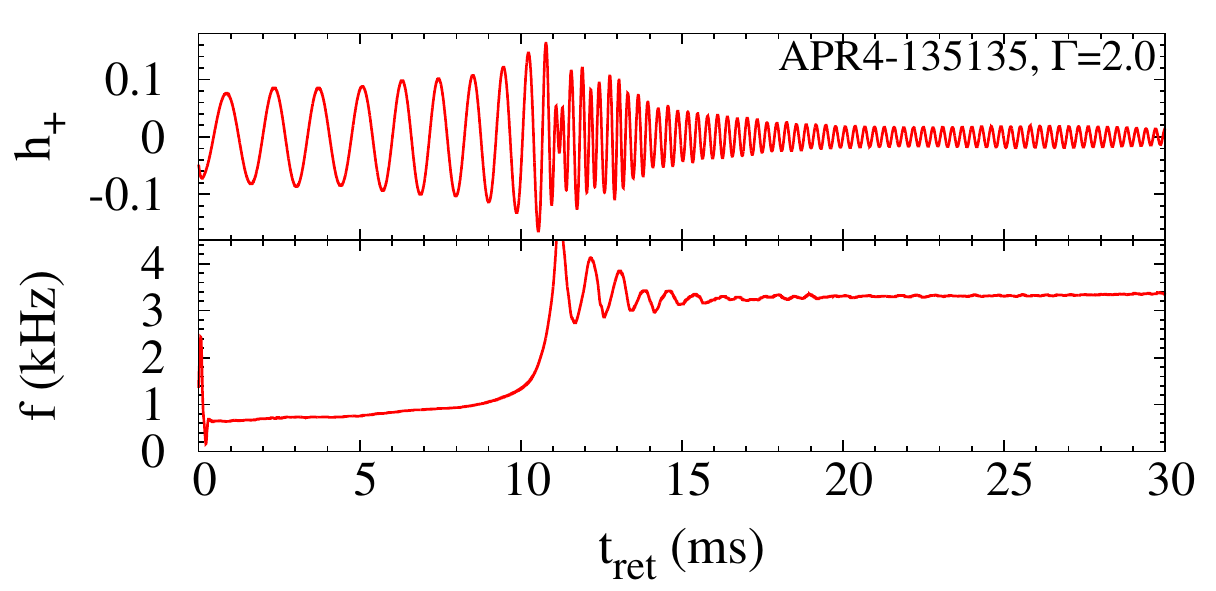}\\
\includegraphics[width=80mm,clip]{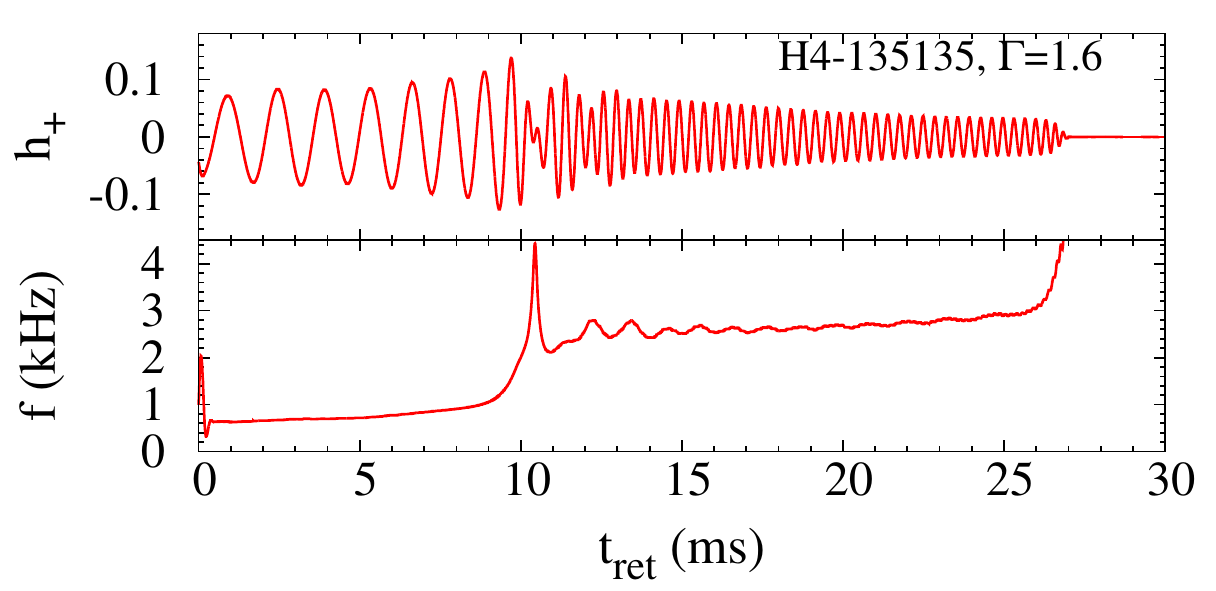}~~
\includegraphics[width=80mm,clip]{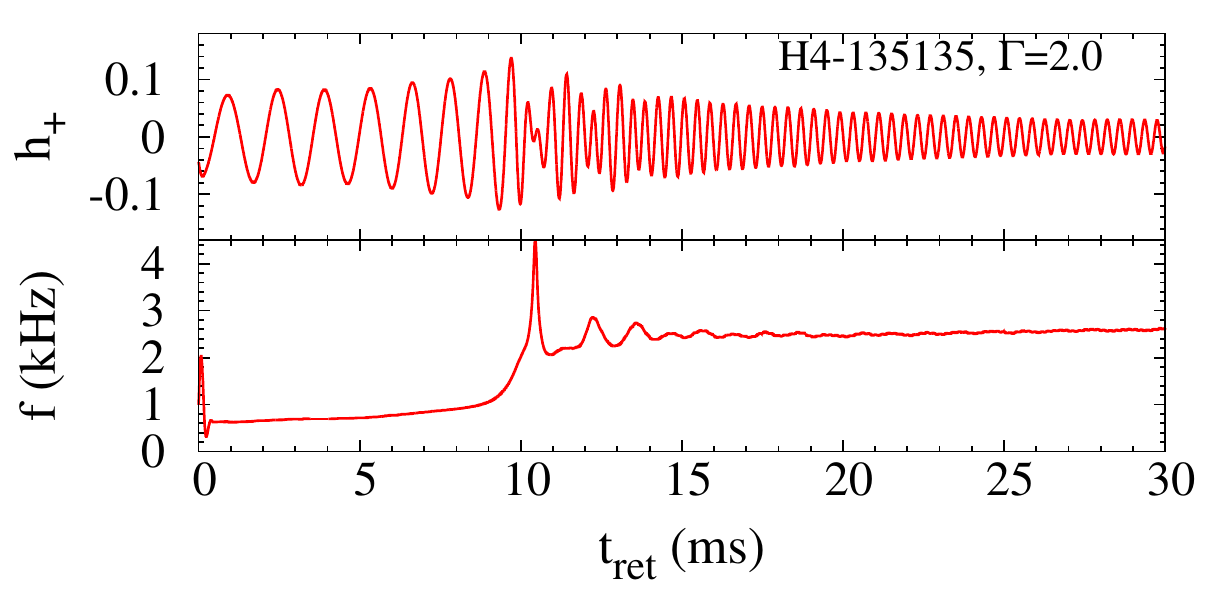}
\caption{The same as Fig.~\ref{figGW1} but for $m_1=m_2=1.35M_{\odot}$
and $\Gamma_{\rm th}=1.6$ (left) and 2.0 (right) with APR4 (upper row)
and H4 (lower row) EOSs. }
\label{figGW4}
\end{figure*}

\begin{figure*}[t]
\includegraphics[width=80mm,clip]{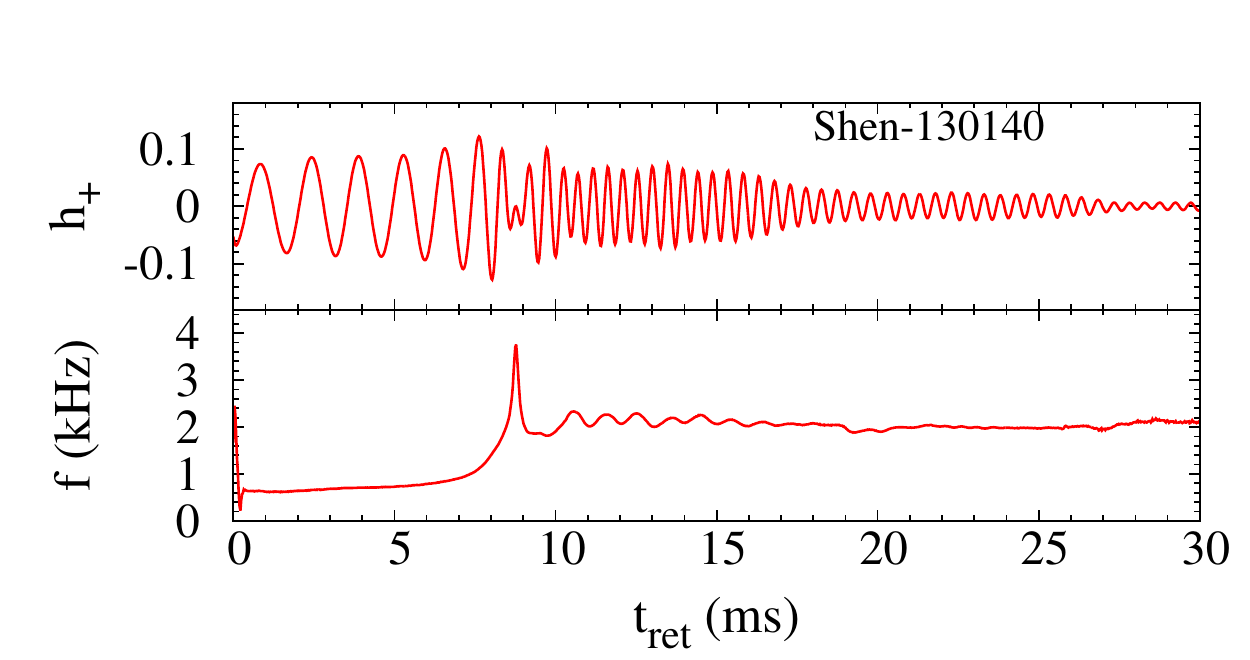}~~
\includegraphics[width=80mm,clip]{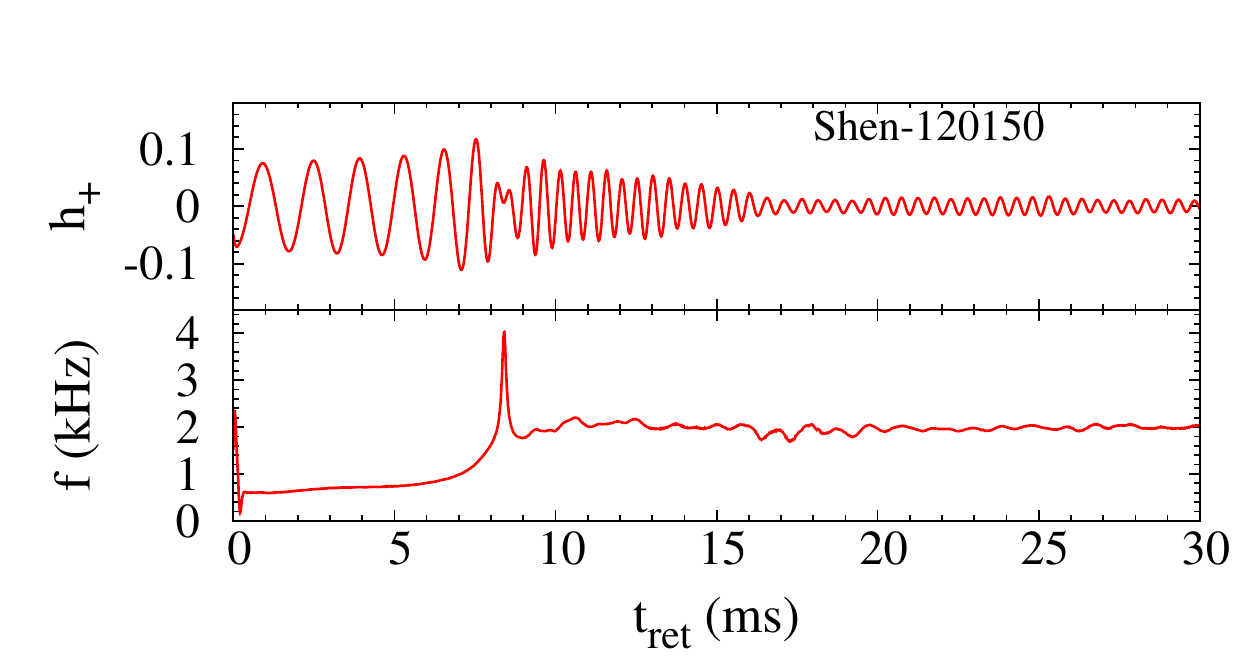}
\caption{The same as Fig.~\ref{figGW1} but for 
Shen EOSs with $(m_1, m_2)=(1.3M_{\odot},1.4M_{\odot})$ 
and $(1.2M_{\odot},1.5M_{\odot})$. }
\label{figGW5}
\end{figure*}

Figures~\ref{figGW1}\,--\,\ref{figGW5} display the plus mode of
gravitational waves, $h_+$, and the corresponding frequencies of
gravitational waves emitted by MNSs for a variety of EOSs and binary
masses. Here, gravitational waves shown are those observed along the
rotational axis which is perpendicular to the binary orbital plane,
and defined by $h_+ D/m$ where $D$ is the distance from the
source. The frequency is determined by the change rate of the phase of
$h:=h_+ - i h_{\times}$ with $h_{\times}$ being the cross mode of
gravitational waves. In this work, we evaluate the frequency
by calculating $\Psi_{4}/\int \Psi_{4}dt$ as employed in Ref.~\cite{yst2008}.
Figures~\ref{figGW1} and \ref{figGW2} display
gravitational waves for $m=2.7M_{\odot}$ with four mass ratios and
five piecewise polytropic EOSs (APR, SLy, ALF2, H4, and MS1);
Figure~\ref{figGW3} displays gravitational waves for the equal-mass
models with $m=2.6M_{\odot}$ and $2.8M_{\odot}$ with five piecewise
polytropic EOSs; Figure~\ref{figGW4} displays gravitational waves for
$m_1=m_2=1.35M_{\odot}$ with APR4 and H4 EOSs and with $\Gamma_{\rm
th}=1.6$ and 2.0.  Figure~\ref{figGW5} displays gravitational waves
for Shen EOSs with $(m_1,m_2)=(1.3M_{\odot},1.4M_{\odot})$ and
$(1.2M_{\odot},1.5M_{\odot})$.  Gravitational waves for Shen EOS are
also shown in Fig.~4 of~\cite{skks2011a} for the equal-mass
models, to which the reader may refer: For this EOS, the gravitational
waveforms are qualitatively similar to those for H4.

Gravitational waves emitted by MNSs are characterized by their
quasiperiodic nature. Namely, the frequency of gravitational waves
remains approximately constant for more than 10 wave cycles.
Nevertheless, the frequency of gravitational waves is not totally
constant and changes with time.  Furthermore, the characteristic
frequency and the time-variation feature of the frequency and
amplitude depend strongly on the EOS, total mass, and mass ratio of
the binary system.  In the following, we summarize the features of
gravitational waves emitted by MNSs in more detail.

\subsection{Amplitude} \label{sec:Amplitude}

Broadly speaking, the amplitude of quasiperiodic gravitational waves
emitted by MNSs decreases with time because the angular momentum of
the MNSs is lost by the hydrodynamical angular-momentum transport
process and gravitational-wave emission.  However, the feature in the
time variation depends on the EOS and mass ratio of the binary.

There are two patterns for the damping process of the
gravitational-wave amplitude. One is that the amplitude decreases
approximately monotonically with time (besides small modulation), and
the damping time scale increases with time. This is the case for the
equal-mass model for all the piecewise polytropic EOSs, APR4, SLy,
ALF2, H4, and MS1 irrespective of the value of $\Gamma_{\rm th}$ and
total mass, as well as for Shen (see Figs.~\ref{figGW1}, \ref{figGW3},
and~\ref{figGW4} as well as Fig.~4 of~\cite{skks2011a}). For APR4
and SLy, an oscillating MNS of nonaxisymmetric shape (dumbbell-like or
ellipsoidal) is formed and quickly loses its angular momentum by the
hydrodynamical angular-momentum transport and gravitational-wave
emission.  The damping time scale of the nonaxisymmetric degree (and
hence the gravitational-wave amplitude) is short, $\sim 10$\,ms,
during the stage that the quasiradial oscillation amplitude of the MNS
is high. Subsequently, the MNS settles to a weakly deformed
quasistationary ellipsoid, and then, the gravitational-wave amplitude
relaxes approximately to a small constant. For H4, MS1, and Shen, by
contrast, the damping time scale of gravitational-wave amplitude is
relatively long for the entire evolution stage of the MNSs. This seems
to be due to the fact that the angular-momentum transport process from
the MNS to the surrounding envelope is not as efficient as in the APR4
and SLy cases. The probable reason for this is that the radial
oscillation amplitude of the MNSs is low for these stiff EOSs, and
thus, a quasistationary nonaxisymmetric MNS is formed in a short time
scale (a few ms) after the onset of the merger. Namely, the merger
proceeds relatively in a mild way, resulting in a long-term
angular-momentum transport process. Figures~\ref{figGW3}
and~\ref{figGW4} show that this fact holds irrespective of the total
mass and the value of $\Gamma_{\rm th}$. For ALF2, the efficiency of
the angular-momentum transport is lower than for APR4 and SLy but
higher than for H4, MS1, and Shen. Thus, the damping time scale of
gravitational-wave amplitude is between two cases.

One point to be noted is that the gravitational-wave amplitude for the
late stage (for $t-t_{\rm merge} \agt 10$\,ms where $t_{\rm merge}$
denotes the time for the onset of the merger) remains high for H4 and
Shen. This seems to reflect the difference in the adiabatic index of
the high-density range; for these EOSs, the central region of the MNS
has low values of $\Gamma_2$ and $\Gamma_3$ (see also
Fig.~\ref{fig1}).  With such relatively small values, we found that a
dumbbell-like structure rather than the ellipsoidal structure is
preserved for the MNS, and hence, the gravitational-wave amplitude is
enhanced.

The second pattern is that the gravitational-wave amplitude damps with
a characteristic modulation.  This pattern is often found for
unequal-mass models, in particular for $q=0.8$ (see
Fig.~\ref{figGW2}). The origin of this modulation is explained as
follows: During the early stage of the MNS evolution, its central
region appears to be composed of a massive core and a deformed
satellite for which the shape varies in the early stage of the
evolution (cf. Sec.~\ref{sec:depQ}). Here, the massive core and
satellite come from the massive and less-massive neutron stars of the
binary, respectively. In the early stage of the MNSs, two asymmetric
cores rotate around each other, and high-amplitude quasiperiodic
gravitational waves are emitted for $\sim 3$\,--\,5\,ms.  Then, the
amplitude once damps to be very small at a moment and subsequently,
long-term quasiperiodic gravitational waves are again emitted. This
feature is clearly seen for APR4-120150, APR4-125145, ALF2-120150, and
H4-120150. The mechanism for producing this pattern is closely related
to the evolution process of the MNSs (see Sec.~\ref{sec:depQ}). For
these asymmetric merger cases, asymmetric double cores are formed as
already mentioned. However, the less-massive core dynamically changes
its shape (like amoeba), and at the moment that the gravitational-wave
amplitude is small, the less-massive core has a highly deformed shape
surrounding the massive core. Namely, at this moment, not double cores
but a single nearly-spheroidal core is formed (see the third panel of
Fig.~\ref{figcont2}). However, after this moment, an asymmetric
double-core structure is formed again (like a hammer-thrower
shape). Because the resulting double-core structure is highly
nonaxisymmetric, quasiperiodic gravitational waves with high amplitude
are emitted.

For MS1-120150 and Shen-120150, the MNS also has an asymmetric
double-core structure which is alive for a long time scale $\gg
10$\,ms. In this case, the MNS never has the moment at which a
spheroidal shape is realized, and hence, the gravitational-wave
amplitude is stably high, although a modulation in the amplitude is
still observed.

\subsection{Frequency}

As already mentioned, the frequency of gravitational waves emitted by
MNSs is approximately constant (see
Figs.~\ref{figGW1}\,--\,\ref{figGW5}). The exception to this occurs
for some models in the very early stage just after the formation of
some of the MNSs in which the frequency oscillates with a dynamical
time scale (this can be observed for all the models to a greater or
lesser degree) or for the stage just before the formation of a black
hole in which the frequency increases steeply with time (see, e.g.,
the results for models SLy-135135, SLy-120150, ALF2-135135,
ALF2-130140, ALF2-140140, H4-135135, and H4-140140). These qualitative
features hold irrespective of the EOSs. 

\begin{figure*}[p]
\includegraphics[width=88mm,clip]{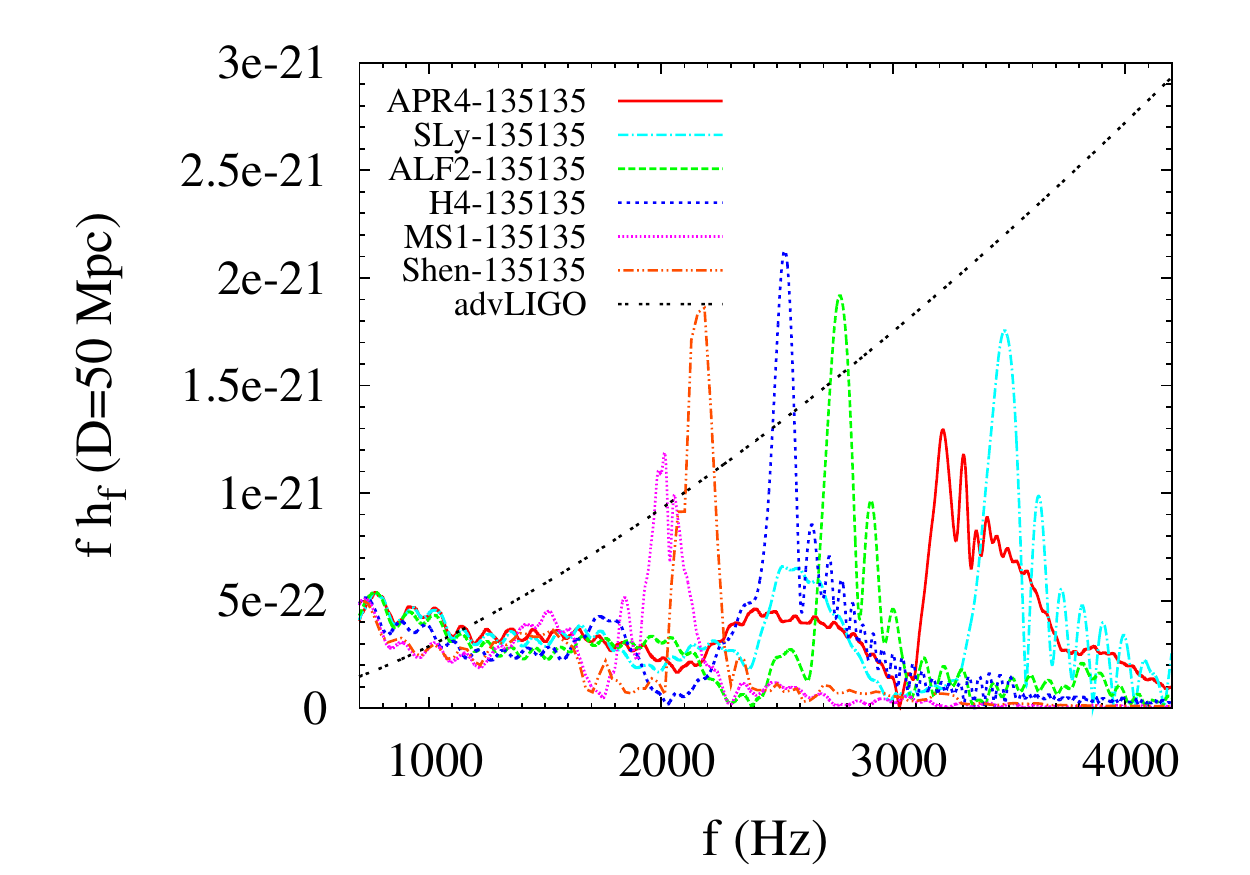}
\includegraphics[width=88mm,clip]{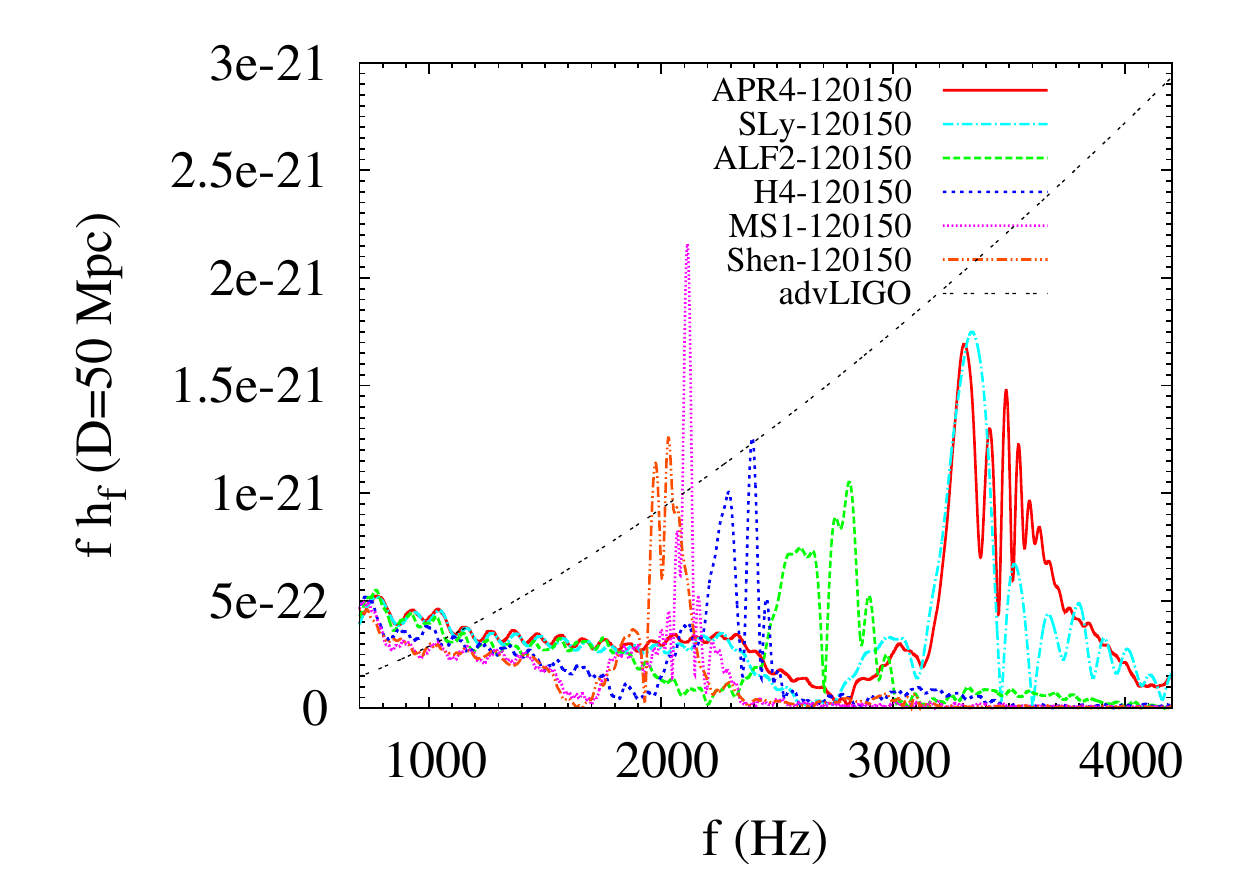} \\
\includegraphics[width=88mm,clip]{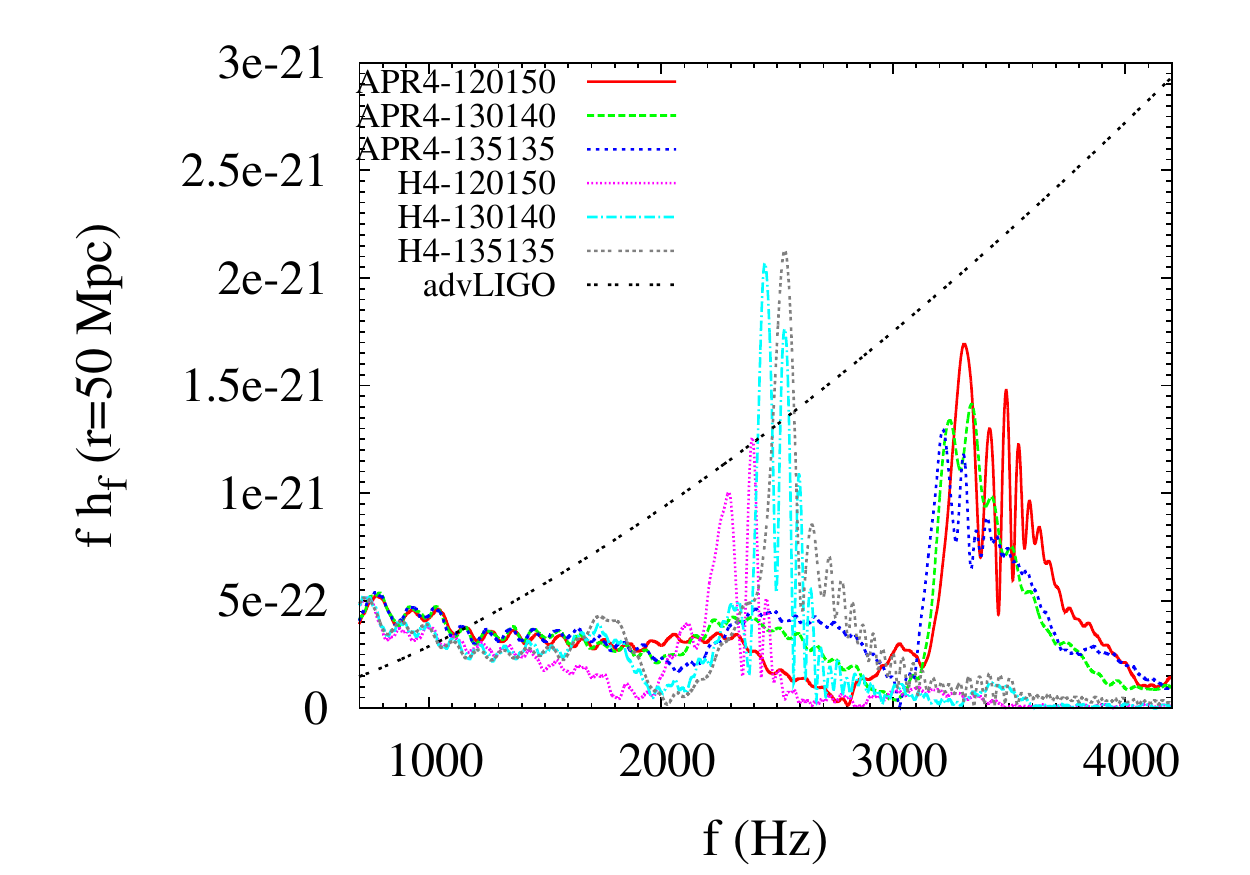}
\includegraphics[width=88mm,clip]{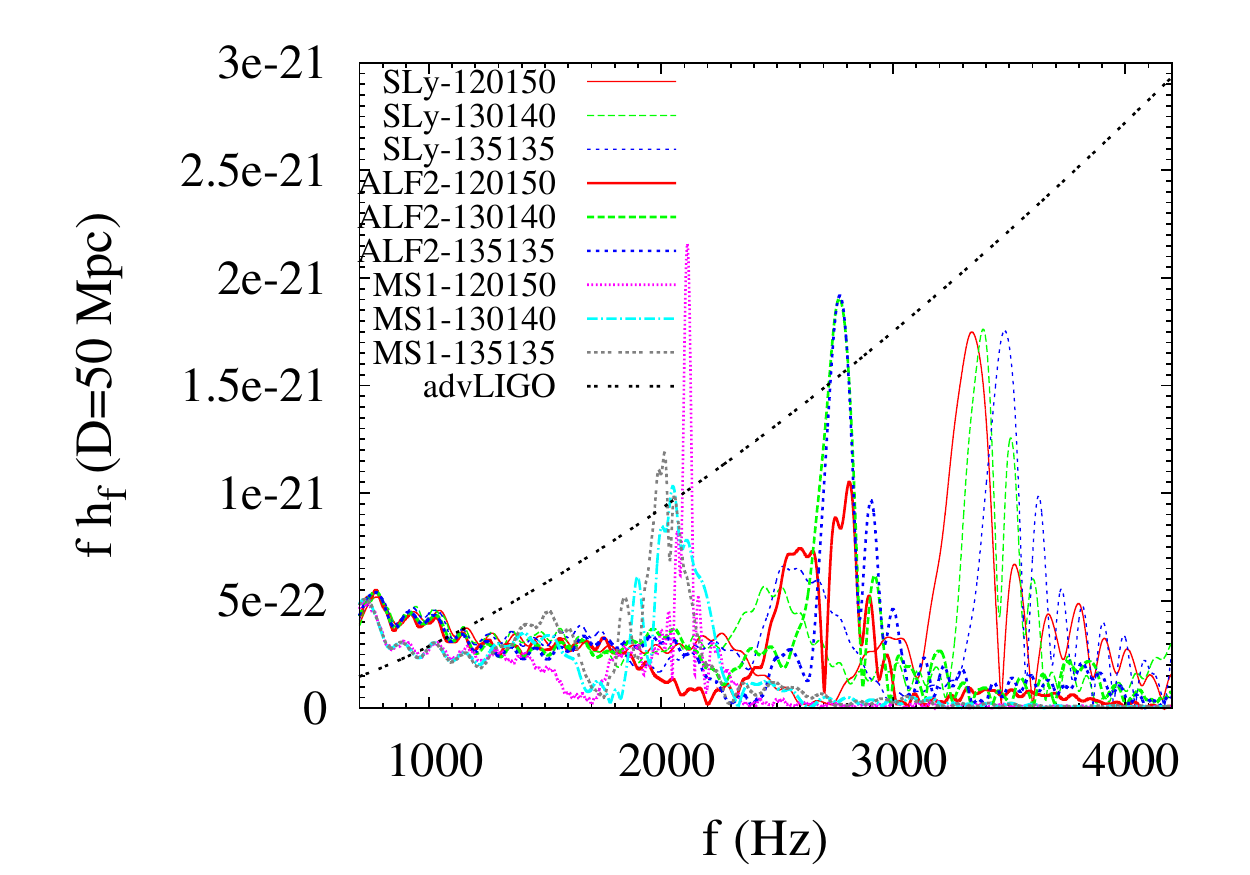} \\
\includegraphics[width=88mm,clip]{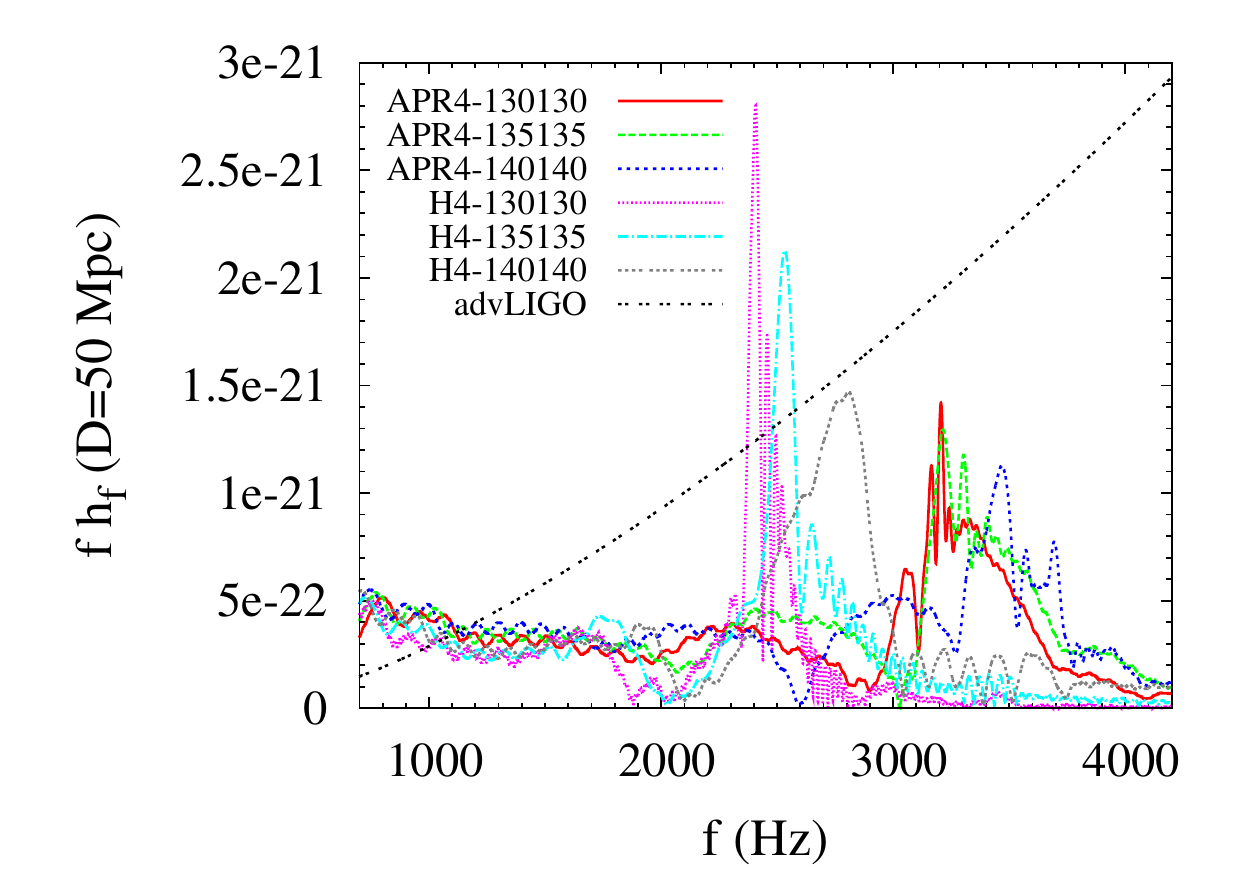}
\includegraphics[width=88mm,clip]{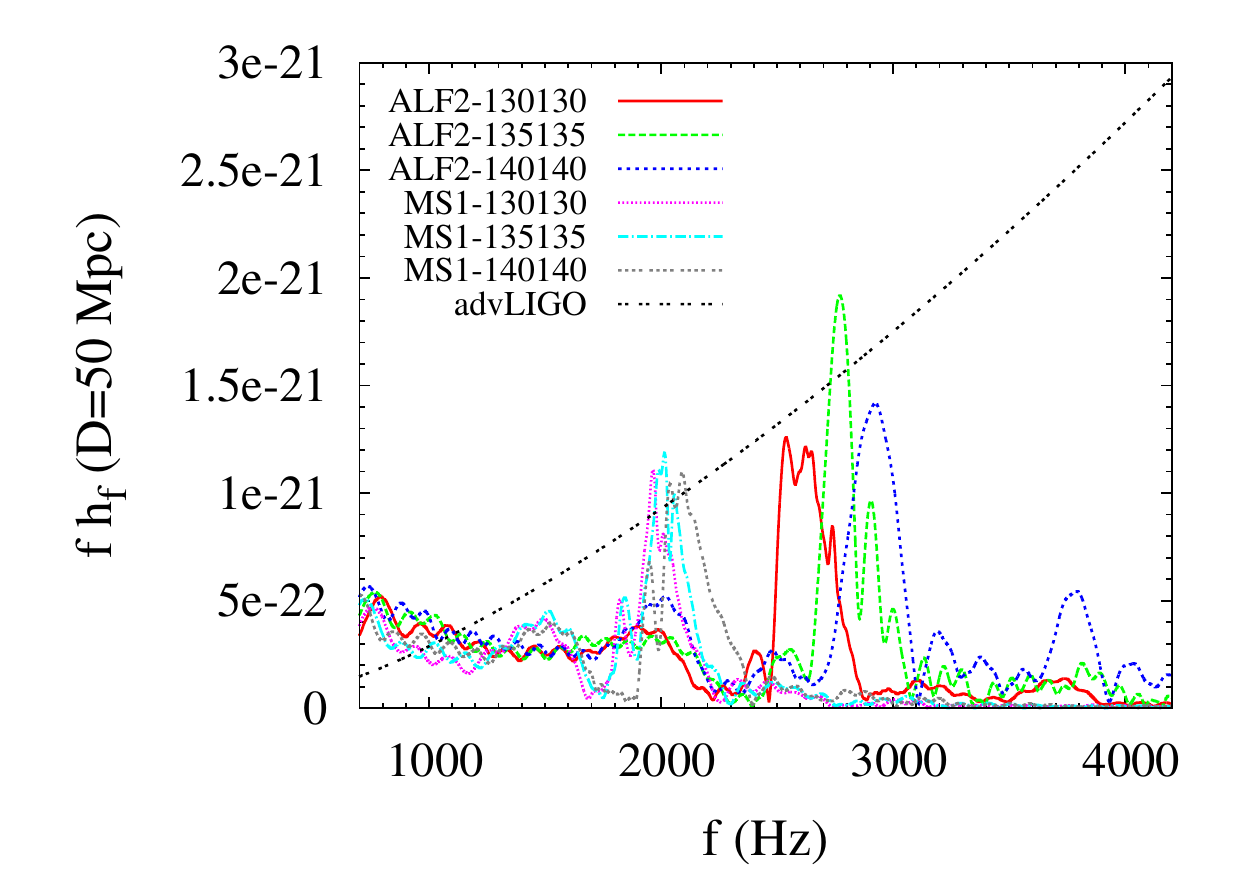}
\caption{Fourier spectra of gravitational waves for some of results
shown in Figs.~\ref{figGW1}\,--\,\ref{figGW5}: Top left; for
$m_1=m_2=1.35M_{\odot}$ with $\Gamma_{\rm th}=1.8$ and with five
piecewise polytropic and Shen EOSs: Top right; the same as top left
but for $m_1=1.2M_{\odot}$ and $m_2=1.5M_{\odot}$: Middle left; for
three mass ratios with $m=2.7M_{\odot}$, $\Gamma_{\rm th}=1.8$, and
with APR4 and H4: Middle right; the same as middle left but for SLy,
ALF2, and MS1: Bottom left; for equal-mass models with $m=2.6$, 2.7,
and $2.8M_{\odot}$, $\Gamma_{\rm th}=1.8$, and with APR4 and H4:
Bottom right; the same as bottom left but for ALF2 and MS1.  The
amplitude is shown for the hypothetical event at a distance of
$D=50$\,Mpc along the direction perpendicular to the orbital plane
(the most optimistic direction). The black dot-dot curve is the noise
spectrum of the advanced LIGO with an optimistic configuration for the
detection of high-frequency gravitational waves (see
https://dcc.ligo.org/cgi-bin/DocDB/ShowDocument?docid=2974).}
\label{figGWF}
\end{figure*}

\begin{table*}
\caption{ Characteristic frequencies of gravitational waves emitted by
  MNSs which are determined by two different methods: Fourier peak
  denotes the peak frequencies of the effective amplitude
  $h(f)f$. $f_{\rm ave, 5 ms}$, $f_{\rm ave, 10 ms}$, and $f_{\rm ave, 20 ms}$ denote the
  results for Eq.~(\ref{eq:fave1}) with 5, 10, and 20\,ms time integration
  after the formation of the MNSs, --- denotes that the lifetime of MNSs is shorter
  than the corresponding integration time.  The deviation of $f_{\rm ave}$
  shown here denotes $\sigma_f$. The multiple values shown for the
  Fourier peak imply that we found many peaks for which their peak
  values of $h(f)f$ are larger than 80\% of its maximum value.
  $f_{\rm ave, 10 ms}^{\rm fit}$ denote the averaged frequency
  calculated from the best-fit results of formulae
  (\ref{eq:totalfit})\,--\,(\ref{eq:freqfit}).  The last four columns
  show the maximum values of ${\cal M}$ for a fitting procedure with the
  number of parameters, $N_{\rm p}=13$, 12, 11, and 10 (see
  Sec.~\ref{sec5}).}
\scalebox{0.9}[0.9]{
{\begin{tabular}{cc|cccc|c|cccc} \hline
Model & $\Gamma_{\rm th}$ &
 Fourier peak (kHz) & $f_{\rm ave, 5 ms}$\,(kHz) &
 $f_{\rm ave, 10 ms}$\,(kHz) &
 $f_{\rm ave, 20 ms}$\,(kHz) &
 $f_{\rm ave, 10 ms}^{\rm fit}$\,(kHz) &
${\cal M}_{N_{\rm p}=13}$ & ${\cal M}_{N_{\rm p}=12}$ &
${\cal M}_{N_{\rm p}=11}$ & ${\cal M}_{N_{\rm p}=10}$
\\ \hline \hline
APR4-130150&1.8&3.40          &$3.48\pm 0.47$&$3.46\pm 0.37$&$3.49\pm0.33$   &$3.39\pm0.29$&0.910&0.894&0.894&0.883\\
APR4-140140&1.8&3.47          &$3.59\pm 0.64$&$3.57\pm 0.58$&$3.53\pm0.44$   &$3.59\pm0.59$&0.968&0.968&0.967&0.965\\
APR4-120150&1.6&3.31 3.54     &$3.47\pm 0.30$&$3.44\pm 0.27$&---             &$3.44\pm0.24$&0.963&0.962&0.959&0.945\\
APR4-120150&1.8&3.31,3.43     &$3.44\pm 0.30$&$3.41\pm 0.24$&$3.41\pm0.21$   &$3.41\pm0.23$&0.959&0.959&0.954&0.951\\
APR4-120150&2.0&3.18          &$3.32\pm 0.32$&$3.27\pm 0.26$&$3.27\pm0.22$   &$3.16\pm0.20$&0.924&0.924&0.919&0.919\\
APR4-125145&1.8&3.23          &$3.36\pm 0.31$&$3.31\pm 0.25$&$3.31\pm0.23$   &$3.20\pm0.18$&0.930&0.930&0.926&0.907\\
APR4-130140&1.8&3.28,3.31,3.40&$3.30\pm 0.29$&$3.27\pm 0.28$&$3.29\pm0.26$   &$3.26\pm0.26$&0.982&0.982&0.981&0.968\\
APR4-135135&1.6&3.45          &$3.46\pm 0.42$&$3.45\pm 0.37$&$3.46\pm0.30$   &$3.44\pm0.33$&0.970&0.967&0.967&0.942\\
APR4-135135&1.8&3.21,3.30     &$3.31\pm 0.41$&$3.28\pm 0.37$&$3.28\pm0.34$   &$3.27\pm0.33$&0.970&0.969&0.969&0.947\\
APR4-135135&2.0&3.29,3.37     &$3.35\pm 0.39$&$3.33\pm 0.34$&$3.33\pm0.29$   &$3.29\pm0.17$&0.952&0.952&0.947&0.947\\
APR4-120140&1.8&3.14          &$3.15\pm 0.21$&$3.13\pm 0.19$&$3.12\pm0.18$   &$3.11\pm0.16$&0.982&0.981&0.981&0.972\\
APR4-125135&1.8&3.20          &$3.22\pm 0.25$&$3.19\pm 0.24$&---             &$3.15\pm0.17$&0.975&0.975&0.974&0.963\\
APR4-130130&1.8&3.21          &$3.22\pm 0.28$&$3.19\pm 0.26$&$3.18\pm0.24$   &$3.18\pm0.28$&0.974&0.970&0.965&0.958\\
\hline
SLy-120150&1.8& 3.34 & $3.31 \pm 0.26$ & $3.35 \pm 0.24$&---             &$3.32\pm0.19$&0.984&0.983&0.979&0.979\\
SLy-125145&1.8& 3.32 & $3.29 \pm 0.32$ & $3.32 \pm 0.27$&---             &$3.31\pm0.28$&0.969&0.957&0.948&0.948\\
SLy-130140&1.8& 3.39 & $3.35 \pm 0.47$ & $3.36 \pm 0.40$&---             &$3.38\pm0.42$&0.965&0.959&0.958&0.913\\
SLy-135135&1.8& 3.48 & $3.41 \pm 0.58$ & $3.46 \pm 0.48$&---             &$3.47\pm0.47$&0.963&0.952&0.946&0.893\\
SLy-130130&1.8& 3.16 & $3.17 \pm 0.34$ & $3.16 \pm 0.29$&$3.18\pm0.26$   &$3.16\pm0.28$&0.988&0.988&0.987&0.972\\
\hline
ALF2-140140&1.8&2.92          &$2.93\pm 0.42$& ---&---                      &$2.90\pm0.34$&0.980&0.955&0.952&0.911\\
ALF2-120150&1.8&2.74,2.82,2.87&$2.70\pm 0.19$&$2.71\pm 0.16$&$2.73\pm0.15$  &$2.61\pm0.20$&0.924&0.916&0.916&0.907\\
ALF2-125145&1.8&2.65          &$2.66\pm 0.14$&$2.66\pm 0.13$&$2.67\pm0.13$  &$2.63\pm0.09$&0.985&0.985&0.985&0.985\\
ALF2-130140&1.8&2.77          &$2.73\pm 0.19$&$2.75\pm 0.17$&---            &$2.75\pm0.12$&0.981&0.979&0.978&0.977\\
ALF2-135135&1.8&2.77          &$2.74\pm 0.17$&$2.76\pm 0.15$&---            &$2.74\pm0.12$&0.989&0.989&0.981&0.981\\
ALF2-130130&1.8&2.54,2.63,2.65&$2.58\pm 0.18$&$2.56\pm 0.16$&$2.56\pm0.15$  &$2.55\pm0.12$&0.978&0.975&0.973&0.972\\
\hline
H4-130160&1.8&2.72          &$2.64\pm 0.26$& ---&---                      &$2.64\pm0.26$&0.973&0.965&0.963&0.942\\
H4-145145&1.8&2.90,2.96     &$2.97\pm 0.56$& ---&---                      &$2.93\pm0.49$&0.965&0.965&0.942&0.941\\
H4-130150&1.8&2.45,2.56     &$2.44\pm 0.17$&$2.45\pm 0.15$&$2.54\pm0.17$  &$2.42\pm0.09$&0.958&0.956&0.956&0.952\\
H4-140140&1.8&2.75,2.81     &$2.63\pm 0.23$&$2.77\pm 0.41$&---            &$2.69\pm0.22$&0.975&0.975&0.966&0.951\\
H4-120150&1.6&2.22,2.32,2.38&$2.28\pm 0.16$&$2.29\pm 0.14$&$2.31\pm0.14$  &$2.30\pm0.03$&0.980&0.977&0.977&0.966\\
H4-120150&1.8&2.29,2.39     &$2.30\pm 0.18$&$2.31\pm 0.15$&$2.33\pm0.14$  &$2.28\pm0.09$&0.955&0.955&0.955&0.939\\
H4-120150&2.0&2.30          &$2.24\pm 0.15$&$2.22\pm 0.14$&$2.26\pm0.12$  &$2.22\pm0.08$&0.983&0.980&0.980&0.973\\
H4-125145&1.8&2.44          &$2.41\pm 0.15$&$2.41\pm 0.13$&$2.44\pm0.11$  &$2.39\pm0.13$&0.981&0.981&0.980&0.980\\
H4-130140&1.8&2.43,2.52     &$2.42\pm 0.17$&$2.42\pm 0.15$&$2.44\pm0.13$  &$2.40\pm0.13$&0.968&0.968&0.967&0.966\\
H4-135135&1.6&2.59          &$2.49\pm 0.19$&$2.54\pm 0.16$&---            &$2.54\pm0.15$&0.985&0.968&0.966&0.960\\
H4-135135&1.8&2.53          &$2.44\pm 0.20$&$2.48\pm 0.16$&$2.54\pm0.17$  &$2.48\pm0.14$&0.984&0.982&0.978&0.963\\
H4-135135&2.0&2.49          &$2.39\pm 0.21$&$2.43\pm 0.17$&$2.47\pm0.15$  &$2.44\pm0.14$&0.977&0.977&0.972&0.972\\
H4-120140&1.8&2.34,2.37,2.43&$2.30\pm 0.15$&$2.30\pm 0.14$&$2.33\pm0.13$  &$2.32\pm0.06$&0.948&0.948&0.947&0.912\\
H4-125135&1.8&2.26          &$2.29\pm 0.17$&$2.27\pm 0.14$&$2.26\pm0.12$  &$2.28\pm0.14$&0.973&0.971&0.971&0.966\\
H4-130130&1.8&2.31          &$2.35\pm 0.18$&$2.38\pm 0.14$&$2.38\pm0.11$  &$2.37\pm0.12$&0.982&0.982&0.980&0.980\\
\hline
MS1-130160&1.8&2.12          &$2.07\pm 0.15$&$2.06\pm 0.13$&---            &$2.02\pm0.14$&0.967&0.967&0.965&0.956\\
MS1-145145&1.8&2.11          &$2.12\pm 0.15$&$2.09\pm 0.13$&---            &$2.09\pm0.12$&0.979&0.979&0.978&0.978\\
MS1-140140&1.8&2.04,2.09     &$2.09\pm 0.14$&$2.07\pm 0.12$&$2.05\pm0.12$  &$2.06\pm0.12$&0.972&0.972&0.968&0.964\\
MS1-120150&1.8&2.11          &$2.08\pm 0.11$&$2.09\pm 0.09$&$2.10\pm0.07$  &$2.08\pm0.10$&0.987&0.987&0.987&0.983\\
MS1-125145&1.8&2.02,2.08     &$2.02\pm 0.14$&$1.99\pm 0.15$&$1.99\pm0.14$  &$1.97\pm0.16$&0.959&0.959&0.955&0.953\\
MS1-130140&1.8&2.05          &$2.06\pm 0.14$&$2.02\pm 0.13$&$2.00\pm0.13$  &$2.03\pm0.11$&0.978&0.976&0.975&0.973\\
MS1-135135&1.8&1.99,2.02,2.05&$1.98\pm 0.17$&$1.96\pm 0.15$&$1.95\pm0.14$  &$2.00\pm0.22$&0.951&0.951&0.941&0.935\\
MS1-130130&1.8&1.96          &$1.94\pm 0.18$&$1.93\pm 0.15$&$1.91\pm0.15$  &$1.96\pm0.22$&0.950&0.950&0.948&0.943\\
\hline
Shen-120150 & --- &1.97,2.03& $2.02 \pm 0.15$ & $2.00 \pm 0.13$&$2.00\pm0.12$  &$2.01\pm0.07$&0.985&0.977&0.977&0.977\\
Shen-125145 & --- &2.10,2.18& $2.15 \pm 0.17$ & $2.17 \pm 0.15$&$2.15\pm0.13$  &$2.16\pm0.12$&0.972&0.966&0.964&0.963\\
Shen-130140 & --- &2.09,2.12& $2.08 \pm 0.18$ & $2.09 \pm 0.14$&$2.06\pm0.15$  &$2.07\pm0.14$&0.972&0.971&0.971&0.967\\
Shen-135135 & --- &2.18     & $2.18 \pm 0.18$ & $2.23 \pm 0.14$&$2.21\pm0.11$  &$2.27\pm0.06$&0.971&0.969&0.967&0.967\\
Shen-140140 & --- &2.28     & $2.29 \pm 0.26$ & $2.28 \pm 0.19$&$2.27\pm0.16$  &$2.31\pm0.05$&0.989&0.989&0.989&0.989\\
Shen-150150 & --- &2.29     & $2.22 \pm 0.24$ & $2.13 \pm 0.18$&$2.11\pm0.16$  &$2.25\pm0.12$&0.989&0.989&0.989&0.984\\
Shen-160160 & --- &2.49     & $2.38 \pm 0.37$ & $2.51 \pm 0.50$&---            &$2.49\pm0.20$&0.943&0.943&0.930&0.919\\
\hline
\end{tabular}
}
}
\label{table:result}
\end{table*}

\begin{figure*}[t]
\includegraphics[width=88mm,clip]{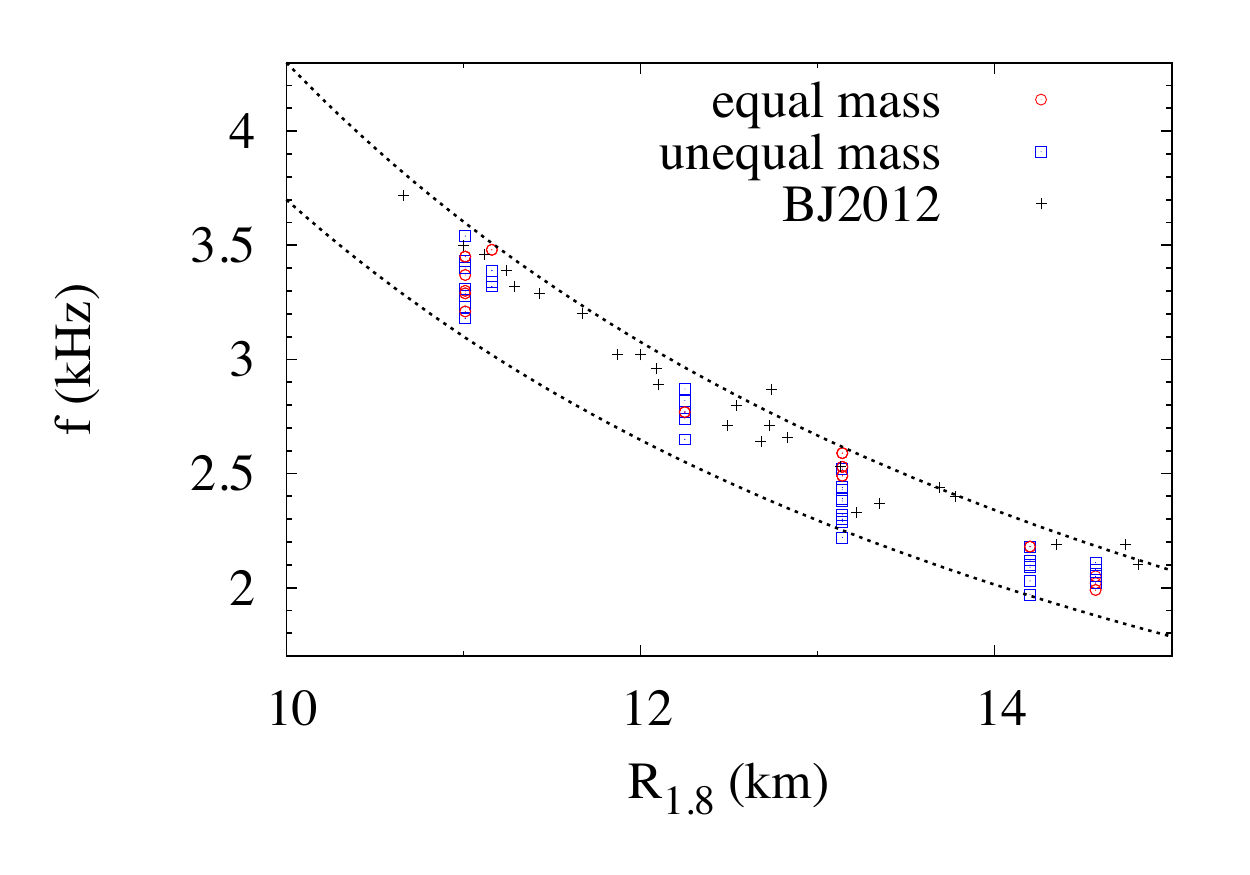}~~
\includegraphics[width=88mm,clip]{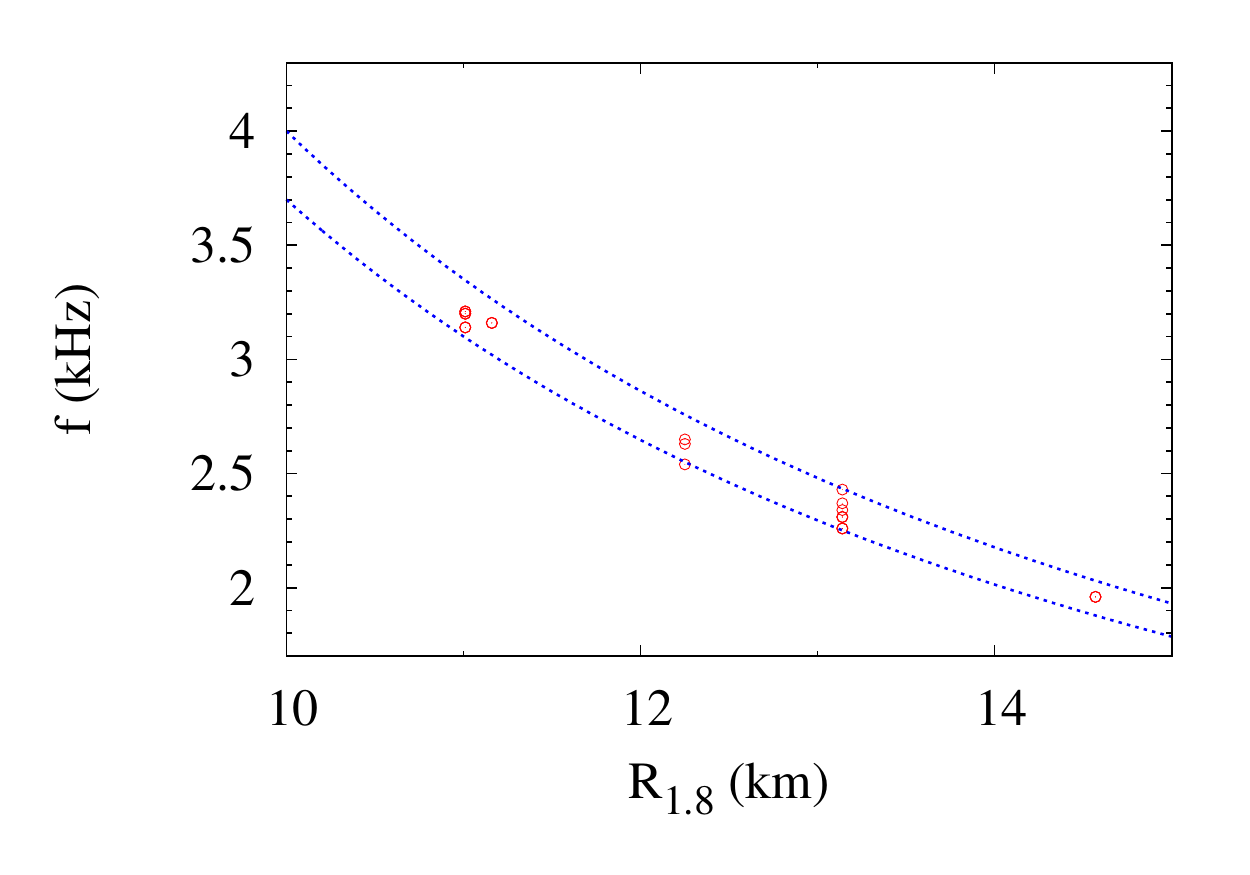} \\
\includegraphics[width=88mm,clip]{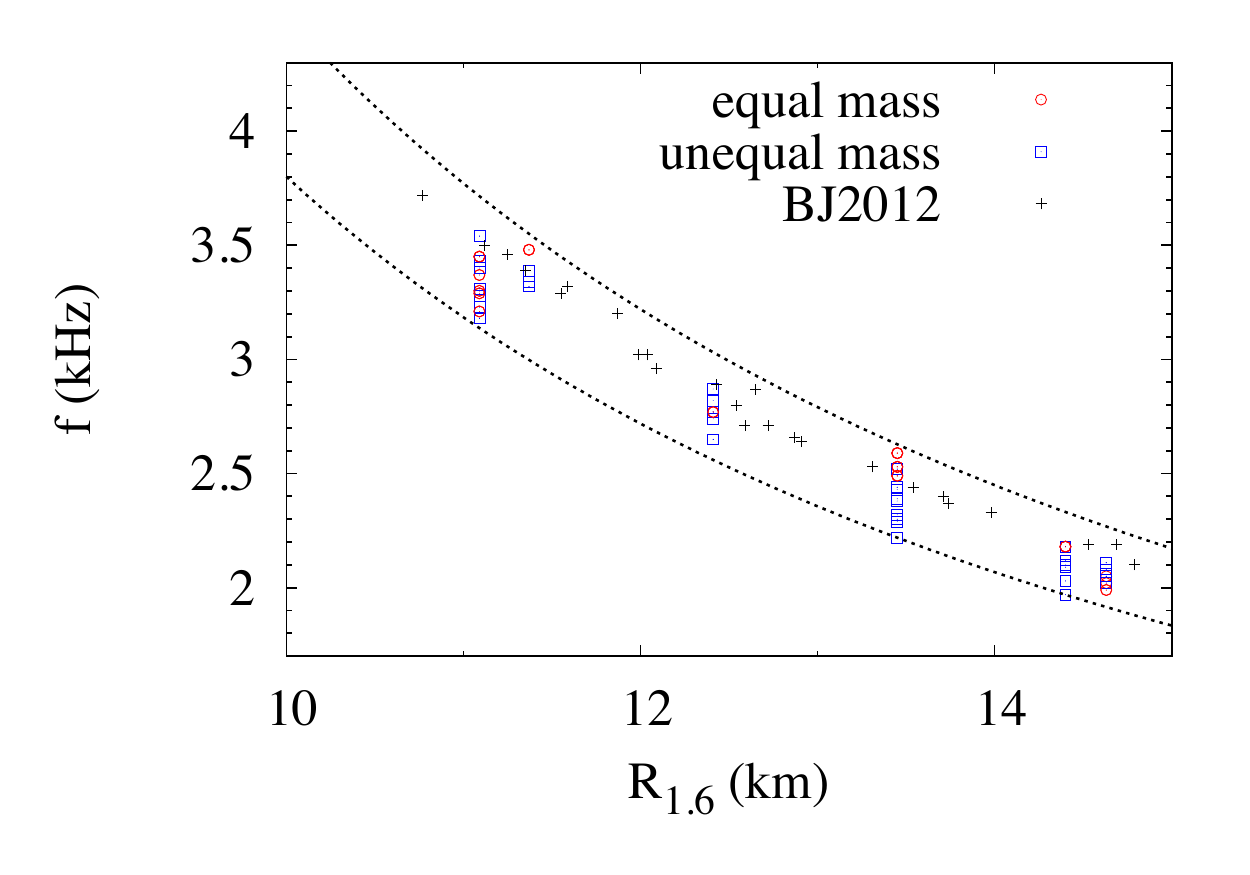}~~
\includegraphics[width=88mm,clip]{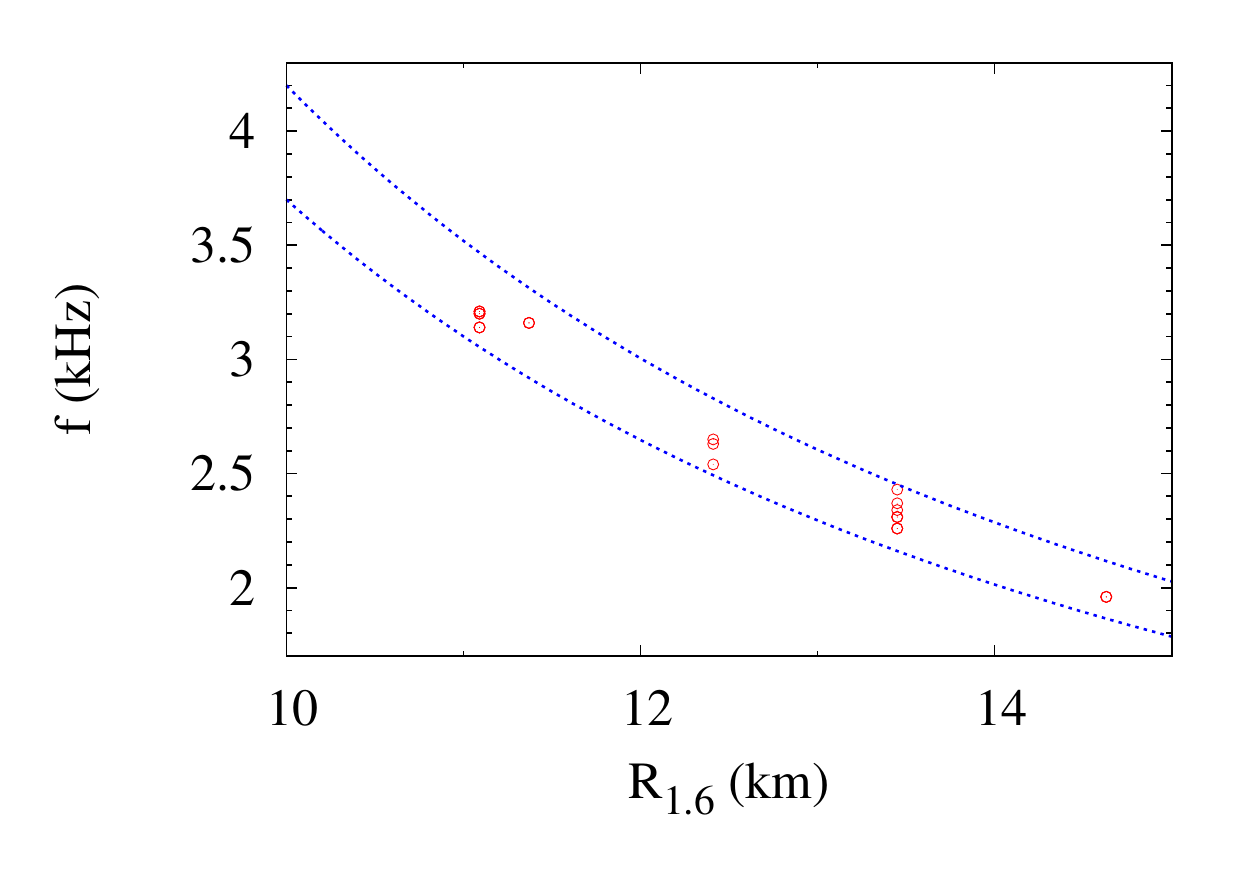} 
\vspace{-5mm}
\caption{The frequency of the Fourier spectrum peak as a function of
the neutron-star radius of $M=1.8M_{\odot}$ (upper panels) and
$M=1.6M_{\odot}$ (lower panels) with a given EOS for $m=2.7M_{\odot}$
(left panels) and $2.6M_{\odot}$ (right panels).  In the right panels,
we plotted all the data (equal-mass and unequal-mass data) using the
same symbol. In the left panels, the cross symbols denote
the data of \cite{BJ2012}.}
\label{figGWFF}
\end{figure*}

Figure~\ref{figGWF} plots the Fourier spectra for some of
gravitational waves displayed in Figs.~\ref{figGW1}\,--\,\ref{figGW5}.
Here, we plot the effective amplitude defined by $|h(f)f|$ as a function
of $f$ where $h(f)$ is the Fourier spectrum of $h_+ - i h_{\times}$.
This shows that there are indeed characteristic frequencies $\sim
2\,{\rm kHz} \alt f \alt 4\,{\rm kHz}$, at which the spectrum
amplitude is high, irrespective of models (see also
Table~\ref{table:result} for the frequency of the spectrum peak). For
a ``soft'' EOS that yields a compact neutron star for the canonical
mass (i.e., APR4, SLy, and ALF2 in this paper), the characteristic
frequency is higher with $f \agt 3$\,kHz, while for other ``stiff''
EOSs that yield a large-radius neutron $R \agt 13$\,km, it is lower
typically as $f \sim 2$\,--\,2.5\,kHz. The reason for this property is
explained as follows: The spin angular velocity of the MNSs is close
to the Kepler velocity, and hence, the characteristic frequencies of
gravitational waves are qualitatively proportional to $(M_{\rm
MNS}/R_{\rm MNS}^3)^{1/2}$ where $M_{\rm MNS}$ and $R_{\rm MNS}$
denote the typical mass and radius of a MNS. Here, the value of
$R_{\rm MNS}$ should be approximately proportional to the radius of
neutron stars of the canonical mass, and hence, it is reflected in the
characteristic frequency.  It should be also mentioned that the
characteristic frequency depends on the value of $\Gamma_{\rm th}$:
For the smaller value of it, the frequency is slightly higher for many
cases, because the effect of shock heating is weaker, and the MNS
becomes more compact. 

As pointed out in~\cite{BJ2012}, we also find that a certain
correlation exists between the characteristic frequency and a stellar
radius of a cold spherical neutron star in isolation.
Figure~\ref{figGWFF} plots the frequency of the Fourier spectrum peak
as a function of the neutron-star radius of mass $1.8M_{\odot}$ (upper
panels) and $1.6M_{\odot}$ (lower panels) for a given EOS (denoted by
$R_{1.8}$ and $R_{1.6}$ in units of km) for $m=2.7M_{\odot}$ (left
panels) and $2.6M_{\odot}$ (right panels)
\footnote{It should be noted that irrespective of the mass 
ratio, the chirp mass defined by $(m_1m_2)^{3/5}m^{-1/5}$ is
approximately equal to the value for the equal-mass system,
$2^{-6/5}m$, within 1\% error for $0.8 \leq q \leq 1$.  Thus, we may
approximately consider that the plots of Fig.~\ref{figGWFF} are the
plots for a given value of the chirp mass. In the data analysis of the
chirp signal of gravitational waves, the chirp mass will be determined
accurately but the mass ratio may not be~\cite{CF1994}.  For this
reason, we generated Fig.~\ref{figGWFF} for a given value of $m$
irrespective of the mass ratio. }. For $m=2.7M_{\odot}$ case,
we also plot the Fourier spectrum peak of the results of~\cite{BJ2012}
for comparison \footnote{ Here we choose only the data of which
the maximum mass of given EOSs is larger than $1.97M_{\odot}$
(see details in Table~II of~\cite{BJ2012}). }. The dotted curves are
\beqn
f=(4.0 \pm 0.3)\,{\rm kHz} \biggl({(R_{1.8}/{\rm km}) -2 \over
8}\biggr)^{-3/2},
\eeqn
for the upper left panel, 
\beqn
f=(3.85 \pm 0.15)\,{\rm kHz} \biggl({(R_{1.8}/{\rm km}) -2 \over
8}\biggr)^{-3/2},
\eeqn
for the upper right panel,
\beqn
f=(4.15 \pm 0.35)\,{\rm kHz} \biggl({(R_{1.6}/{\rm km}) -2 \over
8}\biggr)^{-3/2},
\eeqn
for the lower left panel, and 
\beqn
f=(3.95 \pm 0.25)\,{\rm kHz} \biggl({(R_{1.6}/{\rm km}) -2 \over
8}\biggr)^{-3/2},
\eeqn
for the lower right panel. These curves approximately show the upper
and lower limits of the characteristic frequency. We note that the
subtraction factor of $-2$ for $R_{1.8}$ and $R_{1.6}$ is empirically
needed to capture the upper and lower limits for the star of radius
$\sim 11$\,km. The reason seems to be due to the fact that general
relativistic corrections play an important role for the small value of
the neutron-star radius.  The reason why $R_{1.8}$ is employed is that
we found it a better choice to get a clear correlation between the
peak frequency and a neutron-star radius for our results. 
The choice of $R_{1.6}$ is done following~\cite{BJ2012}. 
In both cases, our results of the correlation
between the characteristic frequency and a stellar radius
are largely consistent
with the results of~\cite{BJ2012} within the uncertainty represented
with the dotted curves.

For the figure, we plotted all the values of the peak frequency when
we found multiple peaks; for some models, we plotted 2 or 3 points.
We also plotted all the data irrespective of the values of
$\Gamma_{\rm th}$ for APR4 and H4: Possible unknown dispersion
associated with the shock heating effect is taken into account in
these plots.  Nevertheless, we still find a fairly clear correlation
for the choice of $R_{1.8}$. In particular for the lower-mass models
($m=2.6M_{\odot}$), the dispersion is quite small.  This figure 
suggests that if we can determine the peak frequency accurately, we
will be able to constrain the radius of the neutron star with the
uncertainty of $\sim 1$\,km.  However, it is also found that it is not
easy to reduce the estimation error to $\ll 1$\,km, because of the
presence of the systematic dispersion of the peak frequency.

The peak frequencies are associated with the major frequencies of the
quasiperiodic oscillation of gravitational waves emitted by the MNSs
as found in Figs.~\ref{figGW1}\,--\,\ref{figGW5}. However, as already
mentioned, the (nonaxisymmetric) oscillation frequency of the MNSs
changes during the evolution due to a quasiradial oscillation (which
changes the peak frequency as $R_{\rm MNS}^{-3/2}$) and to the secular
dissipation processes of their angular momentum, and hence, the major
frequencies change with time, resulting in the broadening of the peak
or appearance of the multiple peaks (e.g., the spectra for APR4-135135,
APR4-130140, H4-120150, ALF2-130130, and ALF2-120150).  This
broadening is not very large for particular models such as equal-mass
models of some EOSs (see a discussion in the last paragraph of this
section), and thus, for these particular cases, the characteristic
frequency may be determined with a small dispersion. However, in
general, the broadening value is $\sim 10\%$ of the peak frequencies which
is $\sim 2$\,--\,3.5\,kHz.  Therefore, it will not be easy to strictly
determine the peak frequencies from the Fourier spectrum.  This
situation will bring a serious problem in the real data analysis, in
which the noise level is by several 10 percent as large as the signal
amplitude; the peak will not be determined strictly due to the
presence of many fake peaks and spurious broadening.

To estimate the possible magnitude of the broadening, we also
determine the average frequency from the results of the frequency $f$
by~\cite{hotoke2012}
\beqn 
f_{\rm ave}:={\displaystyle \int f |h| dt
\over \displaystyle \int |h| dt},\label{eq:fave1} 
\eeqn 
where we used $|h|=(h_+^2+h_\times^2)^{1/2}$ as the weight
factor. Then, we define the physical deviation of the major frequency
by
\beqn
\sigma_f^2:={\displaystyle \int (f-f_{\rm ave})^2 |h| dt \over
\displaystyle \int |h| dt}. \label{eq:fave2} 
\eeqn 
Here, the time integration is performed for 5, 10, and 20\,ms after the
formation of the MNSs, because for each time segment, the frequencies
are changed.  Table~\ref{table:result} lists the average frequency and
the deviation determined for 5, 10, and 20\,ms integration. 


Table~\ref{table:result} as well as Fig.~\ref{figGWF} show that the
values of $f_{\rm ave}$ agree approximately with the peak frequency of
the Fourier spectrum irrespective of the integration time. However, as
expected, the value of $f_{\rm ave}$ changes with time. It is also
found that the magnitude of the deviation, $\sigma_f$, is not
negligible. For APR4 and SLy for which the neutron-star radius is
rather small and the amplitude of a quasiradial oscillation induced at
the formation of the MNSs is rather large, the magnitude of the
deviation is 0.3\,--\,0.4\,kHz. This indicates that for determining
the peak frequency from the Fourier spectrum, the uncertainty of this
magnitude has to be kept in mind.  For other EOSs, the deviation is
relatively small. However, it is still 0.1\,--\,0.2\,kHz. To
summarize, we conclude that the characteristic frequency of
gravitational waves emitted by the MNSs changes with time in general,
and such time variation is the major source of the broadening of the
peak frequency found in Fig.~\ref{figGWFF}.

There is also an uncertainty due to the grid resolution of the simulation.
The averaged value of the frequency converges within $\sim 0.1$\,kHz error.
This error causes an uncertainty of the correlations between the Fourier
peak and the radius of a neutron stars with an error about 0.1 km.
However, the half width of Fourier peaks, which is about $\sigma_f$,
is larger than the uncertainty due to the grid resolution. 
Thus we consider that an uncertainty $\sim 0.1$kHz is not quite significant.

Before closing this section, we summarize several interesting
properties found in the Fourier spectrum.  The first one is that the
peak frequencies vary with the mass ratio even for the same total mass
$m$ (see Table~\ref{table:result}) and that the feature of this
variation depends on the EOS. For APR4, SLy, and ALF2, the frequency
depends only weakly on the mass ratio. For H4 and Shen, the frequency
is lower for the lower values of $q$, i.e., for more asymmetric
system. By contrast, for MS1, the frequency tends to be higher for
more asymmetric system. This property causes a dispersion in the
relation between the peak frequency and $R_{1.8}$ displayed in
Fig.~\ref{figGWFF}.  The second one is that the peak amplitude of the
Fourier spectrum decreases with the decrease of $q$ for ALF2, H4, and
Shen, while it increases with the decrease of $q$ for APR4 and MS1
(see Fig.~\ref{figGWF}). In other words, for ALF2, H4, and Shen, the
spectrum peak is sharper for the equal-mass binaries,
while for APR4 and MS1, it is sharper for the asymmetric binaries.
These properties will be used for constraining the EOS if the peak
frequencies are determined for a variety of binary neutron star
mergers.

\section{Modeling gravitational waveforms from MNS} \label{sec5}

In this section, we attempt to construct a fitting formula of
gravitational waveforms from MNSs.

\def\AP{\mathit{A\hspace{-0.3mm}P}}
\def\PP{\mathit{P\hspace{-0.8mm}P}}

\def\hNR{h_{\mathrm{NR}}}
\def\hfit{h_{\mathrm{fit}}}
\def\APNR{\AP_{\mathrm{NR}}}
\def\APfit{\AP_{\mathrm{fit}}}
\def\PPNR{\PP_{\mathrm{NR}}}
\def\PPfit{\PP_{\mathrm{fit}}}
\def\matching{\mathscr{ M}}

\def\parameter{\mathit{ Q}}
\def\parameters{\vec{\mathrm Q}}

\subsection{Fitting formula}

In this subsection, we describe possible fitting formulae for the waveforms of
quasiperiodic gravitational waves emitted by MNSs. There are two
conflicting requirements for the fitting formulae:

\begin{itemize}
\item On the one hand, we want fitting formulae by which various
numerical waveforms are well fitted universally. This generally
requires more numbers of free parameters.
\item On the other hand, we want fitting formulae that are controlled
by minimal numbers of free parameters, to minimize the computational
costs for the fitting in the parameter survey procedures.  In
addition, with smaller numbers of free parameters, the risk of
unphysical fitting is decreased, and thus, the physical meaning of
each parameter in the best-fitted case will be clearer.
\end{itemize}

Thus, we have to find an optimized fitting formula that is described
by an optimized number of parameters.  To discover the optimized
fitting formula, as a first step, we introduce a formula that
reproduces many characteristic properties of numerical waveforms,
without caring the optimization. Then, we search for the way to reduce
the number of parameters while keeping the matching degree as high as
possible.

The universal features of gravitational waves emitted by MNSs are
described in Sec.~\ref{sec:GW} and summarized as follows:
\begin{itemize}
\item[(i)] The frequency of gravitational waves reaches a peak soon
after the merger sets in (or in other words, in the final moment of
the inspiral phase) and then experiences a damping oscillation for
several oscillation periods, eventually settling approximately to a
constant value, although a long-term secular change associated with the
change of the state of MNSs is always present.
\item[(ii)] Soon after the onset of the merger, the amplitude of 
gravitational waves becomes very low. However, subsequently, the
amplitude steeply increases, and then, it decreases either
monotonically or with modulations.
\item[(iii)] The damping time scale of the amplitude is of the order
of 10\,ms for most MNS models although for some EOSs such as H4 and
Shen, the damping time scale is much longer than 10\,ms.
\end{itemize}
The fact (iii) implies that the emissivity of gravitational waves by
MNSs is in general high for the first $\sim 10$\,ms after their
formation. Hence, to save the search costs, 
we focus on gravitational waves in this time range,
considering the 10\,ms-window function
\beqn
W(t):=\left\{
\begin{array}{ll}
1 & \hspace{1cm}{\rm for}~t_i \leq t \leq t_f, \label{window}\\
0 & \hspace{1cm} {\rm otherwise}, \\
\end{array}
\right.
\eeqn
where $t=t_i$ is the time at which the frequency peak is reached and
$t_f = t_i + 10$\,ms. In the following, we define the origin of the
time by setting $t_i=0$ for simplicity. Beware that this notation of
$t$ is used solely for describing the fitting formulae and is
different from the retarded time, $t_{\rm ret}$, that was used in
Figs.~\ref{figGW1}\,--\,\ref{figGW5}. 

Taking into account the characteristic properties of gravitational
waves listed above, we first introduce a fitting formula that contains
13 free parameters as follows. Using the fact that a complex function
of any gravitational-wave signal, $h(t)$, can be uniquely decomposed
into a pair of real function $\AP(t)$ and $\PP(t)$ as
\begin{eqnarray}
 h(t) = \AP(t) \exp\left[-i\hspace{0.3mm}\PP(t)\right],
 \label{eqDecomposition}
\label{eqTemplate}
\end{eqnarray}
we consider the following forms of fitting formulae, 
\begin{eqnarray}
 \hfit(t) &=& \APfit(t) \exp\left[-i\hspace{0.3mm}\PPfit(t)\right],
\label{eq:totalfit}
\end{eqnarray}
where
\begin{eqnarray}
 \APfit(t)&=& \left[a_1 \exp\left(-\frac{t}{a_{\rm d}}\right) +a_0 \right]
\nonumber \\
  & & \times \left(\frac{1}{1+\exp\left[ (t-a_{\rm co}) / t_{\rm cut}
  \right]}\right)\nonumber \\ 
& & \times \left[1 - \exp\left(-\frac{t}{a_{\rm ci}}\right)\right],
\label{eq:ampfit} \\ 
\PPfit(t) &=& p_0 + p_1 t
  + p_2 t^2 + p_3 t^3 \nonumber \\ 
&+&\exp\left(-\frac{t}{p_{\rm d}}\right)
  \left[p_{\rm c}\cos(p_{\rm f}t) + p_{\rm s}\sin(p_{\rm f}t)\right]. 
\label{eq:freqfit}
\end{eqnarray}
Equation~(\ref{eq:ampfit}) shows that we model the fitting function 
for the amplitude part in terms of three parts: 
\begin{enumerate}
\item $a_1 e^{-t/a_{\rm d}}+a_0$ denotes the evolution for the
amplitude which is assumed to be composed of an exponentially damping
term and a constant term. $a_0$, $a_1$, and $a_{\rm d}$ are free
parameters that should be determined by a fitting procedure.
\item $[1+\exp\{(t-a_{\rm co})/t_{\rm cut}\}]^{-1}$ denotes a cutoff
term that specifies a time interval of a high-amplitude stage with
$a_{\rm co}$ being the center of the cutoff time and $t_{\rm cut}$
being the time scale for the shutdown. $a_{\rm co} < 10$\,ms for the
case that a black hole is formed within 10\,ms after the formation of
a MNS. $a_{\rm co}$ is determined in the fitting process, while
$t_{\rm cut}$ is a parameter that should be manually chosen and fixed;
we here choose it to be 0.1\,ms for simplicity.
\item $1-\exp(-t/a_{\rm ci})$ is a steeply increasing function for $t
\agt t_i$ with $a_{\rm ci}$ being the growth time scale. The reason
for introducing this term is that at $t=t_i=0$, the amplitude of
gravitational waves is universally low, but after this moment, the
amplitude steeply increases and the time scale depends on the total
mass, mass ratio, and EOS (see waveforms in Sec.~\ref{sec:GW}).
\end{enumerate}

We note that in this fitting formula, we do not take into account
the effect of the modulation in amplitude: To do so,
we have to significantly increase the number of fitting parameters.
However these additional parameters increase the search costs, and
hence, in this paper we focus on a relatively simple fitting formula.

Equation~(\ref{eq:freqfit}) shows that we model the fitting function
for the frequency in terms of a secularly evolving term $p_0 + p_1 t +
p_2 t^2 + p_3 t^3$ and damping oscillation term $e^{-t/p_{\rm d}}
[p_{\rm c}\cos(p_{\rm f} t) + p_{\rm s}\sin(p_{\rm f} t)]$.  Here,
eight constants $p_0$, $p_1$, $p_2$, $p_3$, $p_{\rm d}$, $p_{\rm f}$,
$p_{\rm c}$, and $p_{\rm s}$ are free parameters that should be
determined by a fitting procedure.

Thus, in total, there are 13 parameters to be determined, and we
denote them by $\parameters$ in the following.  We here stress that 13
parameters are the maximally necessary ones in our present fitting
procedure. In the following, we ask whether it is possible to reduce
the number as small as possible. 

\subsection{Determining the model parameters} \label{sec:fittingProc}

\subsubsection{Fitting procedure}

First, we focus on the fitting formula that contains 13 parameters and
describe how we determine these parameters. For the determination, it
might be natural to define the following function
\beq
  \matching(\hfit)  :=  \frac{(\hNR,\hfit)}{\sqrt{ (\hNR,\hNR)
  (\hfit,\hfit)}}, \label{eq:fitting}
\eeq
and consider to maximize the absolute value of this function. 
Here, $(~\cdot~,~\cdot~)$ is the
inner product defined in the time domain by
\begin{eqnarray}  
 (a,b)  :=  {\rm Re}\left(\int_{t_i}^{t_f} a(t) b^{*}(t) {\mathit dt}\right),
\end{eqnarray}
and we note that the maximum value of $\matching(\hfit)$ is unity.

One problem to be pointed out is that Eq.~(\ref{eq:fitting}) has a
freedom of the scale transformation as $\hfit \rightarrow C\hfit$ with
$C$ being an arbitrary constant, and hence, the amplitude of the
fitting function cannot be determined by maximizing ${\cal M}$. Thus,
as an alternative, we define the following {\em cost function},
\begin{eqnarray}
C_{\rm C}(\hfit) &:=& - \matching(\hfit) \nonumber\\ &&+ \frac{\left[
  (\hNR,\hNR) - (\hfit,\hfit) \right]^2}{ (\hNR,\hNR)^2}, 
  \label{complexCost}
\end{eqnarray}
and consider to minimize this function. Here, the second term is in a
sense the normalization factor by which the ambiguity in the amplitude
of $\hfit$ is fixed. We note that by the minimization of $C_{\rm C}$,
we can obtain $\hfit$ that maximizes $\matching$ and also the
amplitude of $\hfit$ that agrees approximately with that of $\hNR$. 


We use CMA-ES (covariance matrix adaption evolution search) algorithm
~\cite{CMAES1,CMAES2,CMAES3} to solve the minimization problems.
CMA-ES is a widely-applicable optimization method for $N$-input,
1-output real-valued functions $y = f(\vec x)$. CMA-ES belongs to a
category of stochastic optimization algorithms such as, e.g., Markov
chain Monte Carlo methods and genetic algorithms.  CMA-ES algorithm
keeps track of a multivariate normal distribution ${\cal N} (\mu,
\Sigma)$ from which guess parameters $\vec x$ are generated.  CMA-ES
algorithm proceeds by updating the mean $\mu$ and the covariance
matrix $\Sigma$ according to the values of $f(\vec x)$ for
randomly-sampled $\vec x$.  Due to this, CMA-ES has many
preferable properties: It does not require the information of $\nabla
f(\vec x)$, the values of which are computationally expensive,
inaccurate or inaccessible in many cases; it is robust against noise
in $f$ and/or tiny local minima in $f$; its result is not affected by
composing any increasing function $g$ on the output space $g (f(\vec
x))$; its result is also not affected by Affine transformation in the
input space $f(A \vec x + \vec b)$ if the initial distribution is also
modified by the inverse transformation.

Despite of such properties, minimizing $C_{\rm C}(\hfit)$ is not
straightforwardly achieved by CMA-ES because the function has lots of local minima
in its 13-dimensional parameter space.  Therefore, we resort to
meta-heuristics that decomposes the main problem into multiple
optimization sub-problems, each of which is solved by the CMA-ES
algorithm.

To begin with, we introduce two supplementary cost functions as
\begin{eqnarray}
C_{\rm P}(\hfit) & := & \int_{t_i}^{t_f} \APNR^2 (\PPNR - \PPfit)^2
{\mathit dt},\\ 
C_{\rm A}(\hfit) & := & \int_{t_i}^{t_f} (\APNR - \APfit)^2
{\mathit dt},
\end{eqnarray}
where $\APNR$ and $\PPNR$ are amplitude and phase parts of $\hNR$, 
respectively, which are obtained by the decomposition defined 
in Eq.~(\ref{eqDecomposition}).  Then, instead of
performing the minimization procedure in the 13-parameter space
altogether, we determine a subset of 13 parameters by a minimization
procedure in terms of $C_{\rm P}$ and $C_{\rm A}$ step by step.

\def\o{$\bigcirc$}

\begin{table}[t]
  \begin{tabular}{cc|cccccccc|ccccc}
    \hline
    \multicolumn{2}{c|}{$N_{\rm p}=13$} & $p_{\rm d}$ & $p_{\rm f}$ & $p_{\rm c}$ 
& $p_{\rm s}$ & $p_3$ & $p_2$ & $p_1$ & $p_0$ 
& $a_{\rm ci}$ & $a_{\rm co}$ & $a_{\rm 0}$ & $a_{\rm d}$ & $a_{\rm 1}$ \\
    \hline
      {\tt 1.} & {\tt (P)} &  &    &  &  &  &    &\o&\o&  &    &  &  &  \\
      {\tt 2.} & {\tt (P)} &  &    &  &  &  &  \o&\o&\o&  &    &  &  &  \\
      {\tt 3.} & {\tt (P)} &  &    &  &  &\o&  \o&\o&\o&  &    &  &  &  \\
      {\tt 4.} & {\tt (P)} &\o&  \o&\o&\o&  &    &  &  &  &    &  &  &  \\
      {\tt 5.} & {\tt (P)} &\o&  \o&\o&\o&\o&  \o&\o&\o&  &    &  &  &  \\
      {\tt 6.} & {\tt (A)} &  &    &  &  &  &    &  &  &  &    &\o&\o&\o\\
      {\tt 7.} & {\tt (A)} &  &    &  &  &  &    &  &  &\o&    &\o&\o&\o\\
      {\tt 8.} & {\tt (A)} &  &    &  &  &  &    &  &  &  &  \o&\o&\o&\o\\
      {\tt 9.} & {\tt (C)} &\o&  \o&\o&\o&\o&  \o&\o&\o&\o&  \o&\o&\o&\o\\
      \hline
  \end{tabular}
\caption{A procedure of determining 13 parameters is illustrated.  The
procedure in this example is composed of 9 steps.  In the
first\,--\,fifth steps, we update $(p_0, p_1)$, $(p_0, p_1, p_2)$,
$(p_0, p_1, p_2, p_3)$, $(p_{\rm d}, p_{\rm f}, p_{\rm c}, p_{\rm
s})$, and all 8 parameters of the phase part, $p_i$, using the cost
function $C_{\rm P}$, respectively, and in the sixth\,--\,eight steps,
5 parameters in the amplitude part, $a_i$, is updated using the cost
function $C_{\rm A}$.  At the ninth step, all the parameters are
updated at the same time using the cost function $C_{\rm C}$. At each
step, the values of parameters determined in the previous steps are
used as the initial guess values.  The leftmost letter (P or A or C)
indicates the type of cost function $C_{\rm P}$ or ${C_{\rm A}}$ or
$C_{\rm C}$ used in the corresponding step.  See also the text for
more detailed description.}  \label{table:Strt}
\end{table}

\begin{table*}[t]
\def\e{\times 10^}
\begin{tabular}{cclcl|c|c}
\hline
\multicolumn{5}{c|}{parameter ($\parameter_{\rm std} \pm \sigma_{\rm
   std}$)} & dim. & constraints \\
\hline
\hline
$a_1$ & = & $ 0.113793577854  $ & $\pm$ & $0.2 $ & 1 & $a_1 >0$ \\
$a_{\rm d}$ & = & $ 3.61810442878\e{-3} $ & $\pm$ & $0.01 $ & sec \\
$a_0$ & = & $6.67194850219\e{-3} $ & $\pm$ & $1\e{-3} $ & 1& $a_0 >0 $
\\ 
$a_{\rm co}$ & = & $ 0.012 $ & $\pm$ & $0.01 $ & sec& $3\times10^{-4}<a_{\rm co}<1.2\times10^{-2}$ \\ 
$a_{\rm ci}$ & = & $ 2.28934356228\e{-4}$ & $\pm$ & $5\e{-4} $ & sec \\
\hline                                               
$p_0$     & = & $ 4.12034039189   $ & $\pm$ & $37    $ &   1      \\
$p_1$     & = & $ 1.56534759719\e4   $ & $\pm$ & $4\e3   $ & sec${}^{-1}$  \\
$p_2$     & = & $0                $ & $\pm$ & $3\e5   $ & sec${}^{-2}$    \\
$p_3$     & = & $0                $ & $\pm$ & $1\e7   $ & sec${}^{-3}$    \\
$p_{\rm s}$     & = & $ 0.494703128127  $ & $\pm$ & $1     $  & 1      \\
$p_{\rm c}$     & = & $ -0.116384976953 $ & $\pm$ & $1     $  & 1      \\
$p_{\rm f}$     & = & $ 6.607469741287686\e{3}$ & $\pm$ & $1\e{3}  $  & 
sec${}^{-1}$ & $p_{\rm f} > 0$        \\
$p_{\rm d}$     & = & $ 1.69866461255\e{-3}$ & $\pm$ & $ 2\e3  $  & sec\\
\hline
\end{tabular}
\caption{ The standard values of parameters $\parameters_{\rm std}$,
their deviations $\sigma_{\rm std}$, and their constraints. } \label{table:def}
\end{table*}

Table~\ref{table:Strt} shows an example of the parameter-search
procedure.  In this example, 13 parameters of a waveform are 
eventually updated after the following 9 steps (the step indices
correspond to those in Table~\ref{table:Strt}):
\begin{itemize}
 \item[{\tt 0.}] In the beginning, we give an initial guess values
  (see Sec.~\ref{sec:stdSetParam}) for all 13 parameters.
 \item[{\tt 1.}] In the first step, we update parameters $(p_0,
  p_1)$ by minimizing the cost function $C_{\rm P}$.  
 \item[{\tt 2.}] Using the new set of $(p_0, p_1)$ as a part of the
  updated initial-guess parameters, we update $(p_0, p_1, p_2)$. 
 \item[{\tt 3.}] Similarly, parameters $(p_0, p_1, p_2, p_3)$ are
  updated in the third step.
\item[{\tt 4.}] The parameters of the damping
  oscillation term in phase, $(p_{\rm d}, p_{\rm f} , p_{\rm c},
  p_{\rm s})$, are updated using the cost function $C_{\rm P}$ and
  fixing $(p_0, p_1, p_2, p_3)$ to be the values obtained in the
  previous step.
\item[{\tt 5.}] All of the parameters in phase, ($p_{\rm
  d}$\,--\,$p_{\rm s}$ and $p_{0}$\,--\,$p_{3}$), are updated starting
  from the values obtained in the first\,--\,fourth steps as the
  initial guess and employing $C_{\rm P}$ as the cost function.
\item[{\tt 6.}] We update the exponentially damping
term in amplitude, ($a_{0}, a_{1}, a_{\rm d}$), employing $C_{\rm A}$
as the cost function.  Here, we fix the values of all the eight
parameters in phase, ($p_{\rm d}$\,--\,$p_{3}$), to be those
determined in the fifth step.
\item[{\tt 7.}] We update the parameters for the steeply increasing
  term ($a_{\rm ci}$), with three other parameters in amplitude,
  ($a_{0}, a_{1}, a_{\rm d}$), also subject to change.
\item[{\tt 8.}] We update the parameters for the cutoff term ($a_{\rm
co}$) in a similar manner.
\item[{\tt 9.}] Finally, we minimize the cost function $C_{\rm C}$
using all of the parameters obtained up to the eighth step as the
initial guess.
\end{itemize}
This procedure should be iterated until the sufficient convergence is
achieved. 

The circles in each row of Table~\ref{table:Strt} denote the
parameters that are updated at each step. For the fitting procedure at
each step, we use CMA-ES algorithm.  Note that in the above procedure,
we employ the cost function $C_{\rm P}$ when all the target parameters
belong to the phase part ($PP$), while we employ $C_{\rm A}$ when all
the target parameters belong to the amplitude part ($AP$). We employ
$C_{\rm C}$ only when the parameters in the search include parameters
of both ($AP$) and ($PP$) parts.

\subsubsection{Standard set of the 13 parameters} \label{sec:stdSetParam}

CMA-ES algorithm requires the values and deviations of the initial
guess for all the input parameters.  As the search procedure proceeds,
we can obtain accumulated sets of 13 parameters determined in the
different cycles of search procedures and for additional models. By
taking median over them, we may construct a {\em standard set of the
13 parameters}, which can be used as the initial guess in the search
procedure that follows.  Table~\ref{table:def} denotes the standard
set of the parameters, $\parameters_{\rm std}$, and their deviations,
$\sigma_{\rm std}$, which were obtained from the results of the search
for all the numerical waveforms employed in this paper. Here, we
artificially set $p_2$ and $p_3$ to be zero because we want to always
test the quality of the fitting functions that are linear in time for
the phase part.  Note that this standard set will be improved as the
search procedure proceeds and as more waveforms are involved.
We use the determined 13 parameters for a set of the waveforms as
an initial guess in the next generation of fitting. 


\subsection{Fitting results} \label{sec:fitres}

\begin{figure}[t]
\includegraphics[width=90mm,clip]{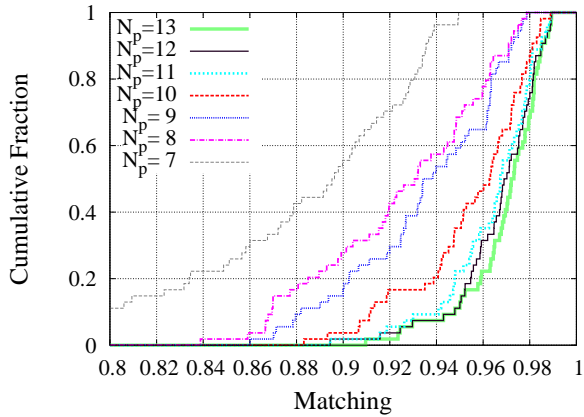}
\vspace{-8mm}
\caption{
Cumulative distribution of the maximum value of $\matching$ for the
parameter fitting with the number of parameters $N_{\rm p}=7$\,--\,13.
The horizontal axis denotes the maximum value of matching, $\matching$, and
the vertical axis shows the cumulative fraction of the models whose
maximum fitting value is less than $\matching$.  }
\label{fig:hist}
\end{figure}

Figure~\ref{fig:hist} shows the cumulative distribution for the
maximum value of $\matching$. Here, see the plot for $N_{\rm p}=13$
that denotes the cumulative distribution for the case of the
13-parameter fitting (see also the column of ${\cal M}_{N_{\rm p}=13}$
in Table~\ref{table:result} which lists the maximum values of
$\matching$ for all the numerical waveforms). This plot shows that for
$\sim 90\%$ of the waveforms, the maximum value of $\matching$ is
larger than 0.95. Also for all the waveforms, the maximum value of
$\matching$ is larger than 0.90.  


\begin{figure*}[t]
\includegraphics[width=80mm,clip]{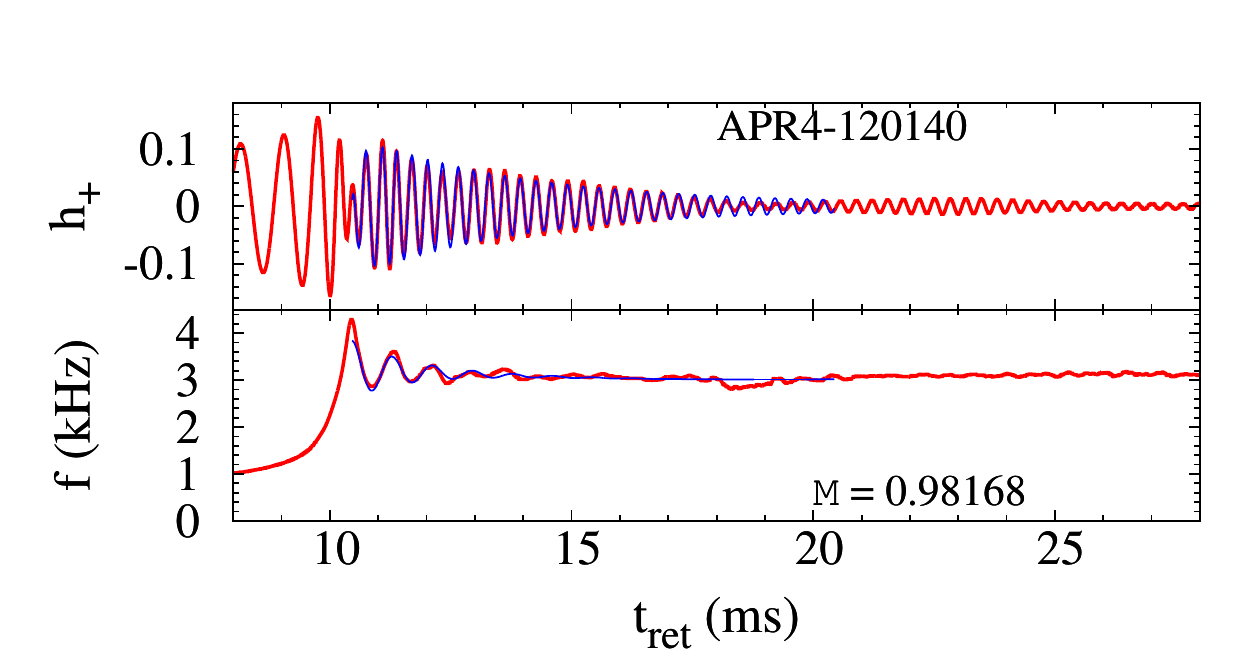} 
\includegraphics[width=80mm,clip]{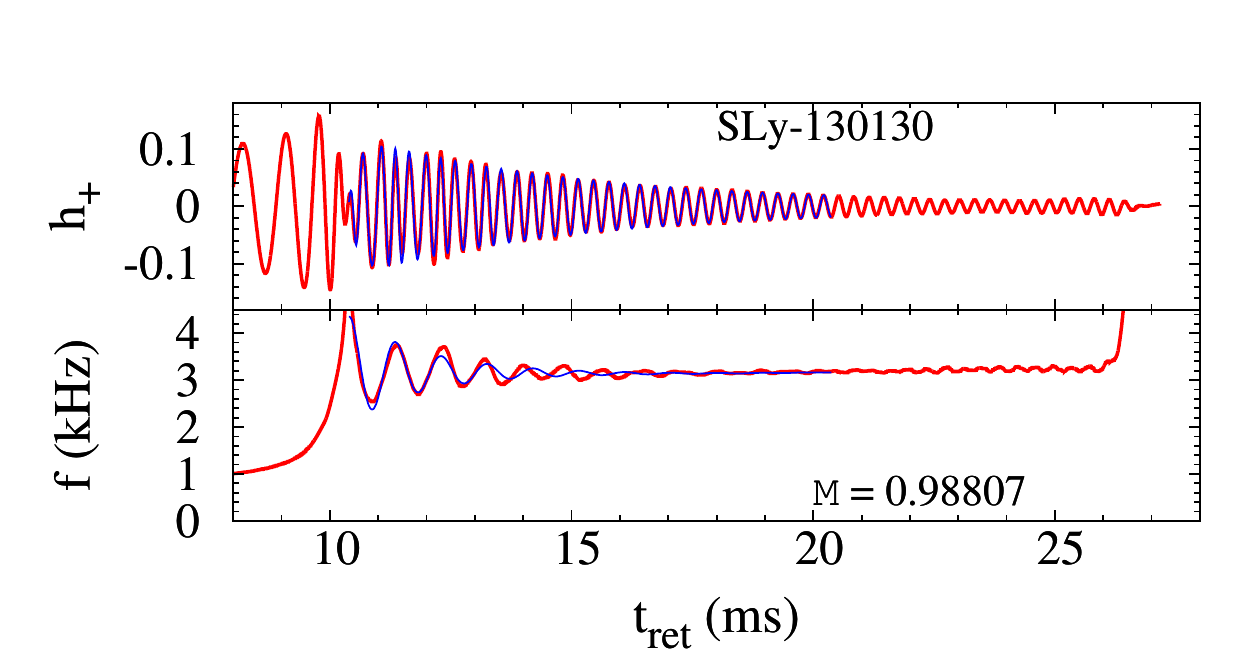} \\
\includegraphics[width=80mm,clip]{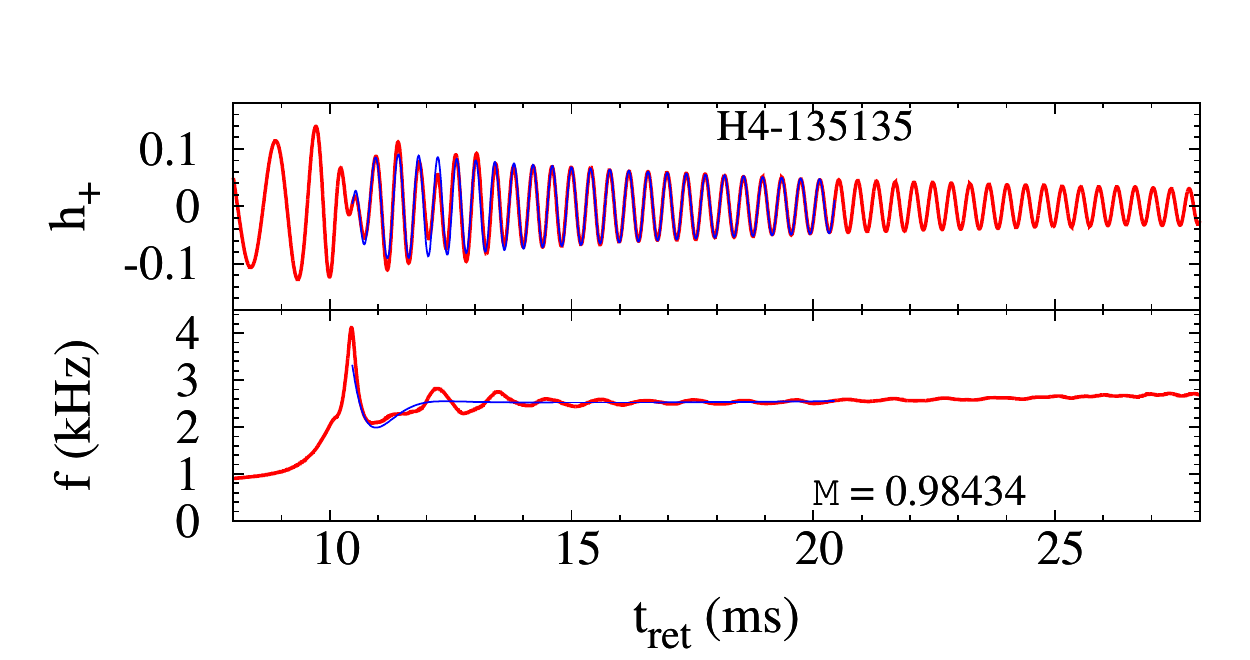} 
\includegraphics[width=80mm,clip]{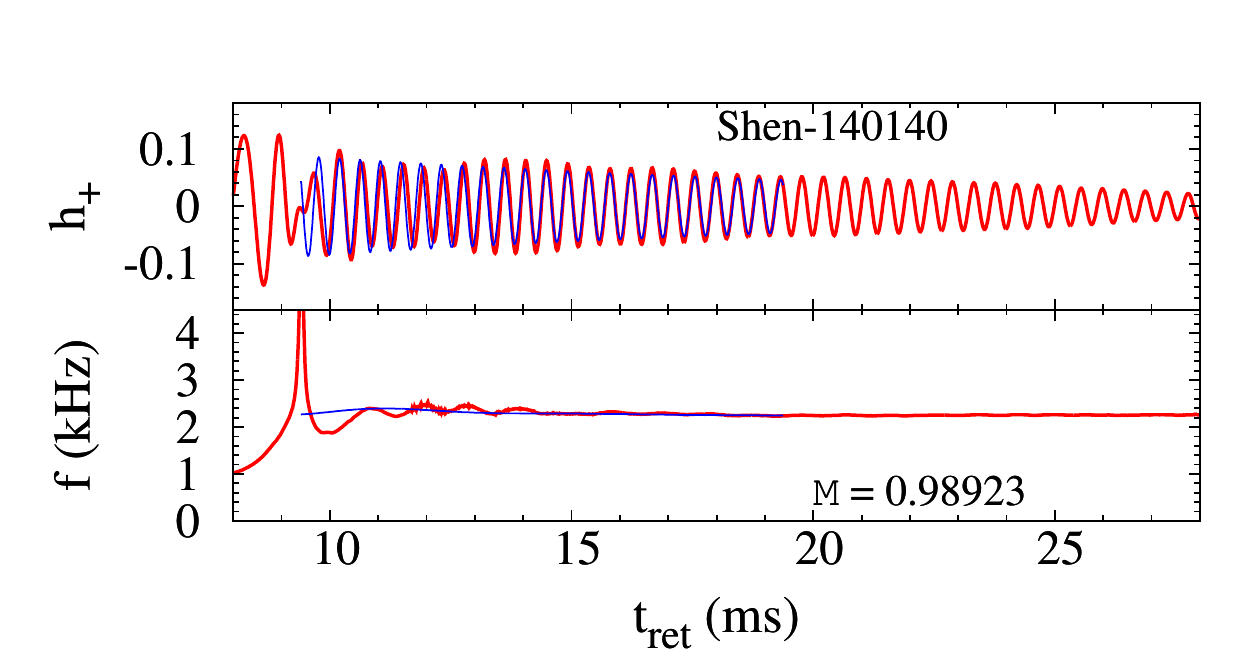} 
\caption{Comparison of numerical waveforms with their fitting results
for APR4-120140, SLy-130130, H4-135135 with $\Gamma_{\rm th}=1.8$, and
Shen-140140.  Gravitational waves ($h_+ D/m$) and the frequency of
gravitational waves $f$\,(kHz) as functions of retarded time are
plotted. The red curves are numerical waveforms, as in
Fig.~\ref{figGW1}, while the blue segment curves are the corresponding
fitting functions with the best-fit parameters. The numerical value
written in the lower right corner of each panel is the value of $\matching$. }
\label{figCompare1}
\end{figure*}

\begin{figure*}[t]
\includegraphics[width=80mm,clip]{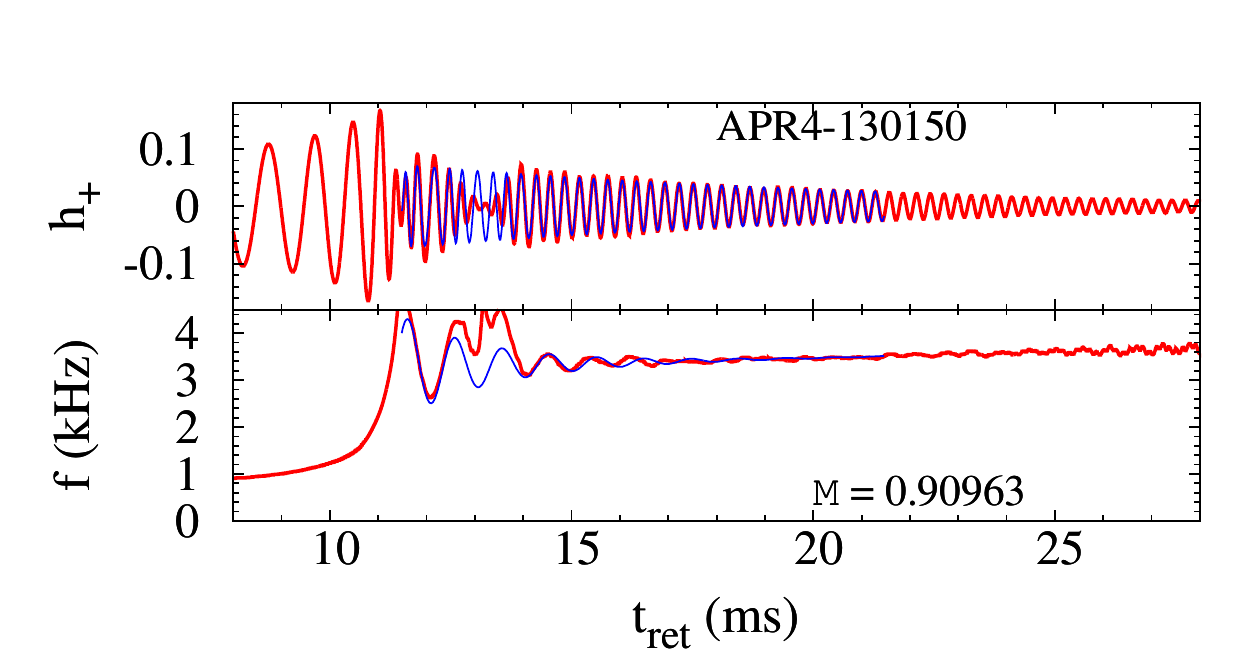} 
\includegraphics[width=80mm,clip]{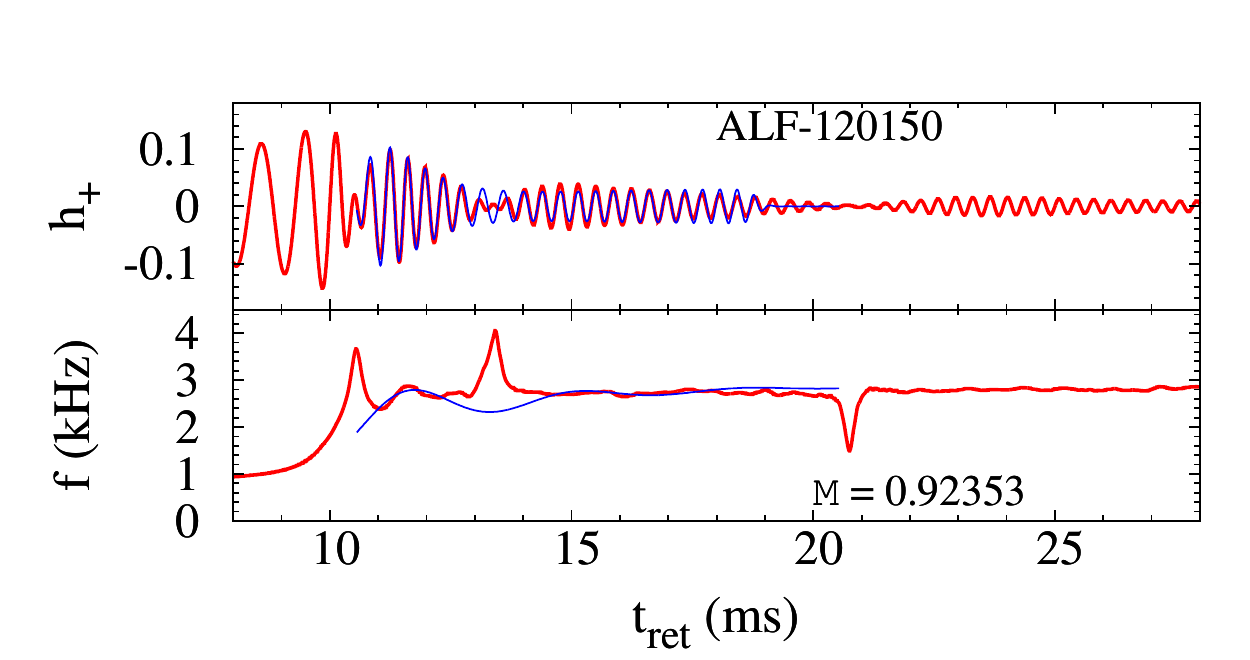} \\
\includegraphics[width=80mm,clip]{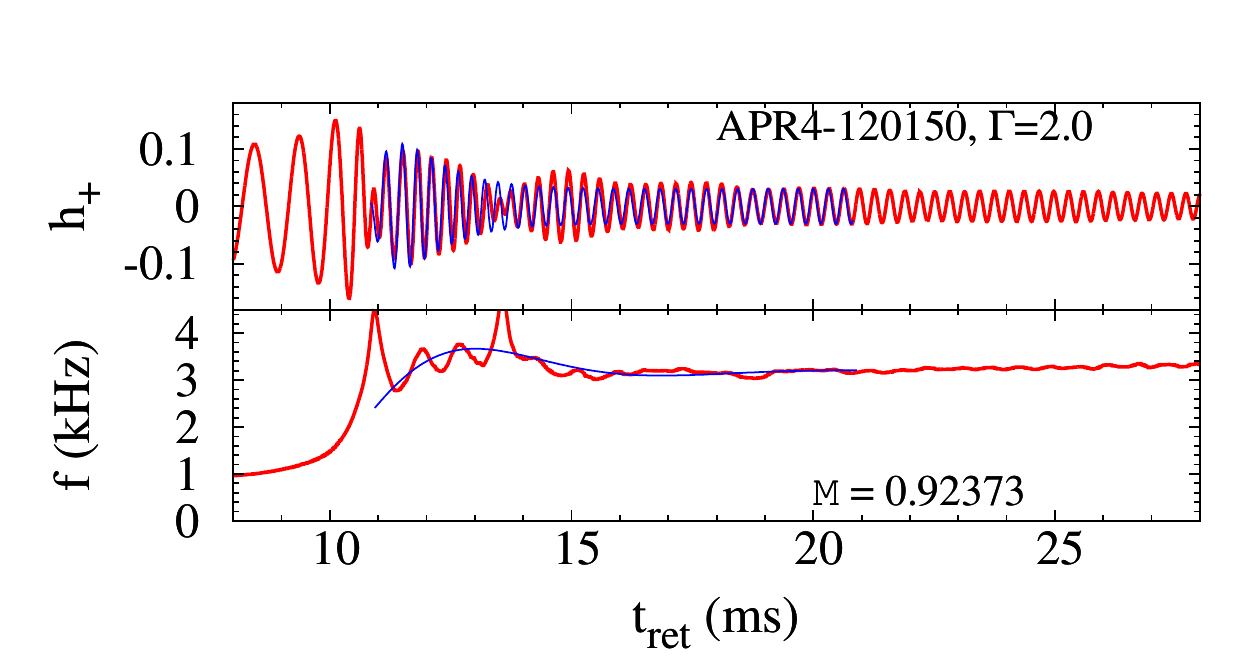} 
\includegraphics[width=80mm,clip]{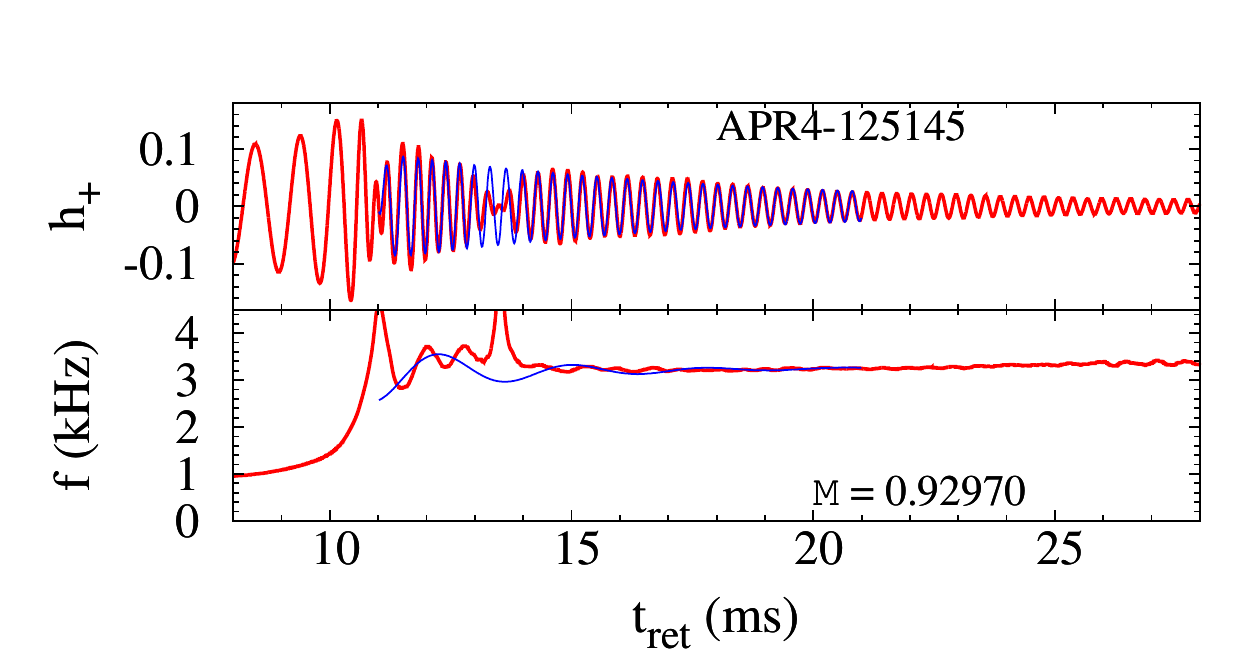}
\caption{ The same as Fig.~\ref{figCompare1} but for APR4-130150,
ALF2-120150, APR4-120150 (with $\Gamma_{\rm th}=2.0$), and
APR4-125145.  This figure shows that our fitting
formula~(\ref{eq:totalfit}) is relatively poor at fitting the
modulating feature in the amplitude.  }
\label{figCompare2}
\end{figure*}


\begin{figure}[t]
\begin{center}
\includegraphics[width=90mm,clip]{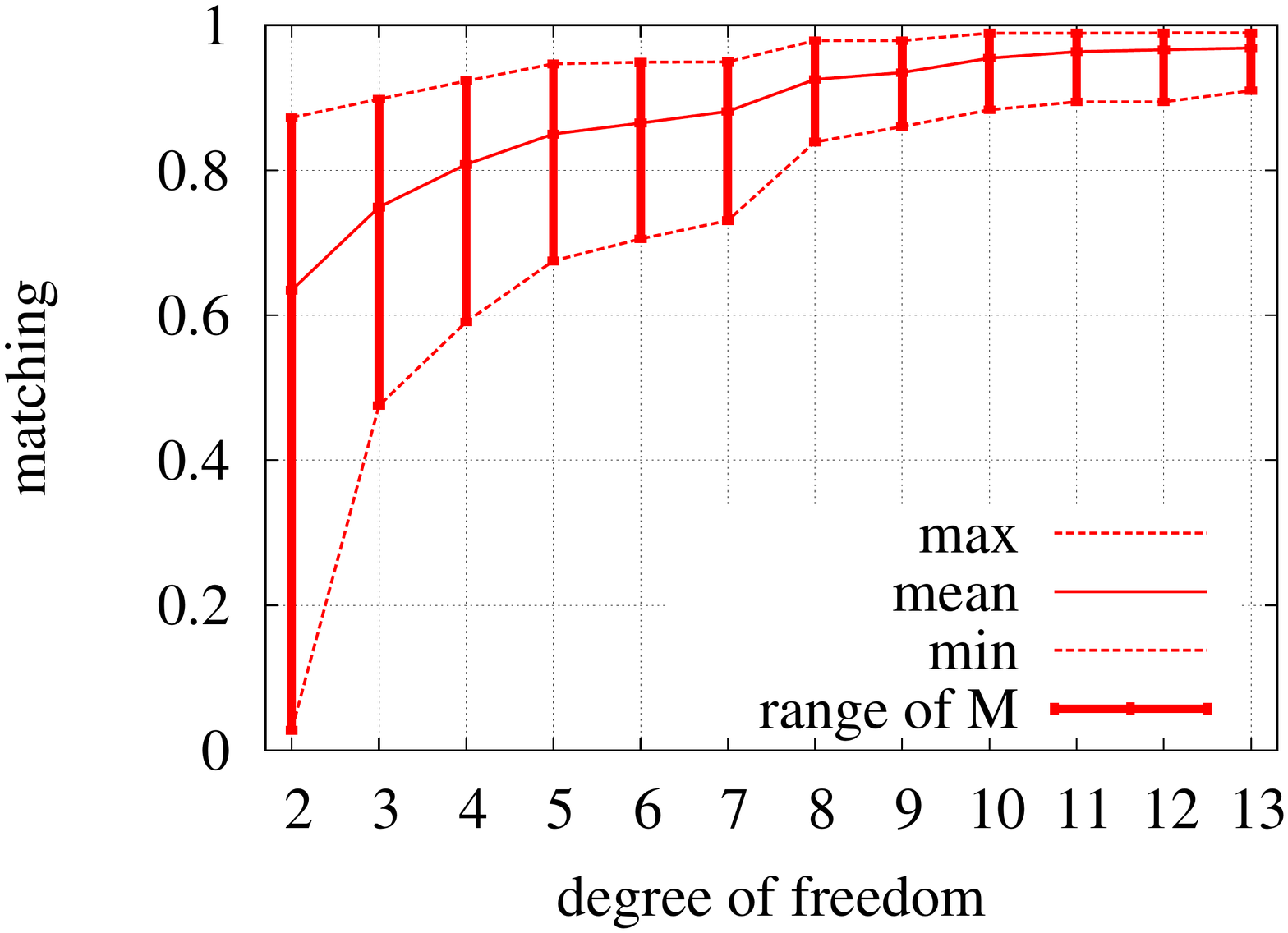}
\end{center}
\vspace{-10mm}
\caption{
Summary of the performance of the best-fitting procedures as a
function of the number of parameters.  The maximum, minimum, and
mean values of $\matching(h)$ for the best-fitting procedures are
shown.  The procedure with the largest quality $q =
\matching(\underset{h}{\rm arg~min}(h))$ is defined as the best one.
}
\label{figTradeOff}
\end{figure}


Figure~\ref{figCompare1} compares numerical waveforms with their
fitting results for four models, for all of which the value of
$\matching$ is larger than 0.98. For these models, the amplitude of
gravitational waves emitted by a MNS decreases monotonically with time
and the corresponding frequency is approximately constant only with
small modulation. For this type of the waveforms, the fitting can be
well achieved in our fitting formula.

By contrast, the fitting is not as well achieved for the case that the
amplitude significantly modulates with time in the early phase of the
MNS evolution.  Figure~\ref{figCompare2} shows the sample for such
cases.  For all the models picked up in this figure, the amplitude
increases with time from $t=t_i$ and reaches a high value. Then, the
amplitude once damps at $t-t_i=2$\,--5\,ms, and subsequently,
quasi-periodic gravitational waves with slowly decreasing amplitude
are emitted. This type of gravitational waveform is often found for
the case of asymmetric binaries with APR4 and ALF2. Nevertheless, the
value of $\matching$ is still larger than 0.9. 

We notice that the fitting formula, which are constructed with
the data for 10\,ms duration,
could reproduce longer data as well (e.g., 20\,ms data). 
Indeed,
the value of $\matching$ for 20\,ms duration data with the fitting formula
is only $1\%$ smaller than the value of $\matching$ listed in Table~\ref{table:result} 
for the compact neutron-star models, APR4, SLy, and ALF2. For the less compact neutron-star models,
the value of $\matching$ is still $1\sim10\%$ smaller than the value listed in Table~\ref{table:result}.
For the softer EOSs, the duration of high-amplitude gravitational waves 
is relatively short as $\alt 10\,ms$. Thus the longer data do not 
contribute
much to the matching. This is the reason that the value of ${\cal M}$ 
depends weakly on the duration of the data. 
For the stiffer EOSs, in particular, for H4, 
the duration of high-amplitude gravitational waves is longer as 
$\agt 10\,ms$. For such gravitational waves, the matching parameters 
have to be reconstructed for the longer-duration data. 

Before closing this subsection, we comment on the convergence of the
value of $\matching$.
For the case that the gravitational waveform has
the modulation of the amplitude, the value of $\matching$ is lower for the
higher-resolution simulations, because the modulation of the amplitude
is more distinctive for the higher-resolution simulations (see details
in Appendix A). Therefore the value of $\matching$ is likely to be
overestimated for the waveforms which have the large modulation.


\def\ttstr{\parameters '}

\subsection{Reduction of the number of the free parameters}

We now explore possible fitting formulae in which the number of free
parameters is smaller than 13; some of 13 parameters are fixed to
particular values. A question to ask in the reduction process is
whether the search output in the fitting formula composed of 13
parameters is insensitive to some of the parameters.  If this is the
case, we could effectively reduce the search space by fixing such
insensitive parameters to standard values.  In fact this is the case:
We can construct fitting formulae that preserve nearly the same
quality even when we reduce the number of free parameters to $\sim
11$, as we can see in Figs.~\ref{fig:hist} and \ref{figTradeOff}.

To access the quality of the reduced fitting formula, in this paper,
we introduce a quality function $q({\ttstr})$ for a subset of
parameters $\ttstr \subset \parameters$.  $q({\ttstr})$ is defined as
the worst value among the maximum values of $\matching$ when the
fitting is performed only for the subset of parameters $\ttstr$:
\begin{eqnarray}
  q({\ttstr}) := \min_{\rm NR~models}[\matching(h_{\rm
  fit})]. \label{eqQuality}
\end{eqnarray}
Then, we consider that the quality of a fitting formula is better if
the values of $q({\ttstr})$ is larger.

We employ a simple heuristics to reduce the number of free parameters
starting from 13 free parameters obtained in the best-fitting formula as
follows:

\begin{enumerate}
  \item
    We can construct a {\em reduced} formula with 12 free parameters,
    by fixing one of the parameters to the standard value
    (listed in Table~\ref{table:def}). 
  \item
    We measure the quality of all the reduced formulae by performing
    fitting procedure described in Sec.~\ref{sec:fittingProc} for each
    of them.
  \item
    We choose the reduced formula with the highest quality and 
    consider it as the best formula in the 12-parameter search.
\end{enumerate}

We further repeat this to construct a reduced formula with 11 free parameters,
starting from the best formula in the 12-parameter search.  In
general, we can define a reduced formula with $(N_{\rm p}-1)$ free
parameters from a $N_{\rm p}$-parameter formula, by choosing the
free parameter that has the least effect on the quality and by fixing
it. In this manner, we constructed 11 reduced formulae with $N_{\rm
p}=2$\,--\,12.

In the above heuristics, we search only for small subspaces of all the
possible parameter reduction spaces. To confirm that our choice for
the parameter reduction would be the best one, we also tested other
heuristics for reducing the number of parameters.  We also tried
several different formulae for the same choice of $\ttstr$.

 Also, it was sometimes found that for certain waveforms increase in
 $N_{\rm p}$ could result in the decrease of $\cal M$. This may seem
 counter-intuitive, but in reality such a miss-fitting is inevitable.
 This is because $\cal M$ is assured to be the increasing function of
 $N_{\rm p}$ only if the global optima are always obtained. However,
 in practice, a fake solution, which falls into a local minimum, is
 often obtained, and hence, it is not easy to achieve the global
 optimization.  When we found such decrease in $\cal M$, we performed
 the fitting for a larger value of $N_{\rm p}$ again, using the
 results of fitting obtained from a lesser value of $N_{\rm p}$ as the
 initial guess.

Update in the standard set of parameters may also result in the
decrease of the quality for some of the reduced formulae. In such
cases we can also re-use the results from the past overwriting the
fitting results for all the waveforms consistently. The number of the
degrees of freedom is still $N_{\rm p}$ after such operations. We can
understand this visually in the following way: What we call a reduced
formula corresponds to an $N_{\rm p}$-dimensional plane embedded in
the 13-dimensional parameter space, and locally adopting a different
``standard set of parameters'' corresponds to a parallel displacement
of the $N_{\rm p}$-dimensional plane in the 13-dimensional space.


\subsection{Results of the parameter reduction}

Table~\ref{table:result} lists the maximum values of ${\cal M}$ for
the cases of $N_{\rm p}=10$ to 13 parameters (see the columns ${\cal
M}_{N_{\rm p}=10}$ to ${\cal M}_{N_{\rm p}=13}$). This table obviously
shows that the maximum values of ${\cal M}$ are approximately
identical for $N_{\rm p}=12$ and $N_{\rm p}=13$ (except for
APR4-130150 and ALF-140140).  Therefore, the cumulative distributions
for these two cases are approximately identical as found in
Fig.~\ref{fig:hist} (see the plots for $N_{\rm p}=12$ and 13).  This
implies that the search procedure may well be performed in the fitting
formula with 12 free parameters fixing $a_{\rm co}$, which might be a
redundant parameter. 



\begin{table*}[t]
\begin{tabular}{c|cccccccc|ccccc|c}    \hline
  & \multicolumn{8}{c|}{phase part} & \multicolumn{5}{c|}{amplitude part} & quality\\
  \hline
 $N_{\rm p}=2$  & & & & & & &$p_1$&$p_0$& & & & & &  0.0274881661358768 \\                                                                                      
 $N_{\rm p}=3$  & & & & & & &$p_1$&$p_0$& & & & $a_{\rm d}$& &  0.4753105281276069 \\                                                                              
 $N_{\rm p}=4$  & & &$p_{\rm c}$& & & &$p_1$&$p_0$& & & & $a_{\rm d}$& &  0.5908178259595284 \\                                                                    
 $N_{\rm p}=5$  & & &$p_{\rm c}$& & & &$p_1$&$p_0$&$a_{\rm ci}$& & & $a_{\rm d}$& &  0.6750096266165332 \\                                                         
 $N_{\rm p}=6$  & & &$p_{\rm c}$& & & &$p_1$&$p_0$&$a_{\rm ci}$& & & $a_{\rm d}$&$a_1$&  0.7053957819403283 \\                                                     
 $N_{\rm p}=7$  & & &$p_{\rm c}$& & &$p_2$&$p_1$&$p_0$&$a_{\rm ci}$& & & $a_{\rm d}$&$a_1$&  0.7305212430130907 \\                                                 
 $N_{\rm p}=8$  & $p_{\rm d}$& &$p_{\rm c}$& & &$p_2$&$p_1$&$p_0$&$a_{\rm ci}$& & & $a_{\rm d}$&$a_1$&  0.8388076222568087 \\                                       
 $N_{\rm p}=9$  & $p_{\rm d}$& &$p_{\rm c}$& & &$p_2$&$p_1$&$p_0$&$a_{\rm ci}$& &$a_0$& $a_{\rm d}$&$a_1$&  0.8600728396291507 \\                                   
 $N_{\rm p}=10$ & $p_{\rm d}$&$p_{\rm f}$&$p_{\rm c}$& & &$p_2$&$p_1$&$p_0$&$a_{\rm ci}$& &$a_0$& $a_{\rm d}$&$a_1$&  0.8831675719064657 \\                         
 $N_{\rm p}=11$ & $p_{\rm d}$&$p_{\rm f}$&$p_{\rm c}$& &$p_3$&$p_2$&$p_1$&$p_0$&$a_{\rm ci}$& &$a_0$& $a_{\rm d}$&$a_1$&  0.8943981384792208 \\                     
 $N_{\rm p}=12$ & $p_{\rm d}$&$p_{\rm f}$&$p_{\rm c}$&$p_{\rm s}$&$p_3$&$p_2$&$p_1$&$p_0$&$a_{\rm ci}$& &$a_0$& $a_{\rm d}$&$a_1$&  0.8943981384792208 \\           
 $N_{\rm p}=13$ & $p_{\rm d}$&$p_{\rm f}$&$p_{\rm c}$&$p_{\rm s}$&$p_3$&$p_2$&$p_1$&$p_0$&$a_{\rm ci}$&$a_{\rm co}$&$a_0$& $a_{\rm d}$&$a_1$&  0.9096283876761225 \\
    \hline
  \end{tabular}
\caption{The sets of free parameters in the chosen fitting formulae
with $N_{\rm p}=2$\,--\,13. The value of $\matching$ is not very
sensitive to $a_{\rm co}$ and $p_s$ if the standard values listed in
Table~\ref{table:def} are assigned. }

  \label{tblTradeOffOrder}

\end{table*}

We further perform the reduction, and construct reduced fitting
formulae with 2 to 11 free parameters ($N_{\rm p}=2$\,--\,11).  We
list the sets of the parameters for $N_{\rm p} = 2$\,--\,13 in
Table~\ref{tblTradeOffOrder}. Figure~\ref{fig:hist} shows the
cumulative distribution of $\matching$ for the formulae with $N_{\rm
p}=7$\,--\,13.  Figure~\ref{figTradeOff} shows the qualities of
reduced formulae as a function of the number of parameters. This
figure indicates that the reduced formula with $N_{\rm p}=11$\,--\,13
have approximately the same quality, giving $\matching>0.9$ for more
than $98\%$ of the waveforms.  The reduced formula with $N_{\rm
p}=8$\,--\,10 have gradually decreasing quality, giving $\matching>0.9$
for $\agt 80\%$ of the waveforms. Then, there is a substantial
quality gap between $N_{\rm p}=7$ and $N_{\rm p}=8$. This 
occurs when $p_{\rm d}$ is fixed to its standard value.  

The last two columns of Table~\ref{table:result} also list the values
of ${\cal M}_{M_{\rm p}=11}$ and ${\cal M}_{M_{\rm p}=10}$.  Comparing
these data with ${\cal M}_{M_{\rm p}=13}$ also shows that the reduced
formula with $N_{\rm p}=11$ has the quality that is approximately the
same as that for $N_{\rm p}=13$.  This implies that we may reduce the
number of the parameters to 11 by fixing the values of the other
parameters to be the standard values.  We may further reduce the
number of free parameters to 10 if we can allow the matching with
$\matching<0.9$ for $\sim 5\%$ of the waveforms.

\section{Summary} \label{sec:summary}

The latest discoveries of high-mass neutron stars with mass $1.97 \pm
0.04M_{\odot}$~\cite{twosolar} and $2.01 \pm 0.04M_{\odot}$~\cite{twosolar2}
constrain that the maximum mass of
(cold) spherical neutron stars for a given hypothetical EOS has to be
larger than $\sim 2M_{\odot}$, and suggests that the EOS of neutron
stars has to be quite stiff. We performed a number of
numerical-relativity simulations employing stiff EOSs with a variety
of the plausible total mass and mass ratio of binary neutron stars. We
found that for the canonical total mass of binary neutron stars $m
\approx 2.7M_{\odot}$, not a black hole but a MNS is the canonical
remnant, and that for many cases, it is a HMNS. The MNSs are rapidly
rotating and nonaxisymmetric, and thus, they are often strong emitters
of quasi-periodic gravitational waves and efficiently exert the torque
to the envelope surrounding them.  We explored the evolution processes
of the remnant MNSs and found that their lifetime is much longer than
the dynamical time scale of the system $\gg 1$\,ms for most
models. Their lifetime also depends strongly on the EOSs and their
total mass, although they should always collapse to a black hole eventually
if they are hypermassive.

We classified the final fate of the MNSs by specifying what determines
their evolution time scale. There are at least four ingredients that
affect the evolution of the MNSs; gravitational-wave emission,
angular-momentum transport via a hydrodynamical process associated
with the nonaxisymmetric structure of the MNSs, angular-momentum
transport process via magnetohydrodynamical processes such as magnetic
winding and MRI, and neutrino cooling. If the gravitational-wave
emission and hydrodynamical angular-momentum transport determine the
evolution of a HMNS, its nonaxisymmetry plays a crucial role and hence
its lifetime will be short $\alt 100$\,ms.  If a HMNS is alive for a
longer time, magnetorotational processes are likely to play an
important role~\cite{DLSSS2006a}: After a substantial amount of
angular momentum is transported outward, the HMNS will collapse to a
black hole. If the system is not massive enough, the angular-momentum
transport alone is not likely to trigger the collapse to a black
hole. For such a system, neutrino cooling will play an important role
(e.g.,~\cite{skks2011a,skks2011b}). If the system is hypermassive but
the thermal pressure significantly contributes to sustaining the
self-gravity of the HMNS, the collapse will occur in the neutrino
cooling time scale of seconds.  If the system is not hypermassive but
supramassive, the SMNS will be alive for a time longer than the
cooling time scale.  Their lifetime will be determined by the
dissipation time scale of angular momentum such as magnetic dipole
radiation.

In the later part of this paper, we studied in detail the properties
of quasiperiodic gravitational waves emitted by MNSs. We found that
the gravitational waveforms well reflect the evolution process of the
MNSs. Basically, the waveforms have the following universal features;
they are quasiperiodic with an approximately constant frequency $\sim
2$\,--\,3.5\,kHz, although the frequency changes with time in
particular in the early stage of the MNSs; the time variation part of
the frequency is composed of an early high-amplitude oscillation and a
subsequent secular variation; the amplitude decreases (approximately)
monotonically with time scale $\agt 10$\,ms which is much longer than
the oscillation period and dynamical time scale of the MNSs. Taking
into account these universal features of the gravitational waveforms,
we constructed a fitting formula that is used for modeling
gravitational waves of a variety of MNSs irrespective of EOSs and the
values of binary mass.  It is found that the waveforms are well fitted
by 13 parameter models with the value of the matching factors $>0.90$
for all the waveforms and $>0.95$ for $\sim 90\%$ of the waveforms.
Even with 11 parameter models, the value of the matching factors is
larger than $0.90$ for 98\% of the waveforms and $\agt 0.95$ for $\sim
75\%$ of the waveforms.

We also found a correlation between the characteristic frequency of
gravitational waves emitted by MNSs and a neutron-star radius, as
found in~\cite{BJ2012}. However, it was also clarified that the
frequency has a systematic dispersion because it changes with time
during the evolution of the MNSs. Due to this systematic component,
the correlation relation is not as sharp as that pointed out
in~\cite{BJ2012}, and thus, we conclude that even if the
characteristic frequency is determined accurately, the systematic
error for the estimation of the neutron-star radius of $\sim 1$\,km
will be inevitable.  Nevertheless, the neutron-star radius is
constrained strongly, and therefore, measuring the characteristic
frequency is an important subject in the future gravitational-wave
observation.

\acknowledgments

We thank M. Ando, T. Nakamura, and H. Tagoshi for fruitful discussions
on the data analysis of gravitational waves.  This work was supported 
by Grant-in-Aid for Scientific Research (21340051, 21684014, 24244028,
24740163), by Grant-in-Aid for Scientific Research on Innovative Area
(20105004, 24103506), and HPCI Strategic Program of Japanese
MEXT. This work was partly supported by "Joint Usage/Research Center
for Interdisciplinary Large-scale Information Infrastructures" in
Japan. The work of Kyutoku is supported by JSPS Postdoctoral Fellowship for
Research Abroad. The work of Hotokezaka was supported by the Grant-in-Aid of
JSPS. 

\appendix

\section{Convergence}

\begin{figure}[t]
\includegraphics[width=84mm,clip]{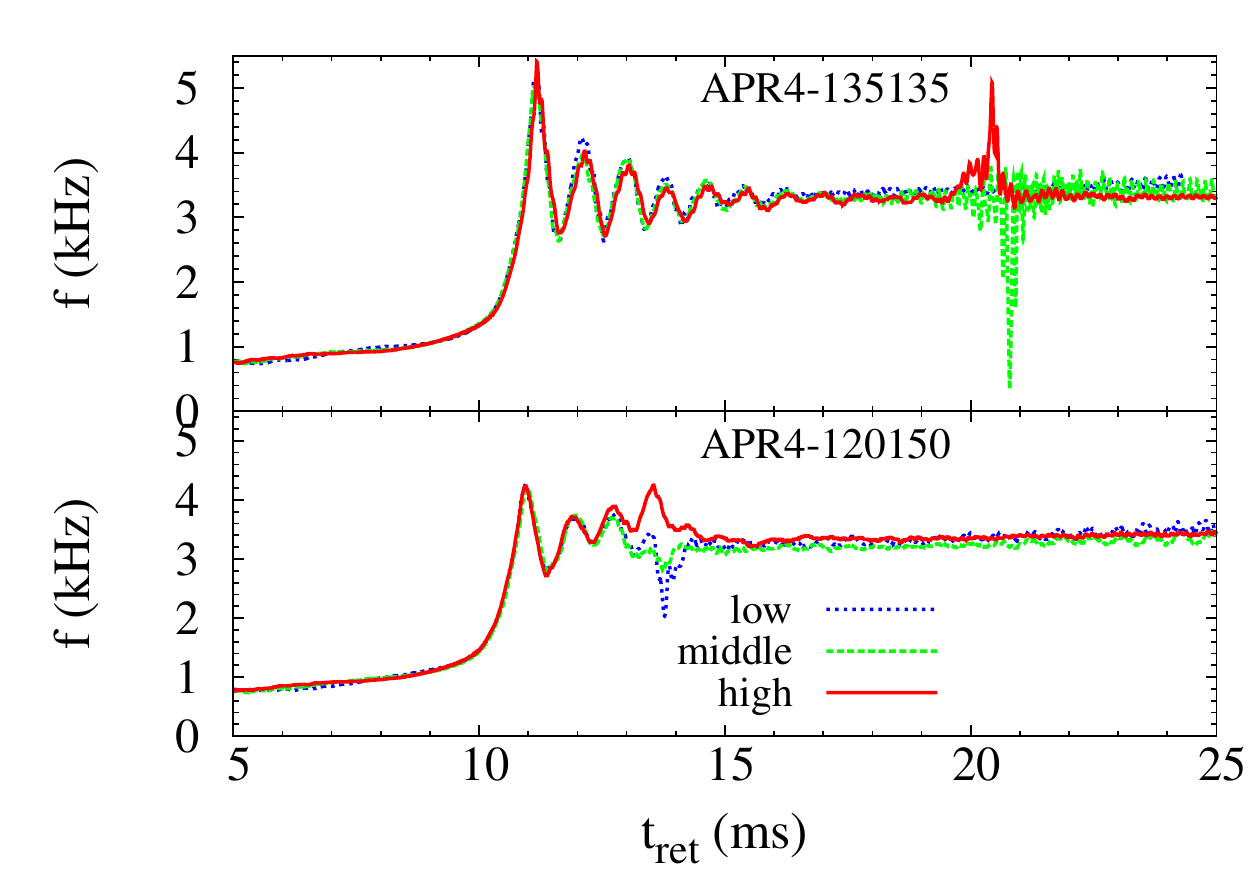}\\
\includegraphics[width=84mm,clip]{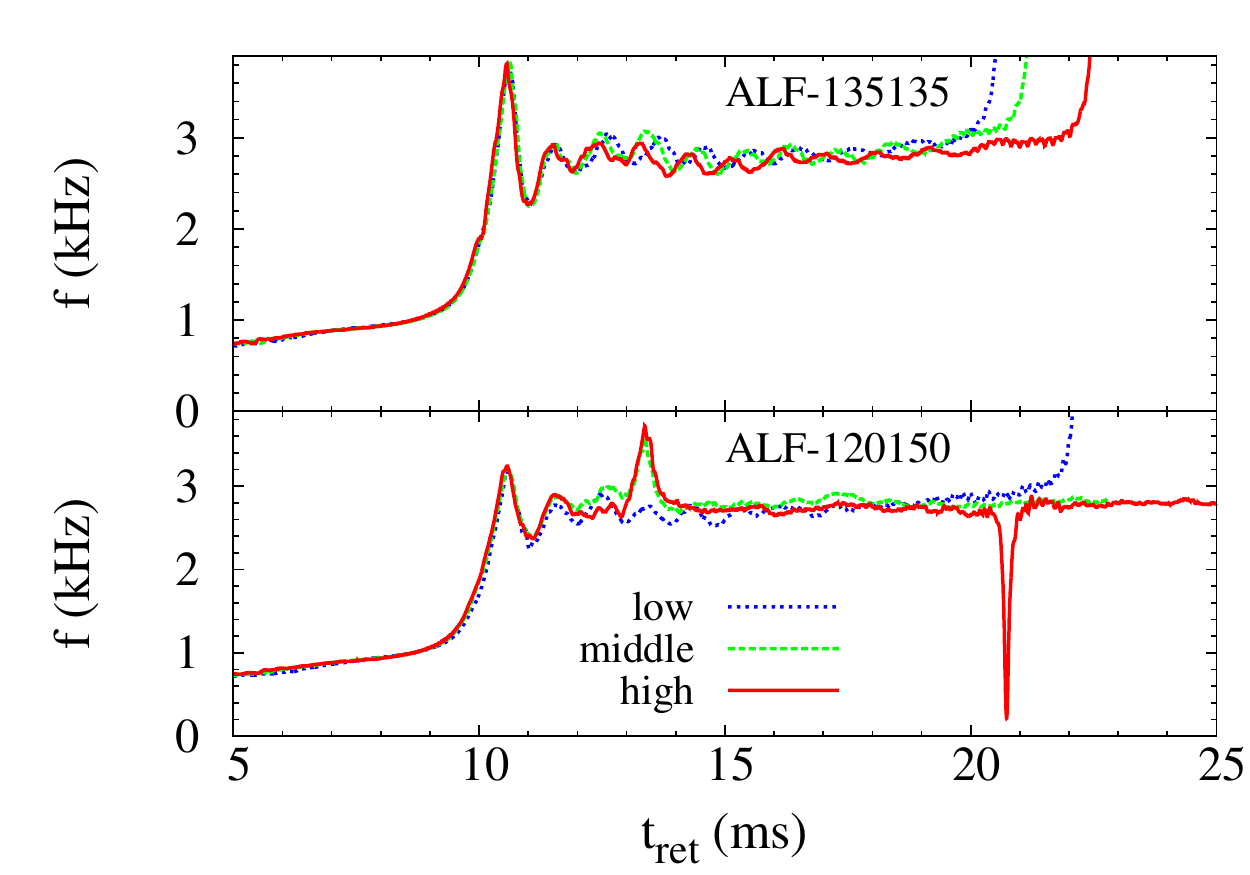}\\
\includegraphics[width=84mm,clip]{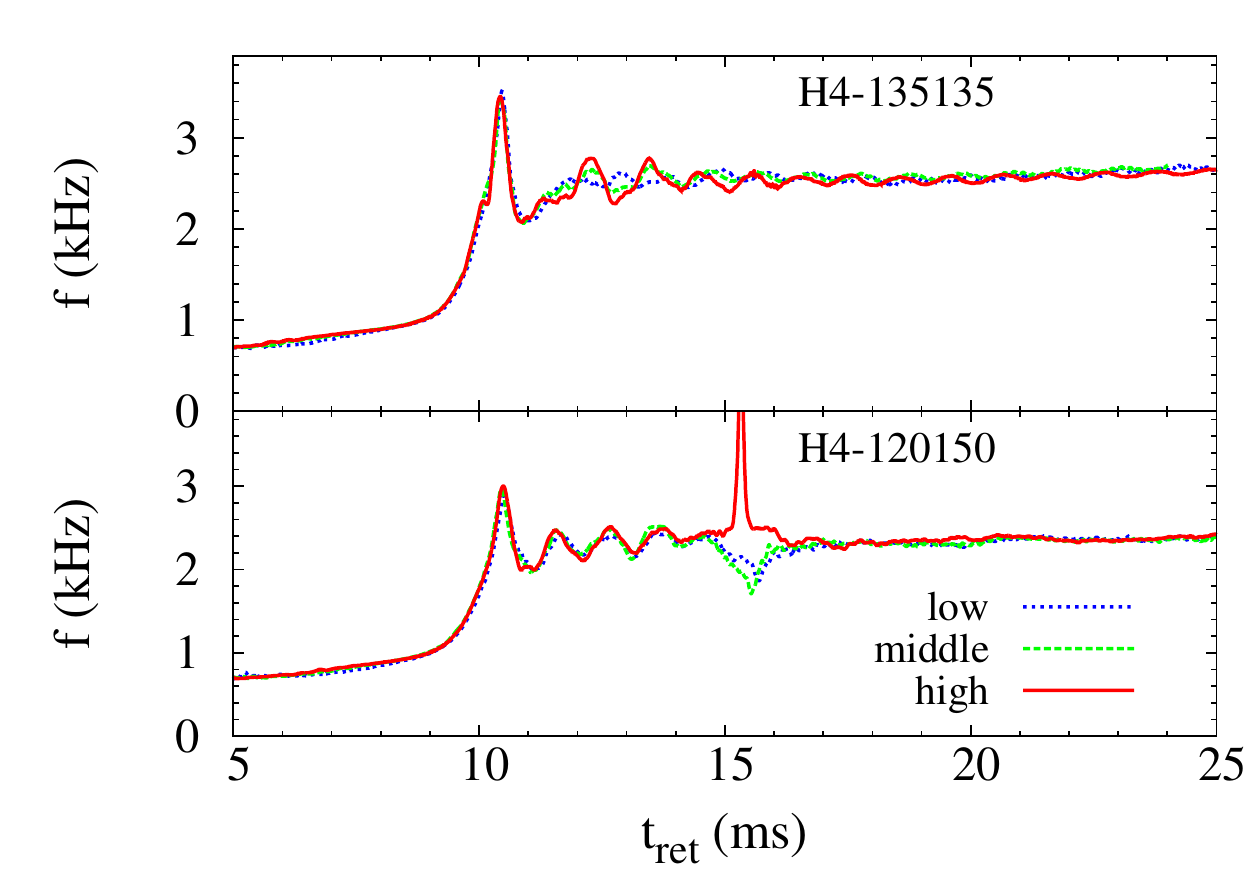}
\caption{The resolution study for the evolution of the frequency of
gravitational waves emitted by MNSs for APR4, ALF2, and H4 EOSs with
$m_1=m_2=1.35M_\odot$ and $(m_1, m_2)=(1.2M_\odot,1.5M_\odot)$.  For
each model, three grid resolutions are chosen; see Sec.~\ref{sec:ID}
and Table V of~\cite{hotoke2012}. For aligning the curve at the onset
of the merger, the time is shifted for the data of low and middle
resolutions.  Spikes found for the plots of APR4-135135, APR4-120150,
ALF-120150, and H4-120150 are not physical; these are generated when
the gravitational-wave amplitude is too low to determine the frequency
accurately.  }
\label{figapp}
\end{figure}

When we model a gravitational waveform in terms of our fitting
formula, the most important quantity is its frequency. We here show
the convergence property for the numerical results of the frequency of
gravitational waveforms emitted by MNSs. Figure~\ref{figapp} plots the
frequency as a function of $t_{\rm ret}$ for APR4, ALF2, and H4 EOSs
with $m_1=m_2=1.35M_\odot$ and $(m_1, m_2)=(1.2M_\odot,1.5M_\odot)$
for typical examples. For each model, three grid resolutions are
chosen (see Sec.~\ref{sec:ID} and Table V of~\cite{hotoke2012}). 
We also provide the average frequency and the maximum value 
of $\cal M$ for those models in Table VII.

As we described in~\cite{hotoke2012}, the peak frequency of the
Fourier spectrum and the averaged value of the
frequency converge within $\sim 0.1\,{\rm kHz}$ error and the error
is smaller than the systematic dispersion $\sigma_f$.  
The convergence for the stiffer EOS such as H4 is better than that 
for the softer EOS such as APR4. The possible reason for this is
that neutron stars for the stiffer EOSs are less compact and shock 
heating effects are weaker that those for the softer EOS. 
(Note that in the presence of shocks the convergence is achieved 
only at the first order.)
Similarly, the
values of the frequency is found to converge to this level at each
stage of MNSs except for the case that a black holes is formed; for
this case, the frequency varies steeply with time and the convergence
at a given moment is relatively poor.  

As already noted in the caption
of Fig.~\ref{figGW1}, spikes found for the plots of APR4-135135,
APR4-120150, ALF-120150, and H4-120150 are not physical; these are
generated when the gravitational-wave amplitude is quite low and hence
the frequency cannot be determined accurately.  Such spikes do not
play a serious role for determining the Fourier spectrum, averaged
frequency, and $\sigma_f$ because the gravitational-wave amplitude
is small at a moment that the spikes are generated. For instance,
the difference in the averaged frequency between H4-120150(low) and
H4-120150(high), which includes a spike, is only 2$\%$. 

For ALF, a black hole is formed in a relatively short time scale; see,
e.g., the plot of ALF-135135. This plot illustrates that for the lower
resolution, the lifetime of the HMNS is shorter. This is also the case
for ALF-120150. As pointed out in the last paragraph of
Sec.~\ref{sec:timescale}, this is an often-found property of our
numerical simulations.

The value of $\matching$ depends on the grid resolution of the simulation
(see Table~\ref{tblconv}). For equal-mass models, the value of $\matching$
varies only about 0.01 depending on the grid resolution. 
For unequal-mass models, the dependence of the value of $\matching$
on the grid resolution is stronger than that of the equal-mass models.
Moreover, the value of $\matching$ is the lowest for the waveform of the
highest-resolution simulations.
The reason is as explained in Sec.~\ref{sec:fitres}: For unequal-mass models, the gravitational-wave amplitude
modulates significantly at 2--5~ms after the merger (see e.g. Fig.~\ref{figGW2}).
The shape of the modulation is more distinctive for the waveform of
the higher resolution simulation. It is difficult for the fitting function Eq.~(15)---(17)   
to deal with such modulation in the gravitational-wave amplitude.
Therefore the value of $\matching$ is low for the highest-resolution simulations.
To increase the value of $\matching$ for these models we would have to increase
the number of the parameters of our fitting formula.

\begin{table*}[h]
\begin{tabular}{c|ccc|ccc}    \hline
   Model & $f_{\rm ave, 10 ms}^{\rm low}$\,(kHz) & $f_{\rm ave, 10 ms}^{\rm middle}$\,(kHz) 
 & $f_{\rm ave, 10 ms}^{\rm high}$ \,(kHz)  
 & ${\cal M}_{\rm low}$ & ${\cal M}_{\rm middle}$ & ${\cal M}_{\rm high}$ \\
  \hline \hline
  APR4-120150   &$3.31\pm0.23$ &$3.28\pm0.23$ &$3.41\pm0.24$ & 0.964 & 0.972 & 0.959 \\ 
  APR4-135135 &$3.40\pm0.36$ &$3.34\pm0.36$ &$3.28\pm0.37$ & 0.970 & 0.981 & 0.970 \\ \hline
  ALF2-120150   &$2.68\pm0.13$ &$2.78\pm0.15$ &$2.71\pm0.16$ & 0.991 & 0.975 & 0.924 \\
  ALF2-135135 &$2.82\pm0.21$ &$2.82\pm0.19$ &$2.76\pm0.15$ & 0.988 & 0.990 & 0.989 \\ \hline
  H4-120150     &$2.27\pm0.12$ &$2.28\pm0.14$ &$2.31\pm0.15$ & 0.986 & 0.984 & 0.964 \\
  H4-135135   &$2.51\pm0.14$ &$2.52\pm0.14$ &$2.48\pm0.16$ & 0.982 & 0.990 & 0.984 \\ 

    \hline
  \end{tabular}
\caption{The resolution study for the characteristic frequency $f_{\rm ave,10ms}$
and the maximum values of ${\cal M}$ for APR4,
ALF2, and H4 EOSs with $m_1=m_2=1.35M_\odot$ and 
$(m_1, m_2)=(1.2M_\odot,1.5M_\odot)$. Here the number of parameters
is set to be $N_{\rm p} = 13$.   }

  \label{tblconv}

\end{table*}

\end{document}